%% file: statedependent.tex
\documentclass[11pt]{article}
\usepackage{jheppubmod}
\pdfoutput=1
\usepackage{mathtools}
\mathtoolsset{showonlyrefs}
\usepackage{psfrag}
\usepackage{array}
\usepackage{amssymb}
\usepackage{amsmath}
\usepackage{amsthm}
\usepackage{graphicx}
\usepackage{caption}
\usepackage[labelsep=quad]{subcaption}
\usepackage{epstopdf}
	
\usepackage{epsfig}
\usepackage[punctsep]{collref}
\collectsep[]{;}	
\def\st{|\Psi \rangle}
\def\stl{\langle \Psi |}
\def\shil{{\cal H}_{\Psi}}
\def\id{{\cal I}}
\def\lvac{\langle 0 |}
\def\rvac{|0 \rangle}
\def\al{A}
\def\op{{\cal O}}
\def\tal{\widetilde{\al}}
\def\tb{\widetilde{b}}
\def\rst{|\Psi_0\rangle}
\def\lst{\langle \Psi_0|}
\def\shil{{\cal H}_{\Psi}}
\def\pb[#1,#2]{\{#1, #2\}}
\def\deb[#1,#2]{[#1,#2]_{\text{D.B.}}}
\def\paul{{\bf s}}
\def\pault{\widetilde{\bf s}} 
\def\paulthat{\hat{\bf s}}
\def\tO{\widetilde{\cal O}}

\def\tr{{\rm Tr}}

\def\w{\omega}

\def\ta{\widetilde{a}}

\def\Or[#1]{{\text{O}}\left({#1}\right)}
\def\dotl[#1,#2]{\left\langle #1,\, #2 \right\rangle}
\def\dotlb[#1,#2]{\left\langle #1,\, #2 \right\rangle}
\def\dotlm[#1,#2]{\left[ #1,\, #2 \right]}
\def\dotp[#1,#2]{(\vect{#1} \cdot\vect{#2})}
\def\aff[#1,#2]{\hat{#1}(#2)}
\def\nc{{\cal N}}
\def\n4sym{{\cal N}=4 SYM}
\def\>{\rangle}
\def\<{\langle}
\def\weight[#1,#2,#3]{\{(#1),#2,#3\}}
\def\ads[#1]{$\text{AdS}_{#1}$}

\newcommand{\be}{\begin{equation}}
\newcommand{\ee}{\end{equation}}
\newcommand{\ba}{\begin{align}}
\newcommand{\ea}{\end{align}}
\newcommand{\bs}{\begin{split}}
\def\sess\end{split}
\newcommand{\vect}[1]{{\boldsymbol{#1}}}

\def\paul{{\bf s}}
\def\pault{\widetilde{\bf s}} 
\title{State-Dependent Bulk-Boundary Maps  and Black Hole Complementarity}
\author[a,b]{Kyriakos Papadodimas}
\author[c]{and Suvrat Raju}
\affiliation[a]{Centre for Theoretical Physics, University of Groningen, Nijenborgh 4, 9747 AG, The Netherlands.}
\affiliation[b]{Theory Group, Physics Department, CERN, CH-1211 Geneva 23,
Switzerland.}
\affiliation[c]{International Centre for Theoretical Sciences, Tata Institute of Fundamental Research, IISc Campus, Bengaluru 560012, India.}
\emailAdd{kyriakos.papadodimas@cern.ch}
\emailAdd{suvrat@icts.res.in}

\date{}

\abstract{
We provide a simple and explicit construction of local bulk operators
that describe the interior of a black hole in the AdS/CFT correspondence. The existence of these operators is predicated on the assumption that the 
mapping of CFT operators to local bulk operators depends on the state of the CFT. We show that our construction leads to an exactly local effective
field theory in the bulk. Barring the fact that their charge and energy
can be measured at infinity, we show that the commutator of local operators inside and outside the black hole vanishes exactly, when evaluated within correlation functions of the CFT. Our construction leads to a natural resolution of the strong subadditivity paradox of Mathur and Almheiri et al. Furthermore, we show how, using these operators, it is possible to reconcile small corrections to effective field theory correlators with the unitarity of black hole evaporation. We address and resolve all other arguments, advanced in arxiv:1304.6483 and arxiv:1307.4706, in favour of structure at the black hole horizon.  We extend our construction to states that
are near equilibrium, and thereby also address the ``frozen vacuum'' objections of arxiv:1308.3697. Finally, we explore an intriguing link between our construction of interior operators and Tomita-Takesaki theory.}

\begin{document}
\maketitle
\input{s_intro.tex}
\input{s_needmirror.tex}
\input{s_constructtilde.tex}

\input{s_resolvepar.tex}
\input{s_noneqscen.tex}
\input{s_tomitatakesaki.tex}
\input{s_conclusions.tex}

\section*{Acknowledgments}

\noindent We would like to thank Ahmed Almheiri, Ofer Aharony, Steven Avery, Souvik Banerjee, Jose Barbon, Jan de Boer, Ramy Brustein, Jan-Willem Bryan, Borun Chowdhury, Bartlomiej Czech, Frederik Denef, James Drummond, Roberto Emparan, Ben Freivogel, Cesar Gomez, Rajesh Gopakumar, Daniel Harlow, Tom Hartman, Gary Horowitz, Norihiro Iizuka, Irfan Ilgin, Dileep Jatkar, 
Jared Kaplan, Andreas Karch, Elias Kiritsis, Hong Liu, R. Loganayagam, Javier Magan, Gautam Mandal,  Samir Mathur, Shiraz Minwalla, Juan Maldacena, Don Marolf, Tim Morris, Rob Myers, Boris Pioline, Andrea Puhm, Joe Polchinski, Eliezer Rabinovici, Slava Rychkov, Shubho Roy,  Aninda Sinha, Ashoke Sen, Kostas Skenderis, Marika Taylor, Sandip Trivedi, David Turton, Erik Verlinde, Herman Verlinde,  Mark van Raamsdonk, Spenta Wadia, Amos Yarom, Xi Yin and especially Lubo\v{s} Motl for very helpful discussions. We are grateful to Rajesh Gopakumar, Shiraz Minwalla, Lubo\v{s} Motl, and Spenta Wadia for comments on a draft of this manuscript.
K.P. and S.R. are grateful to the Seventh regional meeting in String Theory (Crete) for hospitality.  S.R. is grateful to the participants of the
``Cosmological Frontiers in Fundamental Physics'' conference at Perimeter Institute, the ``Fuzz, Fire, or Complementarity'' workshop at KITP (Santa Barbara), and ICTS discussion meeting on the ``Information Paradox, Entanglement And Black Holes'' for useful discussions.  
S.R. is also grateful to the Perimeter Institute, HRI (Allahabad), Princeton University, KITP (Santa Barbara), UCLV (Santa Clara) and ICIMAF (Havana) for their hospitality, while this work was being completed. S.R. is partly supported by a Ramanujan fellowship of the Department of Science and Technology. K.P. is grateful to the organizers and participants of the  XVIII IFT Xmas Workshop (Madrid), the ``Black hole horizons and quantum information'' TH-institute at CERN, the January 2013 Discussion Meeting on String theory in TIFR (Mumbai), the 13th Itzykson Conference (Saclay), the GR20-Amaldi10 conference (Warsaw), the Spanish-Portuguese Relativity meeting 2013 (Benasque). K.P. would like to thank 
the Perimeter Institute, DAMTP (Cambridge), the University of Padova, the University of Southampton for hospitality during the completion of this work. K.P. acknowledges support from the Royal Netherlands Academy of Sciences (KNAW).

\appendix
\addtocontents{toc}{\protect\setcounter{tocdepth}{1}}
\addtocontents{lof}{\protect\setcounter{tocdepth}{1}}

\input{app_alternatepure.tex}

\input{app_choosegauge.tex}
\input{app_measurement.tex}
\input{app_tomitatakesaki.tex}
\input{app_program.tex}

\bibliographystyle{JHEP}
\bibliography{references}

\end{document}

%% file: s_intro.tex
\section{Introduction}

In a previous paper \cite{Papadodimas:2012aq}, we proposed a holographic description of the interior of black holes in anti-de Sitter space (AdS). In this paper we expand on several aspects of our proposal and address the information paradox for black holes in AdS  in the light of the extensive recent 
discussion on the firewall proposal \cite{Almheiri:2012rt, Bousso:2012as, Nomura:2012sw, Mathur:2012jk, Chowdhury:2012vd, Susskind:2012rm, Bena:2012zi, Giveon:2012kp, Banks:2012nn, Ori:2012jx, Brustein:2012jn, Susskind:2012uw, Marolf:2012xe,  Hossenfelder:2012mr, Nomura:2012cx, Hwang:2012nn, Avery:2012tf, Larjo:2012jt, braunstein2009v1, Rama:2012fm, Page:2012zc, Nomura:2012ex, Saravani:2012is, Jacobson:2012gh, Susskind:2013tg,  Kim:2013fv, Park:2013rm, Hsu:2013cw, Giddings:2013kcj,  Lee:2013vga, Avery:2013exa,  Brustein:2013xga,  Kang:2013wda, Brustein:2013qma, Chowdhury:2013tza, Maldacena:2013xja, Page:2013mqa,  Axenides:2013iwa,  Gary:2013oja, Chowdhury:2013mka, delaFuente:2013nba, Almheiri:2013wka, Barbon:2013nta, Nomura:2013gna, Lloyd:2013bza, Hsu:2013fra, Page:2013dx,Giddings:2013noa,Brustein:2013ena,Mathur:2013qda, Verlinde:2013vja,Verlinde:2013uja,Verlinde:2012cy,Harlow:2013tf,  Mathur:2013gua, Almheiri:2013hfa,VanRaamsdonk:2013sza,Marolf:2013dba}. 

The central point that we wish to make in this paper is that the assumption that gravity can be described in a unitary quantum mechanical framework is consistent with {\em the existence of operators $\phi_{\text{CFT}}(x)$
labeled by a point $x$ that can be interpreted as a spacetime point, and low point correlation functions $\langle \Psi| \phi_{\text{CFT}}(x_1) \ldots \phi_{\text{CFT}}(x_n) | \Psi \rangle$, in the black hole state $|\Psi \rangle $ that can be understood as coming from effective field theory.} These low point correlators are the natural observables for a low-energy observer. 
However, if we take the number of points $n$ to scale with the central
charge of the boundary CFT, $\nc$, or take two points to be very close (comparable to $l_{\text{pl}}$), then this effective spacetime description may break down.   
Nevertheless, this breakdown is not
consequential for a low-energy observer, and does not imply the existence of firewalls or fuzzballs, or require any other construction
that radically violates semi-classical intuition.  

A key feature of our description of local operators in this paper is that 
mapping between CFT operators to the bulk-local operator $\phi_{\text{CFT}}(x)$ depends
on the state of the CFT. This is not a violation of quantum mechanics: the operator $\phi_{\text{CFT}}(x)$ is an ordinary operator that maps states to states in the Hilbert space. However, it has a useful physical interpretation
as a local operator only in a given state. 
Said another way, the analysis in this paper relies on the assumption that to obtain a convenient description of the 
physics, in terms of a local spacetime, we need to use different
operators in different states. This issue is related to the issue of
whether it is possible to have ``background independent'' local operators
in quantum gravity. If one gives up the idea of ``background independence'', one is naturally led to the ``state-dependent'' constructions that we discuss here. 


Nevertheless, granting this assumption, we show that our construction resolves all the arguments that have been advanced to suggest that the black hole horizon has structure, or that AdS/CFT does not describe the interior of the black hole. 

In our previous paper \cite{Papadodimas:2012aq}, we had proposed a construction of interior operators by positing a decomposition of the CFT Hilbert space into ``coarse'' and ``fine'' parts. In this paper, we present a refinement of our proposal that does {\em not} rely on any such explicit 
decomposition, although it reduces to our previous proposal in simple cases. The feature of state-dependence of the interior operators carries over from \cite{Papadodimas:2012aq}. But our refined construction removes some of the ambiguity inherently present in our previous proposal, and allows us to write down an explicit formula for interior operators in the CFT, without necessarily understanding the detailed structure of its Hilbert space at strong coupling.

The thrust of our paper is rather simple to summarize. First, we point
out that the issue of 
whether there is structure at the horizon of the black hole, and the
related issue of whether the black hole interior is visible in the CFT,
can be translated to a simple question about CFT operators. It is well
known that local-operators outside the black hole horizon can approximately be mapped to modes of single-trace operators on the boundary, which
we call ${\op}_{\w_n,\vect{m}}^i$, where $i$ labels the conformal
primary and $\w_n,\vect{m}$ are its modes in frequency-space and the angular momentum  on the spatial sphere. To 
describe a smooth interior, we need to effectively ``double'' these
modes and find another set of operators $\tO^i_{\w_n,\vect{m}}$, which
not only commute with the original operators, but are entangled with them
in the state of the CFT. So, within low-point correlators, where
the number of insertions of single-trace operators does not scale
with the central charge $\nc$ ($\nc \propto N^2$ in ${\cal N} = 4, SU(N)$ theory), we require
\be
[\tO^{i_1}_{\w_1,\vect{m}_{1}}, {\op}^{i_2}_{\w_2,\vect{m}_{2}}] {\op}^{i_3}_{\w_3,\vect{m}_{3}} \ldots {\op}^{i_K}_{\w_K, \vect{m}_{K}} |\Psi \rangle = 0, \quad \tO^{i_1}_{\w_1,\vect{m}_{1}} |\Psi \rangle = e^{-{\beta \omega_1 \over 2}} {\op}^{i_1}_{-\w_1, -\vect{m}_{1}} |\Psi \rangle.
\ee

Several authors have pointed out that the CFT does not seem to have enough ``space'' for the existence of the $\tO$ operators. However, our punch-line is as follows. In a
{\em given} state $|\Psi \rangle$, the equations above must hold provided
we do not have too many operator insertions and $K \ll \nc$. The set of 
all possible such insertions is finite, and loosely speaking,
scales like $\nc^K$. So, demanding that $\tO$
has the correct behaviour within low point correlators computed
in a given state simply leads to a set of linear equations for the $\tO$-operators, 
which can be solved in the large Hilbert space of the CFT, which has
a size that scales like $e^{\nc}$ for energies below $\nc$. 
Moreover, as we discuss in detail, these equations are consistent precisely when $|\Psi \rangle$ is close to
being a thermal state.\footnote{In this paper, by ``thermal state'' we mean a typical pure state in the high temperature phase of
the gauge theory.}

This analysis leads to our conclusion that it is possible
to find state-dependent local operators in the bulk that commute
with the local observables outside the horizon. We then proceed to
show that this construction  resolves {\underline{all}} the recent paradoxes associated with black hole information.

First, we describe how our construction of interior operators resolves the strong subadditivity paradox. The resolution is simply that the operators
inside and outside the black hole are secretly acting on the same degrees of freedom. One of the objections to this idea of black hole complementarity, has been that naively, measurements outside the black hole would not commute with those inside. As we describe in great detail, our construction is tailored to ensure that the commutator of local operators outside and inside the black hole---and all of its powers--- vanish {\em exactly} when inserted within low-point correlators.

We turn our attention to some of the more recent arguments of \cite{Almheiri:2013hfa,Marolf:2013dba}, which suggest that the black hole interior cannot be described within the CFT. The authors of \cite{Almheiri:2013hfa} pointed out that the $\tO$ operators behind the horizon 
appear to satisfy the usual algebra of creation and annihilation operators, except that ``creation'' operator maps states in the CFT to those of a lower energy. If this were really the case, it would lead to a contradiction since the creation operator of a simple harmonic algebra always has a left-inverse, and the number of states of the CFT decrease at lower energy. 

Our construction resolves this issue, because the operators behind the horizon behave like ordinary creation and annihilation operators, only when inserted within low-point correlators. Since they satisfy the algebra only in this effective sense, and not as an exact operator algebra, there is no
contradiction with the ``creation'' operator having null vectors. 

We also address the argument of \cite{Marolf:2013dba}, which we call the $N_a \neq 0$ argument. The authors of this paper pointed out, that assuming
that the interior operators were some fixed operators in the CFT, 
the eigenstates of the number  operator for a given mode outside the
horizon would not necessarily be correlated with the eigenstates of 
the number operator for the corresponding mode inside the horizon and 
so the infalling observer would encounter energetic particles at the 
horizon. 
 However, this conclusion fails for state-dependent operators. Our interior operators are precisely designed so that, for a generic state in the CFT and its descendants that are relevant for low-point correlators they ensure 
that the infalling observer sees the vacuum as he passes through the horizon. We describe this in more detail in section \ref{resolvenaneq0}.

After having addressed these issues, we then turn to the ``theorem'' of \cite{Mathur:2009hf} that small corrections cannot unitarize Hawking radiation. We point out that our construction evades the theorem because of two features: the interior of the black hole is composed of the same degrees of freedom as the exterior, and the operators inside that are correlated with those outside depend on the state of the theory. 

This brings us to a final objection that has been articulated against
this state-dependent construction: the ``frozen vacuum'' \cite{Bousso:2013ifa,VanRaamsdonk:2013sza}. Although our construction suggests that the
infalling observer encounters the vacuum for a generic state, it is true
that there are excited states in the CFT, in which we can arrange for the
infalling observer to encounter energetic particles. Our equilibrium
construction already allows us to analyze such time-dependent processes. For example,
we can consider a time-dependent correlation function in an equilibrium state,
and our prescription provides an unambiguous answer. However, in section \ref{noneqscen}, 
we discuss how to adapt our construction to build the mirror operators
directly on non-equilibrium states.  
This extension takes advantage of the fact that it is always possible to detect deviations from thermal equilibrium by measuring low-point correlators of single-trace operators. To perform
our construction on a state that is away from thermal equilibrium, we
``strip off'' the excitations on top of the thermal state, and then perform our construction in this base state. Low point correlators in the excited state are now simply equated with slightly higher point correlators in the base state. We describe this construction in section \ref{noneqscen}.

In section \ref{tomitatakesaki} we discuss a beautiful and intriguing
connection of our construction with the Tomita-Takesaki
theory of modular isomorphisms of von Neumann algebras. We 
start this section by reviewing our construction, but from a slightly
different physical emphasis. 
We then show how our construction can be compactly phrased in the language of Tomita-Takesaki theory.  In this section, we also clearly show how our
construction of the interior in this paper reduces to our previous 
construction \cite{Papadodimas:2012aq} in simplified settings. We hope
to revisit this interesting topic again in future work.

This paper is organized as follows. In section \ref{secneedtildes}, we show
that the issue of whether AdS/CFT describes the interior in an autonomous
manner reduces to the issue of finding operators, which we call the ``mirror'' operators, with certain properties in the CFT. 
After outlining these constraints, we then explicitly construct operators in \ref{sec:three} that satisfy them, when inserted within low-point correlators. 
This central section also contains multiple examples of our construction. We show how our construction works in a general theory, in the CFT,
in a toy-model of decoupled harmonic oscillators, and also in the spin chain. 
In section \ref{resolveallpar}, we then apply this construction to 
the recent discussions of the information paradox, and find that it
successfully addresses each of the recent arguments that have been raised
in favour of structure at the horizon. In section \ref{noneqscen}, we show how to extend our construction to non-equilibrium scenarios, and thereby also resolve the issue of the ``frozen vacuum.''. In section \ref{tomitatakesaki}, 
we explore the link between our construction and Tomita-Takesaki
theory. 
Section \ref{conclusions} contains a summary, and some open questions. The Appendices contain several other details, including a discussion of one of the first ``measurement'' arguments for firewalls articulated in \cite{Almheiri:2012rt}. 

Appendix \ref{appnumerical} may be particularly interesting to the reader, who
wishes to quickly get a hands-on feel for the properties of the mirror operators that we describe. This documents a computer program (included with the arXiv source
of this paper) that numerically constructs these mirror operators in
the spin-chain toy model.
The essential ideas of this paper are summarized in \cite{Papadodimas:2013b}, and the reader may wish to consult that paper first,  and then 
turn here for details.


%% file: s_needmirror.tex
\section{Bulk Locality: Need for the Mirror Operators \label{secneedtildes}}
In \cite{Papadodimas:2012aq}, we discussed how to construct local operators
outside and inside the black hole, by using an integral transform of CFT correlators. We review this construction briefly, and explain the need for the mirror operators.

Consider a generalized free-field operator ${\op}^i(t, \Omega)$ in the conformal field theory at a point $t$ in time and $\Omega$ on the sphere $S^{d-1}$. By definition this is a conformal primary operator of dimension $\Delta$, whose correlators factorize at leading order in the ${1 \over \nc}$ expansion. 
\be
\begin{split}
\label{factorization}
&\lvac {\op}^i(t_1, \Omega_1) \ldots  {\op}^i(t_{2 n}, \Omega_{2 n}) \rvac \\ &= {1 \over 2^n} \sum_{\pi} \lvac{\op}^i(t_{\pi_1}, \Omega_{\pi_1}) {\op}^i(t_{\pi_2}, \Omega_{\pi_2}) \rvac \ldots  \lvac {\op}^i(t_{\pi_{2 n - 1}}, \Omega_{\pi_{2 n - 1}})  {\op}^i(t_{\pi_{2 n}}, \Omega_{\pi_{2 n}}) \rvac + \Or[{1 \over \nc}],
\end{split}
\ee
where $\pi$ runs over the set of permutations.

In this paper, we will be interested in fields with a dimension that is much smaller than $\nc$. 
We remind the reader that, as in our last paper \cite{Papadodimas:2012aq}, by ${\cal N}$, we are referring to the central charge of the CFT, and if the reader wishes to think about supersymmetric $SU(N)$ theory, then she may
take $\nc \propto N^2$.

Now, we take the CFT to be in a state $|\Psi \rangle$ that is in equilibrium and and has an energy $\langle \Psi |H_{\text{CFT}} |\Psi \rangle= \Or[\nc]$. We write $\Or[\nc]$ here, but to be precise, we need to 
take the energy to be much larger than the central charge so that the
theory is unambiguously in the phase corresponding to a big black hole in AdS.

The same generalized free-field now factorizes about this energetic
state as well. Moreover, at leading order in ${1 \over \nc}$, we 
expect that correlators in this state $|\Psi \rangle$ will be the 
same as thermal correlators
\be
\label{thermalapprox}
\langle \Psi | {\op}^{i_1}(t_1, \Omega_1) \ldots  {\op}^{i_n}(t_n, \Omega_n) | \Psi \rangle = Z_{\beta}^{-1}\tr\big(e^{-\beta H} {\op}^{i_1}(t_1, \Omega_1) \ldots  {\op}^{i_n}(t_n, \Omega_n)\big),
\ee
where $Z_{\beta}$ is the partition function of the CFT at the temperature $\beta^{-1}$.

As we showed in \cite{Papadodimas:2012aq} we can use the modes of this
operator to construct another CFT operator that behaves like the local field outside the black hole. The formulas of \cite{Papadodimas:2012aq} were
written for the case of the black-brane in AdS, but here we can 
write down the analogous formulas for the CFT on the sphere to avoid some infrared issues in discussing
the information paradox. 
\be
\label{cftthermal}
\phi_{\text{CFT}}^i (t,\Omega, z) = \sum_{\vect{m}} \int_{\omega>0} {d \omega \over 2 \pi}   
\,\left[ {\op}^i_{\omega, \vect{m}} f_{\omega, \vect{m}}(t, \Omega, z) + \text{h.c}.\right].
\ee
Here ${\op}_{\omega, \vect{m}}$ are the modes of the boundary operators in frequency space and on the sphere respectively, while the sum over $\vect{m}$ goes
over the spherical harmonics.\footnote{These need
to be suitably regulated in frequency space, and we discuss this 
carefully in the section \ref{tildescft}, although this issue is unimportant here.}
What this means is that if we consider the {\em CFT correlators}
\be
\label{lowptcftcorr}
\langle \Psi | \phi^{i_1}_{\text{CFT}} (t_1,\Omega_1, z_1) \ldots  \phi^{i_n}_{\text{CFT}} (t_n,\Omega_n, z_n) | \Psi \rangle,
\ee
then these CFT correlators behave like those of a perturbative field
propagating in the AdS-Schwarzschild geometry.

The analogue of \eqref{cftthermal} in empty AdS had previously been
discussed extensively in the literature \cite{Banks:1998dd, Balasubramanian:1999ri, Bena:1999jv, Hamilton:2006fh,Hamilton:2006az,Hamilton:2005ju,Hamilton:2007wj,VanRaamsdonk:2009ar,VanRaamsdonk:2010pw,VanRaamsdonk:2011zz,Czech:2012bh,Heemskerk:2012mn}.  However, in writing \eqref{cftthermal}, we pointed out that, in momentum space, it was possible to extend this construction in pure-states close to the thermal state. This relies on the fact
that thermal CFT correlators have specific properties at large spacelike momenta, 
and this observation allowed us to sidestep some of the complications that were encountered in \cite{Hamilton:2005ju}.\footnote{It is somewhat delicate to write down the position space version of \eqref{cftthermal}. This is because the position space ``transfer function'' must account for the fact
that it can only be integrated against valid CFT correlators. So, the transfer function must be understood as a distribution that acts as a linear functional on the restricted domain of multi-point CFT correlators. This leads to subtleties in writing it as a simple integral transform.  This observation has led to recent claims that the transfer function does ``not exist'' in the black hole background or, indeed, in any background with a trapped null geodesic \cite{Bousso:2012mh,Leichenauer:2013kaa}.
This statement---which simply refers to the fact explained above --- does not have any significant physical implication; 
the mapping between degrees of freedom between the bulk and the boundary
continues to exist.}

Turning now to the region behind the horizon, effective field theory
tells us that in the analogue of \eqref{cftthermal}, the CFT operator describing the interior must have the form
\be
\label{interiorexpansion}
\phi^i_{\text{CFT}}(t,\Omega,z) =
\sum_{\vect{m}} \int_{\omega>0} {d\omega  \over 2 \pi}\, \left[ {\op}^i_{\omega,\vect{m}}\,g_{\omega,\vect{m}}^{(1)}(t,\Omega,z) + \widetilde{\op}^i_{\omega,\vect{m}} \,g_{\omega,\vect{m}}^{(2)}(t,\Omega,z)+ \text{h.c.}
\right].
\ee
Here $g_{\omega, \vect{m}}^{(1)}$ are the analytic continuations of the left-moving modes from outside, to inside the black hole, while $g_{\omega, \vect{m}}^{(2)}$ are right-moving modes inside the black hole. 

These right-moving modes can be understood in several ways. In Hawking's 
original calculation \cite{Hawking:1974sw}, these modes were the very energetic modes in 
the initial data that can be propagated through the infalling matter 
using geometric optics. In terms of solving wave-equations, the $g_{\omega, \vect{m}}^{(2)}$ modes can also be obtained by analytically continuing the modes from the ``other side'' (region III) of the eternal black hole,
as we discussed in \cite{Papadodimas:2012aq}.  

However, we should caution the reader that while these physical 
interpretations are useful as mnemonics, they are both fraught with
ultra-Planckian problems. This is clear in Hawking's original interpretation, but we also note that while the analytic continuation from region III is easily performed in the free-field theory, mapping the modes at late-times in the black hole, back to region III requires us to go through the ultraviolet regime.

\begin{figure}[!h]
\begin{center}
\includegraphics[height=4cm]{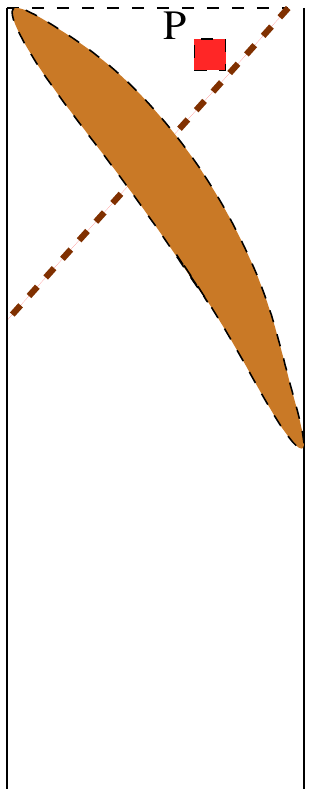}
\caption{\label{figinterestingpatch}{\em A Black Hole is created in AdS by injecting matter from the boundary. We are interested in the red-colored patch $P$, behind the horizon, which is far away
from both the infalling matter and the singularity}}
\end{center}
\end{figure}

We emphasize that neither of these ultra-Planckian problems are relevant
to our discussion. Here, our statement is simply about effective field theory in the patch $P$ that is shown in figure \ref{figinterestingpatch}. In this patch, 
we can {\em locally} expand the field in modes, and we find that to 
get a local perturbative field, we need both left and right moving modes. 
What is important here, though, is the appearance of the modes $\widetilde{\cal O}^i_{\omega_n, \vect{m}}$. First, we need these operators to {\em effectively} commute not only with the ordinary operators of the same species ${\op}^i_{\omega_n, \vect{m}}$, but with other ``species'' of operators ${\op}^j_{\omega_n, \vect{m}}$ that
enter the fields outside the horizon as well
\be
\label{commutetildeord}
[{\op}^i_{\omega_1, \vect{m}_1}, \tO^j_{\omega_2, \vect{m}_2}] \doteq 0.
\ee
The $\doteq$ in \eqref{commutetildeord} indicates that this equation
must hold when this commutator (or a power of this commutator) is inserted
within a low-point CFT correlator like \eqref{lowptcftcorr}, as we discuss in more detail below. As we have mentioned, and will
discuss again below, if we consider a correlator with $\nc$ insertions,
then we should not expect a semi-classical spacetime, or an equation like \eqref{commutetildeord} that expresses locality in such a spacetime to hold.

For the horizon of the black hole to be smooth we require that within
a low-point correlator evaluated in a pure state that is close to a thermal state
\be
\label{otildewithincorr}
\begin{split}
&\langle \Psi| {\cal O}^{i_1}(t_1,\Omega_1)\ldots\widetilde{\cal O}^{j_1} (t_1',\Omega_1')\ldots \tO^{j_l} (t_{l}', \Omega'_{l})\ldots  {\cal O}^{i_n} (t_{n}, \Omega_{n}) |\Psi \rangle \\ &= Z_{\beta}^{-1}{\rm Tr}\left[e^{-\beta H} {\cal O}^{i_1} (t_1,\Omega_1)\ldots{\cal O}^{i_n}(t_n, \Omega_n) {\cal O}^{j_{l}} (t_{l}'+i \beta/2, \Omega_{l}') \ldots {\cal O}^{j_1}\left(t_1'+i\beta /2,\Omega_1'\right)\right], \\
\end{split}
\ee
where $Z_{\beta}$ is the partition function of the CFT at temperature $\beta^{-1}$. The reader should note that the analytically continued operators,
which appear with the index $j_p$ and primed coordinates,  
have been moved to the right of all the ordinary operators, and moreover
their relative ordering has been reversed. 

In momentum space, the equation \eqref{otildewithincorr} can
be translated to
\be
\label{otildewithincorrmom}
\begin{split}
&\langle \Psi| {\cal O}^{i_1}_{\omega_1, \vect{m}_1}  \ldots \widetilde{\cal O}^{j_1} _{\omega_1', \vect{m}_1'} \ldots \widetilde{\cal O}^{j_{l}}_{\omega_{l}', \vect{m}_{l}'} \ldots {\cal O}^{i_n}_{\omega_n, \vect{m}_n}|\Psi \rangle \\ &=  e^{-{\beta \over 2} (\omega_1' + \ldots \omega_{l}')} Z_{\beta}^{-1} \tr\big[e^{-\beta H}  
 {\cal O}^{i_1}_{\omega_1, \vect{m}_1} \ldots {\cal O}^{i_n}_{\omega_n, \vect{m}_n} ({\cal O}^{j_{l}}_{\omega_{l}', \vect{m}_{l}'})^{\dagger} \ldots ({\cal  O}^{j_1}_{\omega_1', \vect{m}_1'})^{\dagger} \big].
\end{split}
\ee
In Fourier transforming from \eqref{otildewithincorr} to \eqref{otildewithincorrmom}, we should keep in mind that while the modes of ${\cal O}^i$ are
defined by $\op_{\omega, \vect{m}}^i = \int \op^i(t, \Omega) e^{i \omega t} Y_{m}(\Omega) d^{d-1} \Omega  d t$, where $Y_m$ is the spherical harmonic on the sphere, the modes of $\tilde{O}^i$ are defined by $\widetilde{O}_{\omega, \vect{m}}^i = \int \op^i(t, \Omega) e^{-i \omega t} Y_m^*(\Omega)d^{d-1} \Omega d t$. This convention simply tells us
that the modes $\widetilde{O}_{\omega, \vect{m}}^i$ have the {\em opposite} energy and angular momentum to the modes $\op_{\omega, \vect{m}}^i$.

To emphasize again, we require operators that when inserted within a state
automatically achieve the ordering within the thermal trace that we have shown here: both in terms of moving to the right of ordinary operators, and 
in terms of reversing their relative positions.

The reader may wish to consult section 5 of our previous paper \cite{Papadodimas:2012aq}, where we showed how the condition \eqref{otildewithincorr}
leads to smooth correlators across the horizon. This is clear, because
in this case, the calculation of correlators across the horizon reduces
to the calculation in the eternal black hole geometry, which is 
clearly smooth. In fact, the converse
also holds: correlators are smooth across the horizon if \eqref{otildewithincorr} holds, at least at leading order in ${1 \over \nc}$. 

\paragraph{${1 \over N}$ corrections\\}
We should point out that the status of the condition \eqref{otildewithincorrmom} (or equivalently \eqref{otildewithincorr}) is quite different
from that of \eqref{commutetildeord} with respect to  ${1 \over \nc}$ corrections. When these are included, we would like \eqref{commutetildeord} to continue to hold at all orders in the ${1 \over \nc}$ expansion and its violations, if any, should be suppressed exponentially in $\nc$. On the other hand \eqref{otildewithincorr} can 
receive corrections at the first subleading order in ${1 \over \nc}$. We can see that such corrections will come about, purely because of differences
between correlators in the  state $|\Psi \rangle$ and the thermal state.
Another source of  ${1 \over \nc}$ corrections, comes from interactions
in the CFT which, in the bulk, corresponds to the back-reaction
of the Hawking radiation on the background geometry. 

\paragraph{Charged States \\}
In writing \eqref{otildewithincorr} we have tacitly assumed that the 
state $|\Psi \rangle$ does not have any charge. In fact, the CFT
contains several conserved charges, which we will generically call $\hat{Q}$. Just as we can associate a temperature $\beta^{-1}$ with the state
$|\Psi \rangle$ using correlation functions (or the growth in entropy
with energy), we can also associate a chemical potential $\mu$ with
a charged state.

In such a state, we need to modify \eqref{thermalapprox} to
\be
\label{thermalapproxcharged}
\langle \Psi | {\op}^{i_1}(t_1, \Omega_1) \ldots  {\op}^{i_n}(t_n, \Omega_n) | \Psi \rangle = Z_{\beta, \mu}^{-1}\tr\big(e^{-\beta H - \mu \hat{Q}} {\op}^{i_1}(t_1, \Omega_1) \ldots  {\op}^{i_n}(t_n, \Omega_n)\big),
\ee
with the same modification in subsequent equations.

In this paper to lighten the notation, we will not write the charge $\hat{Q}$ explicitly. But the reader should note that our entire analysis below goes through with the replacement of $\beta H \longrightarrow \beta H + \mu \hat{Q}$.

\subsection{Comparison with Flat Space Black Holes}
We briefly mention why these mirror operators are also important
in the context of flat-space black holes. The modes in the background of a flat-space black hole have a slightly different structure. Roughly speaking, we can divide the modes into those that are ``ingoing'' and ``outgoing'' near the horizon of the  black hole, and those that are ``ingoing'' and ``outgoing'' at infinity. 

For the familiar case of a scalar field $\phi$ propagating in the 4-dimensional Schwarzschild black hole of mass $M$, we can make this precise by introducing tortoise coordinates $r_{*} = r + 2 M \ln{r - 2 M \over 2 M}$ outside the horizon, and by introducing a second Schwarzschild patch just behind the horizon.  Effective field theory tells us that, in the free-field limit, near the horizon, and at infinity, we can write 
\be
\begin{split}
\phi(r_{*}, t) &= \sum_{l,m} \int {d \omega \over 2 \pi \sqrt{\omega}} 
\left(a_{\omega, l, m} e^{i \omega(r_{*} - t) }  + b_{\w,l,m} e^{-i \omega(r_{*} + t)}  \right) Y_{l,m}(\theta,\phi) + \text{h.c.}, \quad \text{just~outside} \\
\phi(r_{*}, t) &= \sum_{l,m} \int {d \omega \over 2 \pi \sqrt{\omega}}  \left(a_{\w,l,m} e^{i \omega(r_{*} - t)  } + \widetilde{a}_{\w,l,m} e^{i \omega(r_{*} + t)}\right) Y_{l,m}(\theta,\phi) + \text{h.c.}, \quad \text{just~inside} \\
\phi(r, t) &= \sum_{l,m} \int {d \omega \over 2 \pi r \sqrt{\omega}}  \left(c_{\w,l,m} e^{i \omega(r - t)  } + d_{\w,l,m} e^{-i \omega(r + t)}\right)Y_{l,m}(\theta,\phi) + \text{h.c.}, \quad \text{at}~r \rightarrow \infty, \\
\end{split}
\ee
where ``just inside'' and ``just outside'' refers to just inside/outside the horizon.
We have taken the field to be massless, which allows both ingoing and
outgoing modes to exist at infinity for all frequencies. Note the the presence of the
potential barrier between $r = \infty$ and $r = 2 M$ implies that 
the oscillators $d$ and $a$ commute whereas the pairs $a,b$ and $c,d$
have non-trivial commutators. 
Starting with the Schwarzschild vacuum, which is defined by 
\[
a_{\omega} |S \rangle = d_{\omega} | S \rangle = \ta_{\omega} | S \rangle = 0,
\]
the Unruh vacuum is defined by allowing the ``ingoing'' modes at infinity
to remain in their ground state  and by entangling the ``outgoing'' modes at the horizon with their corresponding tilde-partners in a thermofield doubled state
\[
|U \rangle = e^{\int e^{-{\beta \omega \over 2}} a_{\omega}^{\dagger} \ta^{\dagger}_{\omega} \, d \omega} | S \rangle,
\] 
 which leads to $\langle U | a_{\omega'}^{\dagger} a_{\omega} | U \rangle = {e^{-\beta \omega} \over 1 - e^{-\beta \omega}} \delta(\omega - \omega')$.

It is in the Unruh vacuum, that the horizon is smooth. Any significant
deviations from this vacuum will generically lead to a firewall. Hence, we see that we require operators that satisfy the properties \eqref{commutetildeord} and \eqref{otildewithincorrmom}  for flat-space black holes as well to obtain a smooth horizon.

\subsection{Summary}
In this section, we have tried to argue that the issue of whether the
horizon of the black hole is smooth or not has to do with the issue of
whether we can find operators in the CFT that satisfy \eqref{commutetildeord} and \eqref{otildewithincorrmom}. {\em All} the recent discussions
of the information paradox can, essentially, be phrased as questions
about whether such operators exist. We will make this more clear when we discuss these
arguments below. In the next section, we describe how to find operators that satisfy these properties. 

We should mention that, in the argument above, we have pointed out 
the necessity of the mirror operators for generalized free-fields in the CFT that enter the modes of perturbative bulk fields. However, we will
actually succeed in finding mirror operators, for observables in a large class of statistical-mechanics systems. In the case of the CFT, we will succeed in ``doubling'' not only the generalized free-fields but a much
larger class of operators.

We should point out that there are powerful (although, in our opinion, not conclusive) arguments that suggest that one cannot find {\em fixed} (i.e. state-independent) operators that have the correct behaviour specified by \eqref{commutetildeord} \eqref{otildewithincorr} (or \eqref{otildewithincorrmom}) for  an arbitrary given state $|\Psi \rangle$.  However, if we allow the mapping between CFT operators and  local bulk operators to depend on the state itself, then one can indeed find such operators as we show explicitly below. 

Moreover, these operators then resolve {\em all} the recent paradoxes 
that have been formulated to suggest the presence of a structure at the horizon.


%% file: s_constructtilde.tex
\section{Constructing the Operators behind the Horizon \label{sec:three}}

In this section, we will explicitly construct operators behind the horizon.
We will perform this construction in three steps so as to make this section
maximally pedagogical. We start with a description of our idea in a general setting. It is well known that given a limited set of observables,
almost any pure state drawn from a large Hilbert space looks ``thermal'' or equivalently looks {\em as if} it is entangled with some environment.  In the first part of this section, we show how, in this single Hilbert space,
it is possible to construct operators that behave as if they were acting
on the environmental degrees of freedom.

In fact, the operators behind the horizon that we have described above
are precisely of this form. So, in the second and central part of this
section, we go on to describe our construction of these operators
in the CFT. This case comes with a few quirks, including the fact
that the CFT has conserved charges, and so some properties such as
the charge and energy of the mirror operators is still visible outside
the horizon. 

Finally, we descend from this complicated situation and discuss two
toy models in detail. The first is a toy-model of decoupled harmonic oscillators. This captures our ideas in a concrete setting, and has many of the essential features of the CFT, without some of the technical complications. 
The second is a simple spin-chain, which is a popular model --- and probably the simplest available one ---  for considering the information paradox. We describe how the mirror operators can be constructed in this setting
as well.

The reader may choose to read this section in any order, or even jump
directly to the toy models.

\subsection{Defining Mirror Operators  for a General Theory \label{tildegeneral}}
Let us say that we have some system, which is prepared in a pure
state $|\Psi \rangle$ drawn from a large, but finite-dimensional,  Hilbert space ${\cal H}$. We are able to probe the system with a restricted set 
of operators. Let us call the 
\be
\label{setobserv}
\text{set~of~observables:} \quad {\cal A} = \text{span}\{\al_1, \ldots \al_{{\cal D}_{\cal A}}\}.
\ee
As we have written explicitly above, ${\cal A}$ is a linear space
and we can always take arbitrary linear combinations of operators in ${\cal A}$.
However, it is important
that ${\cal A}$ may not quite be an algebra. It may be possible to multiply
two elements of ${\cal A}$ to obtain another operator that also
belongs to ${\cal A}$. In fact, we will often discuss such
products of operators below. However, we may
not be allowed to take arbitrary products of operators in this set. In 
particular, if we try and take a product of $\nc$ operators, it may
take us out of the set ${\cal A}$.

We wish to consider states $|\Psi \rangle$ that satisfy the following
very important property
\footnote{Later, in the discussion
on the CFT, we will consider situations where $|\Psi \rangle$ may be 
an eigenstate of a conserved charge, in which case \eqref{noannihilgeneral} does not hold for certain operators but, for the current discussion, this is an unimportant technicality.}
\be
\label{noannihilgeneral}
\al_p | \Psi \rangle \neq 0, ~ \forall ~ \al_p \in {\cal A}.
\ee
Note that this statement holds for all elements of ${\cal A}$, or equivalently for all possible linear combinations of the basis of observables
written in \eqref{setobserv}.  An immediate corollary of this statement is that the dimension of ${\cal A}$  be smaller than the dimension of the Hilbert space of the theory: 
\be
{\cal D}_{\cal A} \ll  \dim({\cal H}) \equiv {\cal D}_{\cal H}.
\ee
Equation \eqref{noannihilgeneral} also means that if $|\Psi \rangle$ is a state of finite
energy, then the energy of our probe operators in the algebra is
also limited. 

We wish to emphasize that these conditions on the observables we can measure and the state under consideration are physically
very well motivated. For example, if the reader likes to think of 
a spin-chain system, then ${\cal A}$ could consist of all local operators--- the Pauli spins on each site---bi-local operators---which comprise products of local operators at two sites---all the way up to
$K$-local operators, as long as $K \ll \nc$--- the length of the spin chain. 
Generic states in the Hilbert space of the spin-chain now satisfy \eqref{noannihilgeneral}. We work this spin chain example out explicitly in section \ref{dualspinchain}

However, more generally, as the reader can easily persuade herself, if we place a large system in a state that appears to be thermal, and consider some finite set of ``macroscopic observables'' (for example, those that obey the so-called ``eigenstate thermalization hypothesis''), then the condition \eqref{noannihilgeneral} is easily satisfied. In fact, we can consider a larger class of states, which are excitations of
thermal states that are out-of equilibrium.

Now, it is very well known that, given such a set of observables ${\cal A}$, and a pure state $|\Psi \rangle$, we can construct several density
matrices $\rho$, corresponding to {\em mixed states}, which are indistinguishable from $|\Psi \rangle$, in the sense that we can arrange for
\[
\tr(\rho \al_p) = \langle \Psi | \al_p | \Psi \rangle, ~\forall \al_p \in {\cal A}.
\]

Such a density matrix is not unique but the correct way to pick it, 
assuming that the expectation values of $\langle \al_p \rangle$ are all 
the {\em information} we have, is to pick the density matrix that maximizes
the entropy: $S_{\text{th}}=\text{max}\left(-\tr(\rho \ln \rho) \right)$ \cite{jaynes1957information,jaynes1957informationII}.
In fact, this maximum entropy $S_{\text{th}}$ is what should correspond
to the thermodynamic entropy of the system. 
For a generic state $|\Psi \rangle$, we expect to find 
\be
\label{thermaldens}
\rho \approx {1 \over Z} e^{-\beta H},
\ee
up to ${1 \over \nc}$ corrections, where $Z$ is the partition function.\footnote{In an equilibrium state, in any case, we expect off-diagonal terms in the
energy eigenbasis in the density matrix to be strongly suppressed, although
the eigenvalues may be corrected from the canonical ones. For the significance of such corrections, see appendix \ref{tildealternate}, and for non
equilibrium states, see section \ref{noneqscen}.}

It is also well known that the statements above imply that even though the system is in a pure state, it {\em appears} as if the system is entangled with some other heat-bath.
This pure-state in the fictitious larger system is called the ``purification'' of $\rho$. 
This purification is not unique,
even given $\rho$ but given a generic state in which the density
matrix is thermal as in \eqref{thermaldens}, we will pick it to 
be the thermofield doubled state \cite{takahashi1996thermo}.\footnote{In appendix \ref{tildealternate}, we 
discuss other choices of the purification which are, in fact,
required at ${1 \over {\cal N}}$ and this issue of the lack of uniqueness.}
\be
\label{psientangled}
|\Psi \rangle_{\text{tfd}} = {1 \over Z} \sum_{E_i} e^{-{\beta E_i \over 2}} | E_i \rangle |\widetilde{E_i} \rangle,
\ee
where the sum runs over all energy eigenvalues of the system. Note that 
the subscript $\text{tfd}$ emphasizes that this state is {\em distinct} from the pure state $|\Psi \rangle$, and 
lives in a (fictitious) larger Hilbert space.

The new point that we want to make here 
is as follows. In the pure state $|\Psi \rangle$, we can
also {\em effectively} construct the operators that act on the ``other'' 
side of the purification. So, for all practical purposes the thermofield
doubled state and the doubled operators may be realized in the
same Hilbert space!

More precisely, we want the following. For every operator 
acting on the Hilbert space of the system 
\be
\label{almatelem}
\al_p |E_i \rangle = \left(\al_p\right)_{i j} |E_j \rangle,
\ee
we have an analogous operator that acts on the fictitious environment
\be
\label{tfddualdef}
\al_p^{\text{tfd}} |\widetilde{E}_i \rangle = \left(\al_p \right)_{i j}^* |\widetilde{E}_j \rangle.
\ee
The complex conjugation is necessary to ensure that this map remains invariant if we, for example, decide to re-phase the energy eigenstates of the
system by $e^{i \phi_i}$ and those of the environment by $e^{-i \phi_i}$
under which the state \eqref{psientangled} is obviously invariant. 

The operator $\al_p^{\text{tfd}}$ has two other important properties. First, it clearly commutes with the operators $\al_m$, since these act
on different spaces
\be
\label{tfdcommutord}
[\al^{\text{tfd}}_p, \al_m] |\Psi \rangle_{\text{tfd}} = 0, \quad \forall p,m.
\ee
Second, with some simple algebra (see appendix \ref{tildealternate}) we can see that
\be
\label{tfdactpsi}
\al^{\text{tfd}}_p |\Psi \rangle_{\text{tfd}} = e^{-{\beta H \over 2}} \al_p^{\dagger} e^{\beta H \over 2} |\Psi \rangle_{\text{tfd}}.
\ee

We now desire the existence of operator $\tal_p$ that act in the
single Hilbert space ${\cal H}$ and  {\em mimic} the action of \eqref{tfdactpsi} and \eqref{tfdcommutord} while acting on the state $|\Psi \rangle$. Naively, this may seem impossible. For example, if we consider a spin-chain
and the set ${\cal A}$ comprises the set of Pauli-matrices acting on each site, then there is no operator in the Hilbert space that commutes with all the $\al_p\in {\cal A}$. 

However, as we describe here, given a state $|\Psi \rangle$, there is an elegant and almost
unbelievably simple definition of these operators! First, we need to expand the set of observables ${\cal A}$ a little so that for each $\al_p \in {\cal A}$, we adjoin to ${\cal A}$ the element $\hat{\al}_p = e^{-{\beta H \over 2}} \al_p^{\dagger} e^{\beta H \over 2}$. Next, as we mentioned above, while ${\cal A}$ may not be closed under the multiplication of arbitrary pairs, if the product $\al_{p_1} \al_{p_2} \in {\cal A}$, we may also want to include the products $\hat{\al}_{p_1}  \al_{p_2}$ and $\al_{p_1} \hat{\al}_{p_2}$.  We will call this expanded set of observables 
${\cal A}^{\text{exp}}$. If ${\cal D}_{\cal A} \ll \nc$, then the elements of this 
expanded set also satisfy \eqref{noannihilgeneral}.

We want to emphasize that the reader should not get lost in the technicalities of this ``expanded'' set. In fact, in the interesting case of the CFT below, we will see that ${\cal A}^{\text{exp}}$ coincides with ${\cal A}$. This is because in the situation
where the $\al_p$ have some definite energy $\omega_p$, these factors
simply simply invert the energy, and insert a factor of $e^{-{\beta \omega_p \over 2}}$. 

Now, we simply define the mirror operators by the following set of linear equations
\be
\label{tildedefgeneral}
\begin{split}
&\widetilde{\al}_p |\Psi \rangle = e^{-{\beta H \over 2}} \al_p^{\dagger} e^{\beta H \over 2}  |\Psi \rangle, \\
&\widetilde{\al}_p \al_m |\Psi \rangle = \al_m \widetilde{\al}_p |\Psi \rangle,
\end{split}
\ee
where $\al_p, \al_m \in {\cal A}^{\text{exp}}$.
In a given state $|\Psi \rangle$, these two lines together just correspond to $\dim({\cal A}^{\text{exp}})$ equations. Note, that we can write these two lines as the single compact equation
\be
\label{tildedefgeneralcompact}
\widetilde{\al}_p \al_m |\Psi \rangle = \al_m e^{-{\beta H \over 2}} \al_p^{\dagger} e^{\beta H \over 2}|\Psi\rangle,
\ee
but we have written them separately because, as will become clear below, the two lines of \eqref{tildedefgeneral} have different physical interpretations.

Note that $\widetilde{\al}_p$ are linear operators in a Hilbert space of dimension ${\cal D}_{\cal H}$ that we are interested in. The equation \eqref{tildedefgeneralcompact} makes it clear that we are specifying the action of these
operators on a linear subspace, ${\cal H}_{\psi} = {\cal A}^{\text{exp}} |\Psi \rangle$, produced by acting with all elements of the set ${\cal A}^{\text{exp}}$ on the set $|\Psi \rangle$. Equivalently, we are specifying
the action of $\tal_p$ on ${\cal D}_{{\cal H}_{\psi}} = \text{dim}({\cal H}_{\psi})$ basis vectors. It is {\em always} possible to specify the action of an operator on a set of linearly independent vectors that is smaller in size than
${\cal D}_{\cal H}$.

So, the only constraint we have to check is that is that the vector ${\cal A}_p |\Psi \rangle$ produced by acting on $|\Psi \rangle$ are linearly independent i.e. that we cannot find some coefficients $\sum_p \alpha_p {\cal A}_p |\Psi \rangle = 0$. However, \eqref{noannihilgeneral} tells us that
there is no such linear combination.

So, we conclude that, provided \eqref{noannihilgeneral} is met,
we can {\em always} find an operator $\tal_p$ that satisfies
\eqref{tildedefgeneral}. In fact, it is easy to write down an explicit
formula for this operator. Consider the set of vectors
\be
|v_m \rangle = \al_m |\Psi \rangle; \quad |u_m \rangle = \al_m e^{-\beta H \over 2} \al_p^{\dagger} e^{\beta H \over 2} |\Psi \rangle,
\ee
where $m = 1 \ldots \dim({\cal A}^{\text{exp}})$ and the operators
run over {\em any} basis of the set ${\cal A}^{\text{exp}}$. Now, define the ``metric''
\be
g_{m n} = \langle v_m | v_n \rangle,
\ee
and its inverse $g^{m n}$ satisfying $ g^{m n} g_{n p} = \delta^m_p$.  This inverse necessarily exists, because the $|v_m \rangle$ are linearly independent by the conditions above. Now, an
operator $\tal_p$ that satisfies the condition \eqref{tildedefgeneral} above is given by
\be
\label{talexplicit}
\tal_p = g^{m n} |u_m \rangle \langle v_n|,
\ee
where the repeated indices are summed, as usual. Of course, the operator $\tal_p + \tal^{\text{orth}}$, where $\tal^{\text{orth}}$ is any operator that satisfies $\tal^{\text{orth}} |v_m \rangle = 0, \forall m$  also satisfies \eqref{tildedefgeneral}. In \eqref{talexplicit}, we have simply taken $\tal^{\text{orth}} = 0$, but this ambiguity is physically irrelevant.

Furthermore, note that the rules \eqref{tildedefgeneral} also allow us
to build up the action of products of the mirror operators {\em recursively.} For example, notice that these rules lead to
\be
\begin{split}
&\tal_{p_1} \tal_{p_2} |\Psi \rangle = \tal_{p_1} e^{-{\beta H \over 2}} \al_{p_2}^{\dagger} e^{\beta H \over 2} |\Psi \rangle =  e^{-{\beta H \over 2}} \al_{p_2}^{\dagger} e^{\beta H \over 2} \tal_{p_1} |\Psi \rangle = e^{-{\beta H \over 2}} \al_{p_2}^{\dagger} \al_{p_1}^{\dagger} e^{\beta H \over 2} |\Psi \rangle.
\end{split}
\ee
Here in the first equality, we use the first rule of \eqref{tildedefgeneral}. In the next equality we use the second rule to commute $\tal_{p_2}$ 
to the right, and then we use the first rule again to obtain our final 
expression! Notice in particular that
\be
\tal_{p_1} \tal_{p_2} |\Psi \rangle = \widetilde{(A_{p_1} A_{p_2})} |\Psi\rangle.
\ee
Next, note that the rules \eqref{tildegeneral} lead to the result
that {\em acting on the state $|\Psi \rangle$}, the mirror operators
commute with the ordinary operators. For example, consider
the commutator of an ordinary and mirror operator within some product of ordinary and mirror operators acting on $|\Psi \rangle$
\be
\label{provecommutcorr}
\begin{split}
&\tal_{p_1} \al_{p_2} \ldots [\tal_{p_m},  \al_{p_{m+1}}] \ldots \tal_{p_{n-1}} \al_{p_n} | \Psi \rangle  \\
& =\tal_{p_1} \al_{p_2} \ldots \left( \tal_{p_m}  \al_{p_{m+1}} - \al_{p_{m+1}} \tal_{p_m} \right)  \ldots \tal_{p_{n-1}} \al_{p_n} | \Psi \rangle \\
&= \tal_{p_1} \al_{p_2} \ldots \al_{p_{m+1}}  \ldots \tal_{p_m} \tal_{p_{n-1}} \al_{p_n} | \Psi \rangle  - \langle \Psi |\tal_{p_1} \al_{p_2} \ldots \al_{p_{m+1}}  \ldots \tal_{p_m} \tal_{p_{n-1}} \al_{p_n} | \Psi \rangle  = 0.
\end{split}
\ee
Here, the key point is that the second line of \eqref{tildedefgeneral} allows us to move $\tal_{p_m}$ through
$\al_{p_{m+1}}$ and any other occurrences of $\al_p$ operators till the first occurrence of another $\tal_p$
operator. 
In writing these equations, we have tacitly assumed
that we can take the product of the operators $\al_{p_i}$, while remaining within the set ${\cal A}$. This is justified as long as $n \ll {\cal D}_{\cal A}$.

Now, we make a few remarks about correlation functions. First, note 
that by construction we have $\prescript{}{\text{tfd}}{\langle \Psi|} \al_p |\Psi \rangle_{\text{tfd}} = \langle \Psi | \al_p |\Psi \rangle, \forall
\al_p \in {\cal A}^{\text{exp}}$. Within mixed-correlators involving both $\al_p$ and $\tal_p$, we see that we have the following properties
\be
\label{rightcorr}
\langle \Psi |\tal_{p_1} \ldots \tal_{p_m}  \al_{p_{m+1}} \ldots \al_{p_n} | \Psi \rangle = \prescript{}{\text{tfd}}{\langle \Psi|} \al_{p_1}^{\text{tfd}} \ldots \al_{p_m}^{\text{tfd}} \al_{p_{m+1}} \ldots \al_{p_n} | \Psi \rangle_{\text{tfd}}. 
\ee
To show this involves only a small amount of additional work.
First, we see that 
\be
\label{detailedexpl}
\langle \Psi | \tal_{p_1} \ldots \tal_{p_m} \al_{p_{m+1}} \ldots \al_{p_n} |\Psi \rangle = \langle \Psi | \tal_{p_1} \ldots \tal_{p_{m-1}} \al_{p_{m+1}} \ldots \al_{p_n} e^{-{\beta H \over 2}} \al_{p_m}^{\dagger} e^{\beta H \over 2} |\Psi \rangle,
\ee
where we have used the second line of \eqref{tildedefgeneral} to move
the $\tal_{p_m}$ to the right, and then used the first line to substitute
its action on $|\Psi \rangle$. Now, given the right hand side of \eqref{detailedexpl}, we can use the same procedure to move $\tal_{p_{m-1}}$ to 
the extreme right and then substitute for its action. Continuing this, we see that finally 
\be
\langle \Psi | \tal_{p_1} \ldots \tal_{p_m} \al_{p_{m+1}} \ldots \al_{p_n} |\Psi \rangle = 
\langle \Psi| \al_{p_{m+1}} \ldots \al_{p_n} e^{-{\beta H \over 2}} \al_{p_m}^{\dagger} \ldots \al_{p_1}^{\dagger} e^{\beta H \over 2} |\Psi \rangle.
\ee
Now, we we discussed above, correlators of ordinary operators in the set ${\cal A}^{\text{ext}}$
in the state $|\Psi \rangle$ are the same as those in the thermofield
doubled state.  So, we find that
\be
\begin{split}
\langle \Psi | \tal_{p_1} \ldots \tal_{p_m} \al_{p_{m+1}} \ldots \al_{p_n} |\Psi \rangle &= \prescript{}{\text{tfd}}{\langle \Psi|} \al_{p_{m+1}} \ldots \al_{p_n} e^{-{\beta H \over 2}} \al_{p_m}^{\dagger} \ldots \al_{p_1}^{\dagger} e^{\beta H \over 2} |\Psi \rangle_{\text{tfd}}  \\
&= \prescript{}{\text{tfd}}{\langle \Psi|} \al_{p_{m+1}} \ldots \al_{p_n} \al_{p_1}^{\text{tfd}} \ldots \al_{p_m}^{\text{tfd}}  | \Psi \rangle_{\text{tfd}},
\end{split}
\ee
where the reader can easily use the property \eqref{tfdactpsi} 
to verify the second equality.

\paragraph{The $\doteq$ Notation: \\} This feature, where the properties of the $\tal_p$ operators hold only within correlation functions evaluated on a particular state is important enough that we will introduce some special notation for it, which we have already used above, and will use extensively later. We will write
\be
[\tal_{p_m}, \al_{p_{m+1}}] \doteq 0,
\ee
to indicate that \eqref{provecommutcorr} holds, but the operators $\tal_{p_{m}}$ and $\al_{p_{m+1}}$  may {\em not commute} as operators. It is just that this commutator annihilates $|\Psi \rangle$ and its descendants
produced by acting with elements of the algebra $\al_p$.

\paragraph{The space ${\cal H}_{\Psi}$: \\}
Before we conclude this subsection, let us make a comment about 
solving the linear equations \eqref{tildedefgeneral}. We have carefully
argued above that it is possible to find a set of solutions to these
equations. In constructing such solutions, we do not even actually
need to consider the full vector space ${\cal H}$. In fact, it is convenient to consider a slightly smaller vector space
\be
{\cal H}_{\psi} = {\cal A}^{\text{exp}} |\Psi \rangle,
\ee
which is just the space formed by the action of the set ${\cal A}^{\text{exp}}$
on the state $|\Psi \rangle$. In all cases of interest that we will study below,  ${\cal A}^{\text{exp}}$ coincides with ${\cal A}$, and in
these cases we can also write $\shil = {\cal A} |\Psi \rangle$.   We we see \eqref{tildedefgeneral}
is a statement about the action of the operators $\tal_p$ on the
domain $\shil$  and the action of these operators
outside this space is unspecified. In fact, we could even choose $\tal_p$ to annihilate states in the space of vectors orthogonal to $\shil$
without affecting low-point correlators. Note that the definition \eqref{tildedefgeneral}, and the fact that ${\cal A}$ may not be closed under arbitrary pairwise multiplication implies that the {\em range} of $\tal_p$ may differ slightly from $\shil$, even in this case. These ``edge effects'' are usually unimportant, and the physically relevant subspace is $\shil$.

Our construction, as we have presented it here, applies to any
statistical mechanics system. We now specialize to the CFT which,
as we will see, has a few new ingredients. 

\subsection{Mirror Operators in the CFT \label{tildescft}}
We now discuss the construction of the tilde-operators in an interacting CFT. Our construction follows the general method that we outlined above, but this section is written
so as to be self-contained. We will find two new features in the CFT. One is technical and, in our view, not so important: we have to regularize
the modes of the CFT to obtain a finite set of observables ${\cal A}$. The second is also somewhat technical, but a little more interesting. The 
operators that we are constructing are not gauge-invariant, and so, 
while they commute exactly with almost all operators, within correlation functions, they do not commute with the global charges or the Hamiltonian.

To be concrete, we will consider a CFT on $S^{d-1} \times R$. 
The black hole is dual to a state $|\Psi \rangle$ in the CFT, with 
an energy that is much larger than, but of the same order as $\nc$. In 
this section, we will show how to construct the tildes on this state $|\Psi \rangle$.

\subsubsection{Regularizing the Space of Operators}
First, let us discuss the operators that we can use to probe the black hole geometry --- this is the set ${\cal A}$ above. 
We have some number of light operators in the CFT that correspond to the supergravity fields. In addition, we could probe the black hole geometry with excitations corresponding to stringy-states, and perhaps even with brane-probes. In the CFT, all of these can be represented by conformal primary
operators with a dimension that is much smaller than $\nc$. We remind the reader that $\nc$ is the central charge. So, in maximally supersymmetric
$SU(N)$ theory, $\nc \propto N^2$ and even a giant graviton operator
has dimension $\Delta = N \ll \nc$.

It will be convenient for us to discuss the modes of these operators,
which are defined by
\be
{\op}_{\omega, \vect{m}}^i = \int {\op}^i(t,\Omega) e^{i \omega t} Y_{\vect{m}}(\Omega) d^{d-1} \Omega d t ,
\ee
where $Y_{\vect{m}}$ is the spherical harmonic indexed by the $d-1$ integers in the array $\vect{m}$. 

Now, the relevant spacing of the energy levels around energies of order
$\nc$ is actually $e^{-S} \sim e^{-\nc}$. So, the spectrum
of modes of low-dimensional conformal primaries is almost continuous
even when the CFT is on a sphere. 

Now, consider two energy levels $|E \rangle$ and $|E + \delta_{\omega}\rangle$. We can consider the precise mode ${\op}_{\delta \omega, \vect{m}}$ that
causes transitions between these levels. However, if the differences
between energies are non-degenerate, as we expect on general grounds for a ``chaotic'' system, then this mode will
 have a zero matrix element between {\em any other states.} 

So, we need to ``coarse-grain'' these modes a little to come up with
a useful set of operators. We will do this, by introducing
a lowest infrared frequency $\omega_{\text{min}}$, and bin together
the modes of ${\op}^i$ in bins of this width. More precisely, we 
define
\be
\label{coarsening}
{\op}^i_{n, \vect{m}} = {1 \over (\omega_{\text{min}})^{1 /2} } \int_{n \omega_{\text{min}}}^{{(n+1)} \omega_{\text{min}}} {\cal O}^i_{\omega, \vect{m}} d \omega .
\ee
These regularized modes ${\op}^i_{n,\vect{m}}$ have a smooth behaviour
in the Hilbert space, and we might reasonably expect them to obey the 
ETH, as we show in more detail in section \ref{noneqscen}. We will
often use 
\be
\omega_n = n \omega_{\text{min}},
\ee
and correspondingly also write $\op_{\omega_n, \vect{m}}^i$.

We can take $\omega_{\text{min}}$ to go to zero faster than any power of $\nc$, but it must be much larger than $e^{-\nc}$. So, for example, we could take $\omega_{\text{min}} = e^{-\sqrt{\nc}}$. So, the reader may 
wish to think of the $SU(N)$ theory, with an infrared cutoff
that scales like $e^{-N}$. This is certainly adequate for 
all purposes of constructing perturbative fields in the interior.

We have now regulated both the maximum dimension of allowed probe operators, and their modes in the manner above. Let us call these various operators ${\op}^{i}_{n,\vect{m}}$ where $i$ refers to the conformal primary, and $n, \vect{m}$ specify the mode.  We now consider the set formed by taking the span of  arbitrary products of up to $K$ numbers of these operators
\be
{\cal A} = \text{span}\{{\op}^i_{n,\vect{m}},\,\, {\op}^{i_1}_{n_1, \vect{m_1}} {\op}^{i_2}_{n_2, \vect{m_2}}, \ldots, {\op}^{i_1}_{n_1, \vect{m_1}} {\op}^{i_2}_{n_2, \vect{m_2}} \ldots {\op}^{i_k}_{n_K, \vect{m}_K}\}.
\ee
The set ${\cal A}$ is limited by the constraint that each product occurring in ${\cal A}$ satisfies $\omega_{\text{min}} \sum_{i=1}^K n_i  \ll \nc$ which
limits the total energy that can appear in this set.

Note, that as we emphasized in \cite{Papadodimas:2013b}, taking the linear
span of the products of operators above is exactly the same as thinking of ${\cal A}$ as the set of all {\em polynomials} in the modes of the operators $\op^i$
\be
\al_{\alpha} = \sum_{N}  \alpha(N) ({\cal O}^i_{n, \vect{m}})^{N(i,n,\vect{m})},
\ee
with the constraint that
\be
\label{energybound}
\sum_{i,n,\vect{m}} N(i,n,\vect{m}) \omega_{\text{min}} n \leq E_{\text{max}} \ll  \nc.
\ee
We also require that the set cannot be too large: 
\be
\label{dimconstraint}
{\cal D}_{\cal A} = \text{dim}({\cal A}) \ll e^{\nc}.
\ee
The second constraint is automatically satisfied if we also limit the
number of insertions in the polynomials 
\be
\sum_{i,n,\vect{m}} N(i,n,m) \leq K_{\text{max}},
\ee
and do not take $E_{\text{max}}$ to be too large. In fact, there is 
an interplay between the value of $K_{\text{max}}$, $E_{\text{max}}$,and $\omega_{\text{min}}$ so that \eqref{dimconstraint} can be preserved. For example, if we take $\omega_{\text{min}} = e^{-\sqrt{\nc}}$, then we must take $K_{\text{max}} \ll \sqrt{\nc}$ in order to preserve \eqref{dimconstraint}.  If we take $\omega_{\text{min}}$ to scale just as an inverse power of $\nc$, we can take $K_{\text{max}}$ to be larger.

Note that these polynomials, are polynomials in {\em non-commutative}
variables, since the operators do not commute with one another.
 However, there may be operator relations within the CFT, and as a result it may happen that 
some particular set of polynomials vanish because of these relations. In taking the set of polynomials above, we must
mod out by these relations. For
example if for three operators that appear above: $\op^{i_1}_{n_1, \vect{m_1}} \op^{i_2}_{n_2, \vect{m_2}} = \op^{i_3}_{n_3, \vect{m_3}}$, then, the polynomial $(\op^{i_1}_{n_1, \vect{m_1}})^2 \op^{i_2}_{n_2, \vect{m_2}}$ must clearly be identified with the polynomial $\op^{i_1}_{n_1, \vect{m_1}} \op^{i_3}_{n_3, \vect{m_3}}$.

This set ${\cal A}$ consists of all possible probes that we are allowed
to make in the black hole geometry.  We emphasize  that the set of operators in ${\cal A}$ is essentially the largest set of operators, for which one might hope to make sense of a semi-classical geometry. For example,
if we start including products of up to $\nc$ of the conformal
primary modes, then there is no reason at all that expectation values
of such operators should be reproducible by calculations in a 
semi-classical geometry.

In this concrete setting, the reader can also see another feature that we discussed in the 
section above. The set ${\cal A}$ is not quite an algebra, because of the cutoff \eqref{energybound} that have imposed on the energy of the operators that can appear. On the other hand, it is often possible to multiply elements of ${\cal A}$
together to obtain another member of ${\cal A}$. 

Before we proceed to the definition of the mirror operators, we must impose a final technical constraint on the set ${\cal A}$. We do not take the Hamiltonian itself, or any conserved charge (by which we mean any operator, which commutes with the Hamiltonian) to be part of this set. This is equivalent to excluding the {\em zero-modes} of conserved currents. These zero-modes to not correspond to propagating degrees of freedom in the bulk and, in any case, we will deal with them separately below. 

\subsubsection{Defining the Mirror Operators}
We now describe how to define the mirror operators. The CFT in a generic thermal state has the following property. 
\be
\label{noannihil}
\al_p | \Psi \rangle \neq 0\quad ,\quad \forall \al_p \in {\cal A}.
\ee
This is simply the statement that the insertion of a small number of light operators cannot annihilate the generic thermal state. We will work with states that satisfy \eqref{noannihil}. States that do not satisfy this
condition are a measure-0 subset of the set of all states, and as we 
discuss below, they may not have a smooth horizon.

We will now define the tilde operators, by specializing
the rules that we gave above. The mirror operators are defined by two very simple rules.
\begin{align}
\label{tildedefcftactpsi}
\widetilde{\op}^i_{n,\vect{m}} |\Psi \rangle &= e^{-{\beta \omega_n \over 2}} ({\op}^i_{n, \vect{m}})^{\dagger} |\Psi \rangle, \\
\label{tildedefcftcomord}
\widetilde{\op}^i_{n, \vect{m}} \al_p  |\Psi \rangle &= \al_p \widetilde{\op}^i_{n, \vect{m}} |\Psi \rangle, \quad \forall \al_p \in {\cal A}.
\end{align}
As advertised, we do not need to expand the set of allowed observables ${\cal A}$ to ${\cal A}^{\text{exp}}$ in the CFT to define the mirror operators.

Note that \eqref{tildedefcftactpsi} and \eqref{tildedefcftcomord} together give us ${\cal D}_{\cal A}$ {\em linear equations} for the $\widetilde{\op}$. However, $\widetilde{\op}$ can operate in 
a space that is $e^{\nc}$ dimensional! These equations are all internally consistent because of the condition \eqref{noannihil}.  So, there are many possible solutions to these constraints. One explicit solution is shown in \eqref{talexplicit}. 

All these solutions are equivalent for our purposes, since they do not show any difference at 
all, except when inserted in very high-point correlators. As we pointed out above, there is also an, in principle, difference between \eqref{tildedefcftactpsi} and \eqref{tildedefcftcomord}. While \eqref{tildedefcftactpsi} 
needs to be corrected order by order in ${1 \over \nc}$, \eqref{tildedefcftcomord} is already correct at all orders  in the ${1 \over \nc}$ for the correlators that we are interested in.

\subsubsection{Choice of Gauge: Hamiltonian and Abelian Charges}
We now turn to the issue of a choice of gauge.  We are willing to consider cases, where $|\Psi \rangle$ is an energy eigenstate, and certainly it may be possible to put $|\Psi \rangle$ in an eigenstate of some other conserved charge.  We first discuss the inclusion of the Hamiltonian, which corresponds to zero-modes of the stress-tensor, and other Abelian charges, then turn to other kinds of conserved charges including non-Abelian charges in the next subsection.

If $|\Psi \rangle$ is an energy eigenstate, or the eigenstate of some other charge, we still expect it to appear thermal. However, in such cases, we see that we might have 
\be
\left(\hat{Q} - Q \right) | \Psi \rangle = 0,
\ee
where $\hat{Q}$ is the charge operator and $Q$ is the corresponding eigenvalue. 
This is the reason that we cannot include $\hat{Q}$
in the set ${\cal A}$. 
If, with this inclusion, we were to also demand \eqref{tildedefcftcomord}, we would get an inconsistency. 

However, this is quite simple to fix. We set $\tO^i_{n, \vect{m}}$ to have a non-zero commutator with the {\em zero-mode} of the corresponding conserved current. In fact, this zero-mode is not of any interest, except
for the fact that it includes the charge itself. So, we append the
charge to the set ${\cal A}$ and add an additional rule to the set
of rules above. 

First, since the position space operator $O^i(t,\Omega)$ is Hermitian, we need to re-organize its modes ${\cal O}^i_{n,\vect{m}}$ into operators that transform simply under the charger under consideration. If this charge is just the Hamiltonian or the angular momentum on $S^{d-1}$, then the modes already transform in a simple manner. But, in any case, we can construct linear combinations ${\cal O}^{i,q}_{n, \vect{m}}$, which have a well defined charge so that $[\hat{Q}, {\cal O}^{i,q}_{n,\vect{m}}] = q  {\cal O}^{i,q}_{n,\vect{m}}$. The action of the mirror operators 
on the original linear combinations can be constructed by using the anti-linearity of the mirror-map. We now add the following rule to the set of 
rules above
\be
\label{cftgaugeinvar}
\tO^{i,q}_{n, \vect{m}} \al_{1} \hat{Q} \al_2 |\Psi \rangle =  \al_{1} \hat{Q} \al_2 \tO^{i,q}_{n, \vect{m}} |\Psi \rangle + q \al_1 \al_2 \tO^{i,q}_{n, \vect{m}} | \Psi \rangle.
\ee
In the appendix \ref{gaugechoice}, we discuss this issue further. 
We show how a choice of gauge results in these commutation relations, and how they may be interpreted in terms of Wilson lines.
We also explore the fact that these relations already seem to lead to some interesting physical implications. We note, that by virtue of this rule we see that $\tO^i$ does not really correspond to a local field on the boundary, since such a field would have non-zero commutators for other modes of the current as well. Here, this is not a difficulty, since the bulk fields constructed from $\tO^i$ cannot ever be taken close to the boundary to obtain any kind of contradiction. But this also provides a criterion for when
the $\tO^i$ fields can enter bulk operators, and explains why they cannot 
be used in bulk fields below the Hawking Page transition.

Second, notice that since the charge and energy of the $\tO^i_{n,\vect{m}}$
can be measured by the CFT Hamiltonian, this tells us that there is not
really any ``other side'' of the collapsing geometry. We return to
this at greater length in Appendix \ref{gaugechoice}.

\subsubsection{Non-Abelian Charges \label{cftnonabelian}}
We now describe how the mirror-operators act on descendants of the state $|\Psi \rangle$ produced by acting with various non-Abelian charges.\footnote{We thank Rajesh Gopakumar for a discussion on this issue.} The main difference with the analysis for the Hamiltonian and Abelian charges above, is that in this case, we can have other kinds of null-vectors. The analysis of the subsection above is subsumed in the more general analysis of this subsection.

For example,
we might want to consider a Schwarzschild black hole, 
and consider a corresponding ensemble in the CFT, where the states transform in a small representation of some non-Abelian charge, but are yet not charge eigenstates. 
Now, we may have $J_{+}^K |\Psi \rangle = 0$, for some ``raising operator'' $J_{+}$. We wish to ensure that our definition of the $\tO$-operators is correct in this case. Below, we will denote any {\em polynomial} in the charges by ${\cal Q}_{\alpha}$.  The space of physical states is produced by acting with all such polynomials on the base state $|\Psi \rangle$, and then modding
out by the null vectors. The action of $\tO^{i}_{n, \vect{m}}$ must be
correct on this quotient space, in that it must annihilate all null vectors.

\paragraph{The Set of Null Vectors \\}
First, the condition that the action by an observable does not annihilate
the state must be refined in the presence of such charges. We will impose the following condition. Consider a set of charge-polynomials
${\cal Q}_{\alpha_1} \ldots {\cal Q}_{\alpha_m}$. Now, we demand
\be
\label{noannihilnonabelian}
\sum_{i=1}^{m} \kappa_i {\cal Q}_{\alpha_i} |\Psi \rangle \neq 0\,, \,\forall \kappa_i\,\,\Rightarrow \,\,\sum_{i=1}^{m} A_{\beta_i} {\cal Q}_{\alpha_i} |\Psi \rangle \neq 0\,,\, \forall A_{\beta_i}.
\ee
Translated into words, this means that we get various ``descendants'' by
acting on the base state with the charges. If these descendants are linearly-independent, then by acting on them with with our observables, we cannot ``make'' them linearly dependent. This is a very natural generation of \eqref{noannihil} above, and more formally speaking 
the states that do {\em not} satisfy \eqref{noannihilnonabelian} form a
measure-0 space in the Hilbert space. Of course, we can also phrase \eqref{noannihilnonabelian} as
\be
\sum_{i=1}^m \al_{\beta_i} {\cal Q}_{\alpha_i} |\Psi \rangle = 0 \,\,\Rightarrow\,\, \exists \kappa_i \in \mathbb{C}, ~\text{s.t.}~ \sum_{i=1}^m \kappa_i {\cal Q}_{\alpha_i} |\Psi \rangle = 0.
\ee

Now, we want to consider the structure of the quotient space that we can get
by acting {\em both} with the ${\cal Q}_{\alpha}$-polynomials and with the $\al_{\alpha}$-polynomials. First note that by using the commutation relations
of the operators inside $\al_{\alpha}$ with ${\cal Q}_{\alpha}$, we can always
move the ${\cal Q}_{\alpha}$ to the right. So, we start by considering the
module produced by acting freely, first with ${\cal Q}_{\alpha}$ and then with
$\al_{\alpha}$.
\be
{\cal V} = \{ \sum_{i=1}^{{\cal D}_{\cal A}}  \al_{\beta_i} {\cal Q}_{\alpha_i} |\Psi \rangle \},
\ee
where the set is formed by consider all possible combinations of $\al_{\beta_i}$ and ${\cal Q}_{\alpha_i}$.
Some vectors in ${\cal V}$ are null, because the leading charge
polynomials in the expression have annihilated the base state. Say that a basis of polynomials, which annihilate the state, is given by ${\cal Q}_{n_1} \ldots {\cal Q}_{n_{P}}$, all of which satisfy 
\be
{\cal Q}_{n_i} |\Psi \rangle = 0, \quad i = 1 \ldots P.
\ee
For example, we might have null-vectors because $|\Psi \rangle$ is an eigenvector of some charge $(\hat{Q} - q) |\Psi \rangle  = 0$, as we discussed in the previous subsection. Or, as we mentioned earlier, we might have null-vectors because $|\Psi \rangle$ is only finitely-separated from the highest-weight state: $J_{+}^K |\Psi \rangle = 0$, for $K$ greater than some number. 
All of these types are included in the set above.

Then
the set of all null vectors in ${\cal V}$ is given by the set of 
all vectors that are obtained by acting with an element of
the ${\cal A}$ on the null-vectors listed above.
More precisely, the null set in ${\cal V}$ is
\be
\label{nullset}
{\cal N} = \{ \sum_{i=1}^P \al_{\beta_i} {\cal Q}_{n_i} |\Psi \rangle \}, 
\ee
where the set is formed by considering all possible $\al_{\beta_i}$

Let us prove the equivalence of \eqref{noannihilnonabelian} and \eqref{nullset}, which is not immediately obvious. Consider some arbitrary null vector
\be
\label{nullvector}
|n \rangle = \sum_{i=1}^K \al_{\beta_i} {\cal Q}_{\alpha_i} |\Psi \rangle.
\ee
We will now prove that \eqref{noannihilnonabelian} implies that this
 can always be written in the form \eqref{nullset}. We note that 
\eqref{noannihilnonabelian} implies that the set of vectors $\{{\cal Q}_{\alpha_1} |\Psi \rangle, \ldots {\cal Q}_{\alpha_K} |\Psi \rangle\}$ is not linearly independent. For the sake of generality,
we will assume that there are multiple linear dependences in this set, and that some $m$ vector, ${\cal Q}_{\alpha_1} |\Psi \rangle \ldots {\cal Q}_{\alpha_m} |\Psi \rangle$
are linearly independent. However,
\be
\label{individualnullvectors}
|n_j \rangle = {\cal Q}_{\alpha_{j}} |\Psi \rangle - \sum_{i=1}^m \kappa^i_j {\cal Q}_{\alpha_i} |\Psi \rangle = 0, \quad m+1 \leq j \leq K,
\ee
which simply states that ${\cal Q}_{\alpha_{m+1}} |\Psi \rangle \ldots {\cal Q}_{\alpha_K}| \Psi \rangle$ are dependent on the first $m$ vectors. 
Consequently, 
\be
| n \rangle = \sum_{i=1}^{m} \left(\al_{\beta_i} + \sum_{j={m+1}}^K\kappa^i_j \al_{\beta_j} \right)  {\cal Q}_{\alpha_i} |\Psi \rangle.
\ee
From  \eqref{noannihilnonabelian}, we see that for this to hold, each term in the sum over $i$ must vanish individually, and so
\be
\al_{\beta_i} = -\sum_{j={m+1}}^K \kappa^i_j \al_{\beta_j}, \quad 1 \leq i \leq m,
\ee
as an identity. 
This means that we can write \eqref{nullvector} as
\be
|n \rangle = \sum_{j=m+1}^K \al_{\beta_j} \left({\cal Q}_{\alpha_j} - \sum_{i=1}^m \kappa^i_j {\cal Q}_{\alpha_i} \right) |\Psi \rangle.
\ee
From \eqref{individualnullvectors}, we see that we can write this precisely as 
\be
|n \rangle = \sum_{j=m+1}^K \al_{\beta_j} | n_j \rangle,
\ee
which is of the form \eqref{nullset}. This proves what we require. 

\paragraph{The Action of the Mirror Operators \\}
The physical space ${\cal H}_{\psi}$ is given by the quotient
\be
{\cal H}_{\psi} = {\cal V}/{\cal N}.
\ee
Our task is to define the action of $\tO^i_{n,\vect{m}}$ on this space in a natural manner, and also ensure that $\tO^i_{n,\vect{m}}$ annihilates all
elements of ${\cal N}$. 

First, we define the action of $\tO^i_{n,\vect{m}}$ on the space ${\cal V}$. Our intuition is just that, $\tO^{i}_{n, \vect{m}}$ 
should transform the same way as the adjoint of the ordinary operator $({\cal O}^i_{n,\vect{m}})^{\dagger}$. For the ordinary operator,
we have some commutation relations that are imposed by how the operator 
transforms under the algebra. In particular, denoting by $Q^1$ a {\em single charge} (not a polynomial) we have
\be
\label{reptransform}
[(\op^i_{n,\vect{m}})^{\dagger}, Q^1] = t^{i j} (\op^{j}_{n,\vect{m}})^{\dagger},
\ee
where $t^{i j}$ is some matrix that describes the transformation of the operator. Note that, in general, $\op^i_{n,\vect{m}}$ will not transform in 
an irreducible representation, because we have chosen conventions where
the position space operator $O^i(t,\Omega)$ is Hermitian. As we pointed out above, this does not involve
any loss of generality, and the mirror of any operator can be obtained
by means of linear combinations, and the use of the anti-linearity
of the mirror map.

Now, we define the action of $\tO^{i}_{n,\vect{m}}$ on an element of ${\cal V}$ as follows
\be
\label{deftoiaction}
\begin{split}
\tO^i_{n,\vect{m}} \al_{\alpha_1} Q^1 \al_{\alpha_2} {\cal Q}_{\alpha_3} \ldots \al_{\alpha_n}  |\Psi \rangle = 
&t^{i j}\al_{\alpha_1} \tO^j_{n,\vect{m}} \al_{\alpha_2} {\cal Q}_{\alpha_3} \ldots \al_{\alpha_n}|\Psi \rangle \\ &+
\al_{\alpha_1} Q^1 \tO^i_{n,\vect{m}} \al_{\alpha_2} {\cal Q}_{\alpha_3} \ldots \al_{\alpha_n} |\Psi \rangle.
\end{split}
\ee
$Q^1$ is the same charge that appears in \eqref{reptransform}, and  ${\cal Q}_{\alpha_2} \ldots $ are arbitrary polynomials in the charges. 
We have specified how $\tO^{i}_{n,\vect{m}}$
commutes through a single charge, but clearly we can use this definition
recursively to move through the rest of the operators
acting on $|\Psi \rangle$ above, as well, and hence define the action of $\tO^i_{n,\vect{m}}$ on any element of ${\cal V}$.

To ensure that this action is consistent on ${\cal H}_{\psi}$, we simply
notice the following simple fact. The definition \eqref{deftoiaction} implies
\be
\label{actiononn}
\tO^i_{n,\vect{m}} \al_{\beta} {\cal Q}_{\alpha_i} |\Psi \rangle = 
\al_{\beta} e^{-\beta \omega_n \over 2} ({\cal O}^i_{n,\vect{m}})^{\dagger} {\cal Q}_{\alpha_i} |\Psi \rangle.
\ee
In fact, we can  use the commutation relations of the charges with the ordinary operators, to always move all the charges immediately next to $|\Psi \rangle$, so \eqref{actiononn} can be used as an alternate definition of $\tO^i_{n,\vect{m}}$ on ${\cal V}$, as was done in \cite{Papadodimas:2013b}. 

However, what \eqref{actiononn} tells us immediately is that acting on an element of ${\cal N}$, $\tO^i_{n,\vect{m}}$ returns 
another element of ${\cal N}$. Hence, this linear operator $\tO^i_{n,\vect{m}}$  consistently
reduces to a linear operator on the quotient space ${\cal H}_{\psi}$ and transforms in the representation conjugate to ${\cal O}^i_{n,\vect{m}}$.

\subsection{Decoupled Harmonic Oscillators \label{decoupledsho}}
Now, having discussed the construction of mirror operators in a general
theory, and in the CFT, we will move to a simple and concrete example: a set of decoupled harmonic oscillators. 
As we will discuss below, we can even use these decoupled harmonic
oscillators as a model of the s-waves emitted from the black hole. As a result, this model provides us with significant insight into several recent
discussions of the information paradox, some of which \cite{Almheiri:2013hfa,Marolf:2013dba} are basically phrased in this context. We should caution the reader that while this model is very simple and explicit, the 
flip side is that we will need to use states with a small spread in energy, rather than energy eigenstates, to ``mock up'' some of the features of the interacting theory and define the mirror operators.

Consider a collection of harmonic oscillators of different frequencies. If we think of this as a model for the s-waves emitted by the black hole, the lowest frequency $s$-wave is inversely proportional to the Page Time. We will write
\be
\omega_{IR} = {2 \over 3} M_{pl} \left({M_{pl} \over M} \right)^3, 
\ee
where the coefficient of ${2 \over 3}$ has been chosen by hand, for reasons that will be apparent below. The ``gas'' of oscillators consists
of frequencies $p \omega_{\text{IR}}$ for positive integers $p$.

As introduced above, $M$ is just a parameter that controls this lowest frequency, but we will take it to be the average total energy in the harmonic oscillators. For consistency with the notation above, in this section, we will
use 
\be
\nc = \sqrt{M \over \omega_{IR}}.
\ee

\begin{figure}[!h]
\begin{center}
\includegraphics[height=4cm]{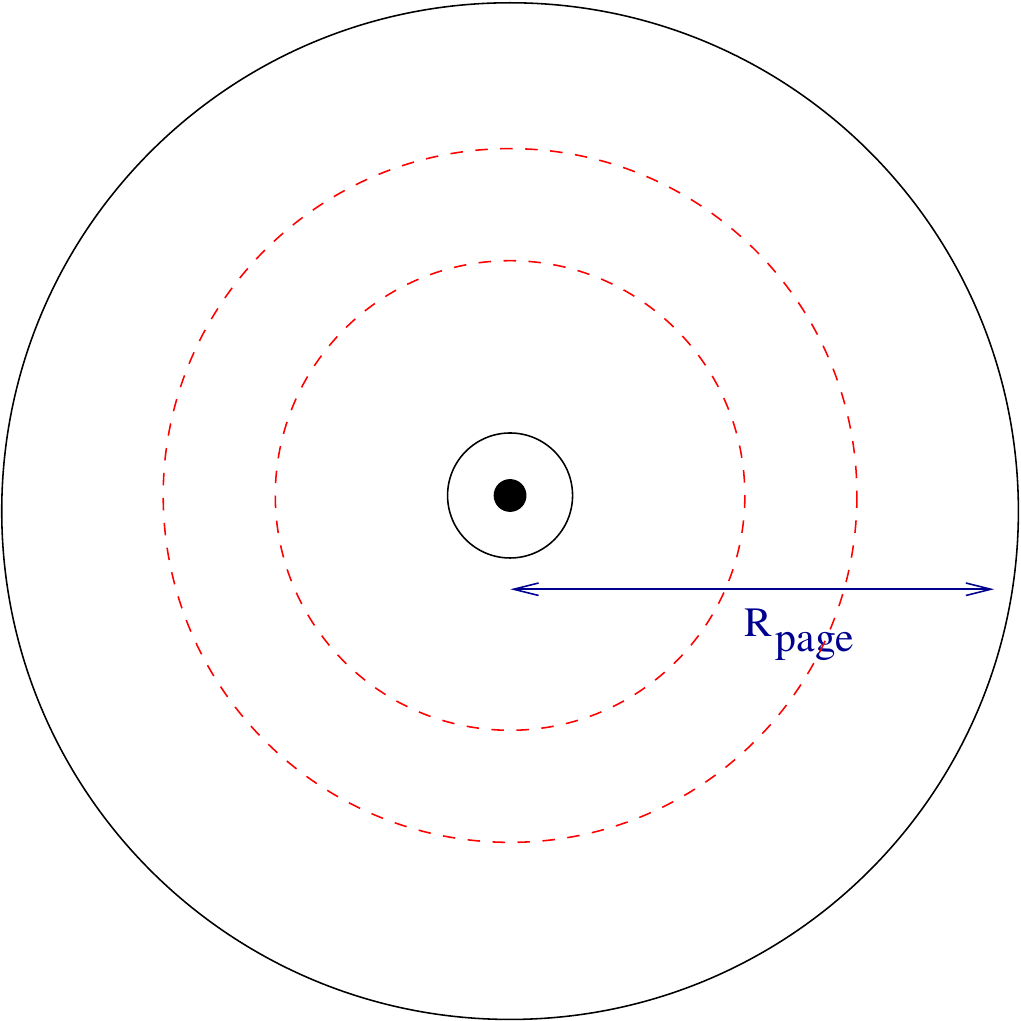}
\caption{\em A toy model of a black hole (small black-circle in the center) emitting Hawking radiation. We are quantizing a massless field between the two solid circles, one of which is a Page distance away. The emission is mostly in s-waves if the inner cutoff is far enough from the horizon.\label{pagetimebox}}
\end{center}
\end{figure}

Now, notice that there is another physical interpretation of this
gas of decoupled harmonic oscillators. Let us say that
we quantize a massless 
field outside the black hole with one boundary condition at the Page length $R_{\text{page}}$ and another boundary condition on the field that can be placed a few Schwarzschild radii
away from the horizon of the black hole. The exact position of the inner
cutoff is not important, and it can be placed far enough from the black hole, that most of the radiation is in outgoing s-waves.
We have shown our setup schematically in Fig \ref{pagetimebox}.
This would automatically lead to the set of frequencies that we have above.

Now, consider a configuration of these oscillators with total energy that in a small band: $[M-\Delta, M + \Delta]$ where, in this analysis,
we will have to take $\Delta \propto \nc$, for a reason that
we explain below. 

We see that the number of such configurations is given by
the number of sets $\{n_p\}$ that satisfy the integer equation 
\be
M - \Delta < \sum_{p} p n_p \omega_{IR} 
< M + \Delta.
\ee
Since $\Delta \ll M$, 
to leading order, we are just counting the number of solutions to the Diophantine equation  $\sum_{p=1}^{\nc^2} p n_p = \nc^2$.
The log of the leading term in the number of solutions, $N_{\text{sol}}$ is given by Cardy's formula
\[
\log(N_{\text{sol}}) = 2 \pi \sqrt{\nc^2 \over 6} = \pi {M^2 \over M_{pl}^2} \equiv S.
\]
So, this gas of s-waves has the right entropy up to a numerical factor that we have inserted by hand in choosing the lowest IR frequency.

Now, let us consider the field outside the black hole. Again, neglecting
the higher angular momenta, this has an expansion in terms of outgoing $s$-waves and can be written as
\be
\label{fieldexpansion}
\phi(t,r) = \sum_{p=1}^{\nc^2} \left[{a_p \over 2 \pi \sqrt{p}}  {e^{-i \omega_p (t - r)}  \over r} + {b_p \over 2 \pi \sqrt{p}}{e^{-i \omega_p(t + r)} \over r} + \text{h.c} \right],
\ee
where, $\nc$ is defined above. The modes $a_p$ correspond to the outgoing
modes, and $b_p$ correspond to the "ingoing"
modes.

We want to consider an excitation of this field, that comprises
purely outgoing modes. So, we consider a {\em pure state} in our gas of decoupled harmonic oscillators made out of the states in the energy band that we discussed above. 
\be
|\Psi \rangle = \sum_{n_p} \alpha(n_p) |n_p \rangle, \quad M - \Delta < \sum p  n_p \omega_{IR} < M + \Delta, \quad |n_p \rangle \equiv \prod_p {(a_p^{\dagger})^{n_p} \over \sqrt{n_p!}} |\Omega \rangle,
\ee
where the $\alpha(n_p)$ are some randomly chosen coefficients,\footnote{The band defines a finite Hilbert space, and we can choose the $\alpha(n_p)$
by using the Haar measure on this space.}
and the sum runs over all states that live in this energy band.  We can
associate a ``temperature'' to the state $|\Psi \rangle$, take $\beta = {\partial S \over \partial M} = {2 \pi M \over M_{\text{pl}}}$

We want to find the mirror operators $\widetilde{a}_p$ and $\widetilde{a}^{\dagger}_p$. We could simply take an appropriate state from the ensemble
discussed above, and follow the general rules for defining the mirror operators. 
However, this model is so simple that it
is useful to derive them from scratch.

We would like operators that act in this theory, but which``mimic'' the thermofield state
\be
|\Psi \rangle_{\text{tfd}} = Z_{\beta}^{-1}\sum_{\{n_p\}} e^{-{\beta \over 2} \sum_p n_p \omega_p}  |\widetilde{n_p} \rangle |n_p \rangle,
\ee
where the sum is taken over all functions $n_p$, and $Z_{\beta}$ is 
the partition function, which normalizes the state.

The operators that act on the ``other side'' of this entangled-state are
\be
\begin{split}
\widetilde{a}_{m}^{\text{tfd}} |\Psi \rangle_{\text{tfd}} &= Z_{\beta}^{-1}\sum_{\{n_p\}} e^{-{\beta \over 2} \sum_p n_p \omega_p}  \sqrt{n_m} |\widetilde{n_p} - \delta_{p m} \rangle |n_p \rangle \\
&= Z_{\beta}^{-1} \sum_{\{n_p'\}} e^{-\beta {\omega_m \over 2} -{\beta \over 2} \sum_p n_p' \omega_p}  \sqrt{n_m' + 1 } |\widetilde{n_p'}  \rangle |n_p' + \delta_{p m} \rangle \\
&= e^{-\beta \omega_m \over 2} a_m^{\dagger} |\Psi \rangle_{\text{tfd}}.
\end{split}
\ee
In the second line above, we changed the sum from $n_p$ to $n_p' = n_p - \delta_{p m}$, which allowed us to notice that the action of $\widetilde{a}_m^{\text{tfd}}$ on this state was simply related to the action of $a_m^{\dagger}$.
We can easily derive the same result by following the prescription of 
\eqref{tildeactdef1}.

Using a very similar calculation, we find that
\be
\widetilde{a}_{m}^{\dagger, \text{tfd}} |\Psi \rangle_{\text{tfd}} = e^{\beta \omega_m \over 2} a_m |\Psi \rangle_{\text{tfd}}.
\ee

For each pair of operators $a_m, a_m^{\dagger}$, we now {\em define} the operators
\be
\label{tildeadefshm}
\begin{split}
\widetilde{a}_m |\Psi \rangle &= e^{-{\beta \omega_m \over 2}} a_m^{\dagger} |\Psi \rangle, \\
\widetilde{a}_m a_{m_1} \ldots a_{m_{n_1}} a^{\dagger}_{m_{n_1+1}} \ldots a^{\dagger}_{m_{n_2}} |\Psi \rangle &= a_{m_1} \ldots a_{m_{n_1}} a^{\dagger}_{m_{n_1+1}} \ldots a^{\dagger}_{m_{n_2}}  \widetilde{a}_m |\Psi \rangle.
\end{split}
\ee
We define $\widetilde{a}_m^{\dagger}$ in a similar manner
\be
\label{tildeadaggerdefshm}
\begin{split}
\widetilde{a}_m^{\dagger} |\Psi \rangle &=  e^{\beta \omega_m \over 2} a_m |\Psi \rangle, \\
\widetilde{a}_m^{\dagger} a_{m_1} \ldots a_{m_{n_1}} a^{\dagger}_{m_{n_1+1}} \ldots a_{m_{n_2}} |\Psi \rangle &= a_{m_1} \ldots a_{m_{n_1}} a^{\dagger}_{m_{n_1+1}} \ldots a_{m_{n_2}}  \widetilde{a}_m^{\dagger} |\Psi \rangle.
\end{split}
\ee
In the formulas above, the product of operators in the second line of both \eqref{tildeadaggerdefshm} and \eqref{tildeadefshm} is, as usual, limited to cases where $n_2 \ll \nc$. 

We see, once again, that these equations are consistent provided that 
the set of products that we consider must has the property that no linear combination of these products must annihilate the state $|\Psi \rangle$. Otherwise, we run into the difficulties mentioned above, and the linear equations defining $\widetilde{a}_m, \widetilde{a}_m^{\dagger}$ may fail to have a solution.

We now see the importance of the band $\Delta$. It serves to ensure
that the operator 
\be
\left( \sum m  a^{\dagger}_m a_m - \nc^2 \right) | \Psi \rangle \neq 0.
\ee
In fact to annihilate $|\Psi \rangle$, we need to take a product
of $\nc$ such operators, which is the width of the energy band. In the CFT, we required no such restriction
because even an energy eigenstate in the CFT has a spread of occupation
numbers of single-trace operators.

With these restrictions, the tilde operators can be used
to construct a mirror ``field''
\be
\label{fieldexpansiontilde}
\widetilde{\phi}(t,r) = \sum_{p} {\widetilde{a}_p \over 2 \pi \sqrt{p}} {1 \over r} {e^{i \omega_p (t - r)}  \over r} + \text{h.c.}
\ee
Note that we cannot reconstruct the $\widetilde{a}^{\dagger}_p$ for very high $p \propto \nc^2$ very well, because acting even a few times with the corresponding
$a_p$ can annihilate the state. However, these operators are negligible
within correlation functions.

The field \eqref{fieldexpansiontilde} commutes with the ordinary field in \eqref{fieldexpansion}, within low-point correlators evaluated on $|\Psi \rangle$ and has the same correlators as one expects from the thermofield doubled state up to corrections that are expected in changing ensembles. Note, as usual, that the
wave-function multiplying $\ta_p$ has been conjugated.

\subsection{Mirror Operators in the Spin Chain \label{dualspinchain}}
To aid the reader, we finally describe our construction in  a second simple example: a simple spin-chain model. In Appendix \ref{appnumerical},
we present a numerical computation of the 
mirror-operators in this
model. The reader may choose to directly consult that Appendix
and the included computer program to see how the various features 
of the mirror-operators work out in an absolutely concrete setting.

We considered this toy model first in  \cite{Papadodimas:2012aq}. Consider a spin chain consisting of $\nc$ spin $1/2$ particles labeled by $i=1,\ldots,\nc$.  Each spin $i$ has a set of associated spin observables, $\paul^i_a$, which satisfy 
\be
\label{paulspin}
[\paul^i_a ,\paul^j_b] = {1\over 2}i \epsilon_{a b c}\delta^{ij} \paul^i_c.
\ee
The simultaneous eigenstates of the $\paul^i_z$ operators  
in this theory can be specified in terms of a single numbers from $0$ to $2^{\nc-1}$ using the eigenvalues of the operator $B = \sum_{i=1}^{\nc} \left(\paul^i_z + {1 \over 2} \right) 2^{i - 1}$. In this basis of $B$-eigenstates, satisfying $B |n \rangle_B = n |n \rangle_B$, consider a state
\be
\label{stspinchain}
|\Psi \rangle = \sum \alpha_n |n \rangle_B,
\ee
where the $a_n$ can be picked randomly using the Haar measure on ${\cal C}P^{2^{\nc}-1}$. 

One commonly considered model of Hawking evaporation has been to imagine
these spins ``breaking off'' from the spin-chain one by one to constitute the outgoing Hawking radiation. This model should not be taken too seriously, but we will use it to illustrate our ideas. 

The key issue in Hawking radiation is that bits are emitted in ``pairs''. After $p$ bits have evaporated, the ``outside observer'' can make measurements involving
the $\pault^1_a \ldots \pault^p_a$ operators.  For each such measurement that the outside observer can make,  there is a commuting measurement $\widetilde{B}$ that the ``inside'' observer can make, and moreover the results of the two experiments are exactly correlated.

Here, we are interested in identifying the mirrored measurements. This means that we would like to find operators $\pault^i_a$, which also satisfy 
\be
[\pault^i_a, \pault^j_b] \doteq {1 \over 2} i \epsilon^{a b c} \delta^{i j} \pault^i_c,
\ee
where, as we have mentioned above, the $\doteq$ indicates that this algebra will be satisfied in the state of the theory, and not as an operator algebra. Moreover, we would like
\be
\label{demand1}
[\pault^i_a, \paul^j_b] \doteq 0,
\ee
and that, in the state under consideration, (and its descendants obtained
by acting with these Pauli-spin matrices) these measurements be perfectly correlated
\be
\label{demand2}
\langle \Psi| \pault^i_a \paul^j_b | \Psi \rangle = -\delta^{i j} \delta_{a b}.
\ee

These conditions together imply that for measurements of low-point correlators of $\paul^i_a$ and $\pault^j_b$, a given state $|\Psi \rangle$,
on which they are defined, {\em looks like} the thermo-field doubled state
\be
\label{tfdspinchain}
|\Psi \rangle_{\text{tfd}} = \sum_B |B \rangle |2^{N} - 1 - B \rangle,
\ee
in the notation above. This thermofield doubled state
is just a direct-product of $\nc$-entangled EPR pairs: $|\Psi \rangle_{\text{tfd}} =  \left(|0 \widetilde{1}  \rangle + |1 \widetilde{0} \rangle \right)^{\nc}$, where the exponentiation by $\nc$, means we need to take the
direct-product of this state with itself $\nc$-times.

We now show how the operators $\pault^i_a$ can be obtained very simply in a given state.  As usual, we define  these operators by specifying their action on a set of vectors. First, we describe how $\pault^i_a$ acts on $|\Psi \rangle$. \footnote{The careful reader may have noticed that we have switched conventions a little from the setting of section \ref{tildegeneral}, by inserting the additional minus sign in \eqref{rule1}. This is because the
  thermofield doubled state, we are mimicking here, has anti-correlated
  eigenvalues. It also allows us to ensure that the mirror operators
 obey the same, rather than the conjugated algebra. In the spin-chain
  setting, the convention we use here is more natural, and we hope
 that this will not confuse the reader.}
\be
\label{rule1}
\pault_a^i |\Psi \rangle = -\paul_a^i |\Psi \rangle.
\ee
Next, we describe how it acts on states that differ from the action of $|\Psi \rangle$ by an action of up to $K$-ordinary $\pault^i_a$ operators. For any product of operators, where $p$ below satisfies $p < K$, we demand
\be
\label{rule2}
\pault_a^i \prod_{j=1}^p \paul_{a_1}^{i_1} \ldots \paul_{a_p}^{i_p} |\Psi \rangle = \left(\prod_{j=1}^p \paul_{a_1}^{i_1} \ldots \paul_{a_p}^{i_p} \right) \pault_a^i |\Psi \rangle.
\ee
Note that $\pault_a^i$ can be a $2^{\nc} \times 2^{\nc}$ matrix, and to describe the operator, we need to specify its action on $2^{\nc}$ linearly independent vectors.  The rules \eqref{rule1} and \eqref{rule2} together specify the action of $\pault_m^z$ on 
\be
{\cal D}_{\cal A} = \sum_{j=0}^K \begin{pmatrix} \nc \\ j \end{pmatrix} 3^j,
\ee
basis vectors. Provided that we do not take $K$ to scale with $\nc$, we have $n_K \ll 2^{\nc}$. In fact, the precise condition we need in order to
be able to construct the mirror operators is just ${\cal D}_{\cal A} < 2^{\nc}$. 
So, there is a $(2^{\cal N} - {\cal D}_{\cal A})^2$-parameter family
of choices of operators $\pault_a^i$ that satisfy \eqref{rule1} and \eqref{rule2}. If we like, we can restrict this ambiguity by increasing $K$, 
but the action of all of these operators coincides 
{\em exactly} within low-point correlation functions. 

This prescription guarantees the correct behaviour of $\pault_a^i$ within low point correlators, with up to $K$ insertions,  as specified by \eqref{demand1} \eqref{demand2}.  For example, \eqref{rule2} tells us that within a low-point correlator $\pault_a^i$ commutes {\em exactly} with $\paul_b^j$, $\forall a,b,i,j$.

Note that the fact that our operators are state-dependent is quite important here. For example, it is easy to prove that there is no operator $\pault_b^j$ in the theory, except for the identity operator, with commutes exactly with all the $\paul^i_a$ matrices. Our point is that, within low-point correlators, we can produce operators that {\em look} like they achieve this zero commutator.


%% file: s_resolvepar.tex
\section{Resolving Various Paradoxes \label{resolveallpar}}
We now explain how our construction of the previous section resolves {\underline{all}} the recent paradoxes that have been brought up in the recent
literature on the information paradox. In particular, we resolve the following issues in order.
\begin{enumerate}
\item
The strong subadditivity paradox in section \ref{resolvestrong}.
\item
The apparent issue of non-vanishing commutators between the early radiation
and measurements inside the black hole in section \ref{resolvevanishcommut}.
\item
The apparent problem with the lack of a left-inverse for ``creation'' operators inside the black hole in section \ref{resolveleftinv}.
\item
The apparent argument that the infalling observer measures non-zero 
particle number in section \ref{resolvenaneq0}.
\item
The apparent ``theorem'' that small corrections cannot unitarize
Hawking radiation in section \ref{resolvesmallcorr}.
\end{enumerate}

\subsection{Resolution to the Strong Subadditivity Puzzle \label{resolvestrong}}
We now describe how our operators resolve the strong subadditivity
puzzle of Mathur \cite{Mathur:2009hf} and AMPS \cite{Almheiri:2012rt}. 
This puzzle was proposed, while tacitly keeping in mind the picture of the spin-chain,
where Hawking evaporation is understood to be simply the detachment of
individual spins from the chain. So, we first resolve the puzzle in this context. 

We then try and formulate the strong subadditivity paradox, as carefully as we can, in terms of 
CFT correlators --- something that, to our knowledge, has not been done so far.  We then resolve it in this more precise context.

\paragraph{\bf Summary of Resolution \\}

Before we proceed to our detailed resolution, let us summarize the naive formulation of the strong subadditivity
puzzle. We think of three subsystems: (1) the early radiation $E$, and a Hawking pair that is just being emitted , which consists of (2) $B$ --- the particle just outside the horizon and (3) its $\widetilde{B}$ --- the particle just inside the horizon. For an old enough black hole, $B$ must be entangled with $E$, and for the horizon to be smooth $B$ is entangled with $\widetilde{B}$. 

Our resolution to this paradox is simple: the system $\widetilde{B}$ is {\em not} independent of $E$. However, this overlap is cleverly designed 
so that commutators of operators on the early radiation $E$ and the bit $\widetilde{B}$ vanish $[E, \widetilde{B}] \doteq 0$, when inserted within a low-point correlation function evaluated on the state of the system $|\Psi \rangle$.

We have already discussed this resolution in our previous paper \cite{Papadodimas:2012aq}. The key new point is to ensure that such a resolution does
not lead to a situation, where the observer outside can transmit messages
to the observer inside, within the regime where effective field theory should be reliable. Our construction, where the $\widetilde{B}$ operators are cleverly constructed to have exactly vanishing commutator with the
operators in $E$, within all low-point correlators, ensures this.

We now discuss the resolution of the strong subadditivity puzzle
in the spin chain, where the puzzle can be formulated most clearly. 
We then discuss some of the subtleties of formulating the puzzle in the
CFT, attempt to formulate it as precisely as possible, and then resolve
it in that context. 

\subsubsection{Resolution to the Strong Subadditivity Puzzle in the Spin Chain
\label{spinstrongsub}}
\begin{figure}[!h]
\begin{center}
\includegraphics[width=0.5\textwidth]{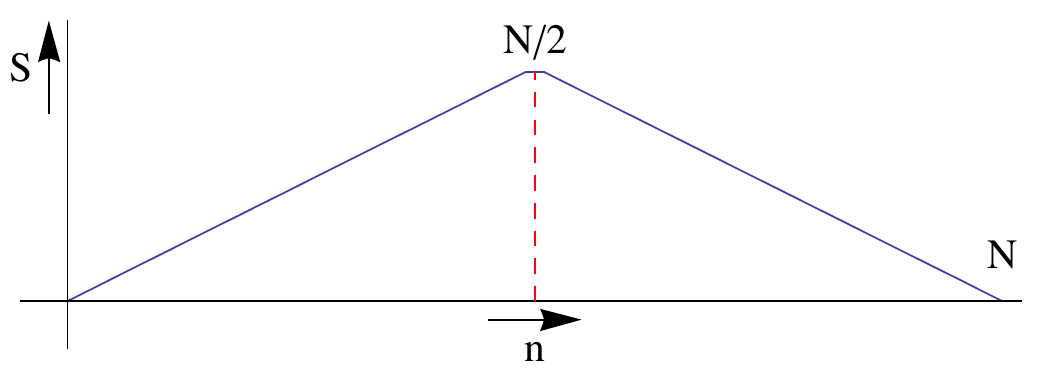}
\caption{\em Behaviour of the entanglement entropy $S_n$ with
$n$.}
\label{entangentr}
\end{center}
\end{figure}

Let us first describe how how the ``strong subadditivity''
of entropy puzzle of \cite{Mathur:2009hf,Almheiri:2012rt} is resolved in the spin chain.  Given the state $|\Psi \rangle$ in \eqref{stspinchain}, if we consider the reduced density matrix of the first $n$-qubits then for most choices of the coefficients $a_i$, we expect that \cite{Page:1993df} 
\be
S_n = -\tr(\rho_n \ln \rho_n) = \left[n \theta\left({\nc \over 2} - n \right) + (\nc - n) \theta \left(n - {\nc \over 2} \right) \right] + \Or[2^{-{\nc \over 2}}]  \ln 2. 
\ee
This curve is shown in Fig. \ref{entangentr}.

The well-known interpretation of this equation is as follows.
 Consider the case where $k-1 >  {\nc \over 2}$ bits have evaporated and we are considering the evaporation of the $k ^{\text{th}}$ qubit. Then it is possible to find a set of operators ---which we will call $\paulthat^{k}_a$ --- that obey the usual $SU(2)$ algebra and satisfy 
\be
\label{paulthatprops}
[\paulthat^k_a, \paul^k_b] = 0, \quad \paulthat^k_a |\Psi \rangle  = -\paul^k_a |\Psi \rangle,
\ee
Hence, these $\paulthat^k_a$ operators effectively realize the algebra of the $k^{\text{th}}$ spin, without acting on that spin at all. This is the statement
that these operators are ``entangled'' with the $k^{\text{th}}$ spin.

The choice of $\paulthat^k_a$ as operators on the first $(k-1)$ bits is not unique: we can take it to act on any selection of ${\nc \over 2}$ qubits in these $(k-1)$-qubits. Nevertheless, the strong subadditivity condition, in this context, can be stated by saying that any such operator that we find {\em cannot} commute with the spin-operators on the first $(k-1)$-bits:
\be
[\paulthat^k_a, \paul^m_b] \neq 0, \quad \text{for~some~m~with}~1 \leq m \leq k-1.
\ee

How is this consistent with our explicitly constructed operator $\pault^k_j$, which {\em appears} to commute with all the ordinary spins.  The point is that, as an {\em operator}, it is indeed true that $[\pault^k_j, \paul^m_i] \neq 0$, for some $m \in {1 \ldots k-1}$. But, nevertheless , this commutator annihilates the state $|\Psi \rangle$, and its descendants produced by acting with the insertion of up to $K$-ordinary and mirror operators, 
\be
\label{commutvanish}
[\pault^k_i, \paul^m_j] \paul^{m_1}_{a_1} \ldots \paul^{m_p}_{a_p} |\Psi \rangle = 0, \quad \forall \{m,j,i, a_1, \ldots a_p, m_1, \ldots m_p\}.
\ee
The equation continues to be true if we replace either some or all of the ordinary $\paul^{m_p}_{a_p}$ matrices with the tilde-counterparts.

Thus, within this model, our construction provides a precise realization of black hole complementarity. After the Page time, the operators in the interior of the black hole secretly act on the early radiation as well. Nevertheless, this action is exactly ``local'' within $K$-point correlators because of the vanishing of the commutator, as displayed in \eqref{commutvanish}. The physical interpretation is that locality can be preserved exactly 
unless we try and consider correlators with $\Or[\nc]$ insertions.

\subsubsection{Resolving the Strong Subadditivity Puzzle in the CFT}
We now resolve the strong subadditivity puzzle within the CFT. First,
we need to formulate the puzzle precisely, and even this exercise 
suffers from some subtleties as we describe here. One possible precise formulation is to use the
``plasma-ball'' construction of \cite{Aharony:2005bm}, and this is what we use.
After formulating the paradox in terms of plasma-ball evaporation
in the boundary CFT, we then describe a resolution that is identical
in spirit to the resolution for the spin-chain demonstrated above.

\paragraph{\bf Subtlety in formulating the strong subadditivity paradox in quantum gravity}
Making the strong-subadditivity paradox precise within quantum gravity is actually
somewhat subtle.\footnote{We are grateful to Ashoke Sen for emphasizing this point to us, in several discussions.}  We summarize
this difficulty and then attempt to reformulate the strong subadditivity
paradox in terms of the CFT in an independent manner, and resolve it
in that context.

The naive formulation of strong subadditivity relies on the idea that
$S_E$ rises, and then falls to zero. Within local quantum field theory,
$S_E$ could be defined as the entanglement entropy between the region ``outside'' and ``inside'' an imaginary barrier that is placed at a fixed distance from the black hole. 

The subtlety is that the entanglement entropy of these regions even in the vacuum is infinite. This may not be the case in a fully theory of quantum gravity, but we do not understand
how quantum gravity effects automatically resolve this divergence, in any detail. One could try and define a ``renormalized'' entanglement entropy, by considering the ``excess'' entanglement entropy in the state $|\Psi \rangle$ over the vacuum
\[
S_E^{\text{ren}} = S_E^{\Psi} - S_E^{\Omega}
\]
However, now we run into the following difficulty: the definition
above is very sensitive to the precise definition of the region $E$,
since both terms on the right hand side are divergent. Since the metric is changing as the black hole
evaporates, it does not make sense to define $E$ to be the region inside
a given coordinate distance. Depending on how precisely we define the
region, we can make $S_E^{\text{ren}}$ increase, decrease or stay
constant, even as we cross the Page time.

It may be possible to avoid this subtlety by defining the entanglement
entropy on ${\mathcal I}^+$, but below we explore an alternate
formulation, which avoids having to go to asymptotic infinity.

\paragraph{Formulation of the Strong Subadditivity Paradox in terms of Plasma Ball Evaporation\\}

So, we now describe an alternate formulation that helps us sidestep this issue of the back-reaction of Hawking radiation on the metric.\footnote{We are grateful to Shiraz Minwalla for emphasizing the
utility of this plasma ball perspective to us.}
Let us imagine a plasma-ball in the conformal field theory, which is a lump
of a deconfined phase  that is localized in the boundary directions. Gravity solutions corresponding to such a configuration were found in \cite{Aharony:2005bm}. To consider
these solutions, we switch back to the picture of the CFT living on $R^4$.

The picture of Hawking radiation is shown in the figure below.
\begin{figure}[!h]
\begin{center}
\includegraphics[width=0.6\textwidth]{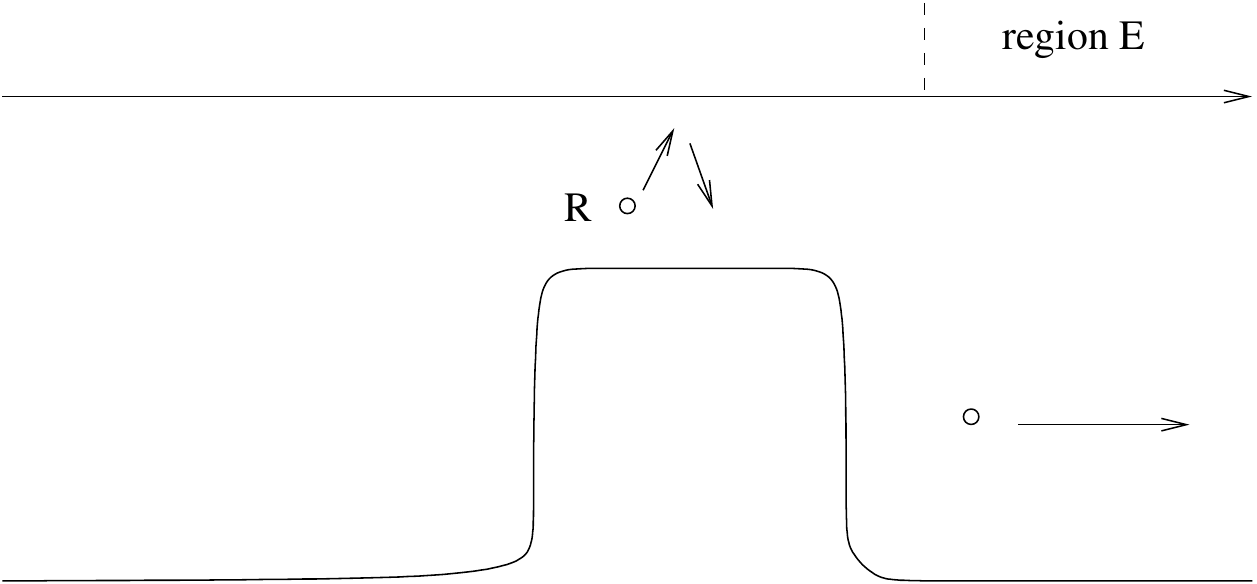}
\caption{\em Modeling Hawking Radiation in a ``Localized Black Hole'': Hawking
Quanta $R$ emitted towards the boundary reflect and fall back, but quanta emitted in the non-compact directions escape to infinity. \label{plasmaballconfig}}
\end{center}
\end{figure}
On the boundary, we expect that the quark-gluon plasma will decay
via the emission of glueballs. These glueballs propagate away
freely from the original plasma ball.  

An {\em intuitive} way to think of this process, which is valid at
large $N$ is to imagine a ``plasma ball operator'' ${\cal P}_M$
which creates a Plasma ball of energy $M$, but no glueballs. We can
now consider operators that create wave-packets of glueballs
\be
\label{ofdef}
O^i(f) = \int O^i(\vect{x}) f(\vect{x}) d^d \vect{x},
\ee
where $f(\vect{x})$ is some function that controls the profile of the 
wave-packet on the boundary.
At large $N$, we can imagine that the evaporation of the Plasma
ball can be thought of as the Schrodinger evolution of the 
state from a pure Plasma-ball state to a Plasma ball of lower energy $M'$
and some glueball wave-packets.
\be
{\cal P}_M |\Omega \rangle \longrightarrow  \sum_{\{N\}} \alpha_{\{N\}}(t) (O^i(f_n))^{N(i,n)}  |{\cal P}_{M'} \rangle,
\ee
where the sum runs over all functions $N(i,n)$. The functions $f_n$
are some suitably regularized basis of glueball wave-packet profiles,  and the $\alpha$ are 
some coefficients. 
We caution the reader again that the equation above is valid only 
at large $\nc$, where we can clearly differentiate the Plasma Ball
from the glueballs, and we provide it only for intuition. 

The advantage of formulating this puzzle in the field theory, is that we can make a much more precise statement. Consider the entanglement entropy 
of the region $E$ on the {\em boundary.}  We can regulate this entanglement entropy in 
some time-independent way, without having to worry about
gravitational-back reaction. Then we can consider the behaviour
of the entanglement entropy of region $S_E$ with time,
 and we expect that this has the form expected from Page's general analysis, which is shown in Figure \ref{entangentr}.

Now, we can rephrase the strong subadditivity paradox as follows in 
this setting. We can, as usual, map the boundary fields to bulk fields, and  construct the mirror operators inside the black hole, and 
use them to construct the bulk fields. Now, we can consider
some ``wave-packet'' of the bulk-operator 
\be
\phi_{\text{CFT}}^i(g) = \int \phi_{\text{CFT}}^i(z, \vect{x}) g(z, \vect{x}) d^{d+1} \vect{x},
\ee
where $g$ is a function in $d+1$-dimensions with support in some region
entirely inside the bulk black-hole, behind the horizon.
We might expect that this corresponds to some localized excitation
entirely inside the Plasma-ball. However, let $i'$ run over the set of glueball primaries that correspond to the
supergravity modes. Then, strong subadditivity implies that
\be
\label{sseopversion}
\exists~i', h(x)~\text{such~that} \quad [\op^{i'}(h), \phi_{\text{CFT}}^i(g)] \neq 0,
\ee
where $O^{i'}(h)$ is defined by \eqref{ofdef}, and the function $h$ 
on the boundary has the property that it vanishes everywhere {\em inside} the (past) light cone of the domain of $g$. 

We can also phrase this as a property of the function $\tO^i(\vect{x})$. Consider a ``wave-packet'' of this mirror operator on the boundary, $\tO^i(g_{\text{bound}})$ defined by \eqref{ofdef}, where $g_{\text{bound}}$ 
has support on a region after the ``Page time'' of the plasma-ball.  Then we find that strong subadditivity implies that
\be
\exists i', h', ~\text{such~that}~[\op^{i'}(h'), \tO^i(g_{\text{bound}})] \neq 0,
\ee
where $h'$ is localized
on some region that is spacelike separated
from the domain of $g_{\text{bound}}$
So, $\tO^i(\vect{x})$ cannot be a local operator on the boundary.

In words, this is telling us that the operators inside the black hole after the Page time $\tO_{\vect{m_0}, \vect{n_0}}$ must act on the glueball modes in the region $E$. Since these glueball modes also constitute the Hawking
particles outside the horizon, we see that we have a precise version of the
statement that the interior of the black-hole has support on the degrees
of freedom outside.

As we have seen several times above, however, the statement \eqref{sseopversion}
is an operator statement. What we really want is that within low-point
correlators built on the state $|\Psi \rangle$, 
\be
[O^{i'}(h), \phi_{\text{CFT}}^i(g) ] \doteq 0,
\ee
and there is absolutely no contradiction between this statement and
\eqref{sseopversion}.

\paragraph{\bf Distilling the Entangled Bit?}
It is worthwhile to briefly comment on another version of the 
strong subadditivity paradox. Could the observer outside ``distill''
the part of the outgoing radiation that is entangled with the near-horizon mode, and then jump into the black hole to obtain a contradiction.

It is simple to see, as we now show,  that to ``distill'' the entangled bit, the 
infalling observer has to measure a correlator where the energy of the 
insertions scales with  $\nc$. 
First note that to ``distill'' the entangled bit, is the
same as finding an operator that is a  polynomial in the ordinary operators $\op^{i}_{n, \vect{m}}$, but one that does {\em not} commute with the operators $\tO^i_{n, \vect{m}}$. More precisely, calling 
this extraordinary operator ${\cal E}$, we need 
\be
\label{extraorddef}
\begin{split}
&{\cal E} = P \left({\op}^i_{n, \vect{m}} \right), \\
&[{\cal E}, \tO^{i_0}_{n_0, \vect{m_0}}] |\Psi \rangle \neq 0, \quad \text{for~some~} i_0,n_0,\vect{m_0},
\end{split}
\ee
where $P$ is some  polynomial.

 Now, in \eqref{tildescft}, we have already ensured that the operators $\tO^{i_0}_{n_0, \vect{m_0}}$ commute with all elements of ${\cal A}$, while acting on the state
$|\Psi \rangle$. In particular, this includes all polynomials, in which the the energy of every
monomial does not scale with $\nc$. 
So, we see that the polynomial $P$ in \eqref{extraorddef} must include a term, that violates \eqref{energybound} to have 
a non-trivial commutator with $\tO^{i_0}_{n_0, \vect{m}_0}$ when
acting on the state $|\Psi \rangle$. 

This could happen, if for example, we consider a measurement that has
$\Or[\nc]$ insertions of supergravity fields. Or else, we could
run into this difficulty if we take $\Or[\sqrt{\nc}]$ insertions of 
fields with an energy $\Or[\sqrt{\nc}]$ each. Translated to the supersymmetric $SU(N)$ theory, these measurement correspond to correlators with $N^2$ insertions of supergravity fields, or  $N$ insertions
of giant graviton operators. 

Our point is that these {\em correlators do not have an interpretation
in terms of fields propagating on a perturbative spacetime}, and so it
is not surprising that our intuitive concepts of spacetime---such
as the idea that the interior and exterior of the black hole are distinct and well separated regions--- break down for such correlators.

We would like to make a few more comments on this issue. One way of 
getting around the difficulty above has been to ``couple'' the CFT
to another large system, and then perform the measurement in that large
system. This does not affect our conclusions here at all. To the extent
that the CFT coupled to the large system has a spacetime interpretation,
this interpretation breaks down for measurements in this extended system
that correspond to inserting $N^2$ supergravity fields.

We emphasize that our argument here is entirely independent of the 
bounds from quantum computing that have been discussed in this context \cite{Harlow:2013tf}. This argument has been criticized in 
the later versions of \cite{Almheiri:2012rt}, and we refer the reader to that paper. But, in any case we do not feel that these bounds are crucial to the discussion on the information paradox.

Finally, it is amusing to note that, in any case, ``distilling'' the entangled bit requires a state-dependent measurement (see Appendix \ref{appmeasurement}). 
Hence, if state-dependent
measurements are disallowed even in principle then an observer who is part of the 
bulk-spacetime in the first place, and then evolves autonomously
with this spacetime,  cannot make the required measurement.

\subsection{The $[E,\widetilde{B}] \neq 0$ Paradox \label{resolvevanishcommut}}
An immediate objection to the picture of ``complementarity'' that we
have outlined above is that the commutator of measurements on the 
radiation outside, and on measurements inside will not vanish. This
is based on the observation that generically the commutator of
two qubits is $O(1)$. 

Let us briefly explain this objection although, it obviously does not apply to our construction.
The point is that if we take the operator ${\op}^i_{n_0, \vect{m_0}}$ and ``scramble''
it using some generic $e^S \times e^S$ unitary matrix $U_{\text{scram}}$ then it is generically true that
\be
[U_{\text{scram}} {\op}^i_{n_0,\vect{m_0}} U_{\text{scram}}^{\dagger}, {\op}^i_{n_0, \vect{m_0}}] \sim \Or[1],
\ee
in the sense that the generic size of the eigenvalue of the matrix on the left is $\Or[1]$. This non-zero commutator can be detected within low-point correlators.

We emphasize that our construction of the mirror operators is {\em not} of this sort, and so the argument above fails completely. As we have emphasized many times above, our entire construction is designed to ensure that the commutator $[\widetilde{\op}^i_{n_0,\vect{m_0}}, {\op}^i_{n_1, \vect{m_1}}]$ is undetectable within low-point correlators.

So, our version of complementarity cannot be used to send messages across
spacelike distances, at least within the approximation that the spacetime
geometry makes sense at all.

\subsection{The Lack of a Left Inverse Paradox \label{resolveleftinv}}
Now, let us turn to some of the other arguments of \cite{Almheiri:2013hfa}. One of these arguments goes as follows. Consider some conformal primary corresponding to a supergravity field, and 
consider the action of $\tO^i_{-n, \vect{m}}$ on the pure state $|\Psi \rangle$, where $n$ is any positive integer. 
This operator acts like a ``creation'' operator for the field behind the horizon. For this subsection, we
adopt the following shorthand notation
\be
\begin{split}
&G_{\beta}(n, \vect{m}) = \langle \Psi | [{\op}_{n, \vect{m}}^i, ({\op}_{n, \vect{m}}^i)^{\dagger}] |\Psi \rangle,  \\
&b = {1 \over \sqrt{G_{\beta}(n,\vect{m})}} {\op}^i_{n, \vect{m}}; \quad b^{\dagger} = {1 \over \sqrt{G_{\beta}(n,\vect{m})}} ({\op}^i_{n, \vect{m}})^{\dagger}, \\  &\widetilde{b} = {1 \over \sqrt{G_{\beta}(n, \vect{m})}} \tO^i_{n,\vect{m}}; \quad \widetilde{b}^{\dagger} = {1 \over \sqrt{G_{\beta}(n, \vect{m})}} (\tO^i_{n,\vect{m}})^{\dagger}.
\end{split}
\ee
However, by the
relation \eqref{cftgaugeinvar}, we have $[H_{\text{CFT}}, \widetilde{b}^{\dagger}] = -\omega_n \widetilde{b}^{\dagger}$
so the action of this operator {\em lowers} the energy in the CFT and maps a state of average energy $E$
to a state of average energy $E - \omega_n$. (Recall that $\omega_n$ was defined to be $n \omega_{\text{IR}}$ in section \ref{tildescft}.) 
Nevertheless, some simple algebra shows us that
this operator satisfies the following
relation to leading order in the ${1 \over \nc}$ expansion
\be
\begin{split}
& \lst  \left( \widetilde{b} \widetilde{b}^{\dagger} - \widetilde{b}^{\dagger} \widetilde{b}\right) \rst  = \lst \left( \widetilde{b} e^{{\beta \omega_n \over 2}} b - \widetilde{b}^{\dagger} e^{-\beta \omega_n \over 2} b^{\dagger} \right) \rst \\
&= \lst \left(b b^{\dagger} - b^{\dagger} b \right) \rst \\
&= 1 + \Or[{1 \over N}].
\end{split}
\ee
This allows us to write
\be
\label{tildealgebraexpected}
\widetilde{b} \widetilde{b}^{\dagger} \doteq 1 + \widetilde{b}^{\dagger} \widetilde{b}.
\ee
We have been careful to put a $\doteq$ above, once again indicating that this relation holds within
low-point correlation functions. 

The "lack of a left-inverse paradox" is simply the claim that had we had a true operator equality in \eqref{tildealgebraexpected} then since the right hand side is a manifestly positive operator, $\widetilde{b}^{\dagger}$ should have a left-inverse. But this appears to be impossible, since there are fewer states in the smaller energy range.

Of course, we do not have any contradiction with our state-dependent construction, where \eqref{tildealgebraexpected} is not satisfied as an operator equation, but as a relation that holds within low-point correlators
constructed on $|\Psi \rangle$. 

In fact, we can choose the $\widetilde{b}^{\dagger}$ operators to be rather sparse on the full Hilbert space. This is because the linear equations of \eqref{tildedefcftactpsi} and \eqref{tildedefcftcomord}  are not in contradiction with multiple null-vectors for $\widetilde{b}^{\dagger}$. For example, as we pointed out 
in section \ref{tildegeneral}, we could choose $\widetilde{b}_m$ so that it obeys the equations \eqref{tildedefcftactpsi}, \eqref{tildedefcftcomord} within the space $\shil$, but annihilates all vectors that are
orthogonal to this subspace.
By construction, this would not create any contradiction with low-point correlators. 

What the argument of \cite{Almheiri:2013hfa} tells us is whatever action we choose for $\widetilde{b}^{\dagger}$ outside the space $\shil$ this operator must have null vectors. However, within low-point correlators, these null vectors are completely unobservable and it appears that these operators obey the algebra \eqref{tildealgebraexpected}.

\begin{figure}[!h]
\begin{center}
\resizebox{0.4\textwidth}{!}{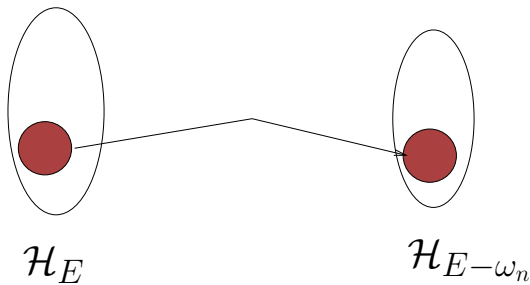}
\caption{\em $\widetilde{b}^{\dagger}$ is a sparse operator, and it maps the intersection of $\shil$ with the
space of states of average energy $E$ to the intersection of $\shil$ with the states of average energy $E - \omega_n$. The precise domain and range depend on the base state $|\Psi \rangle$. \label{figwidetilddeact}}
\end{center}
\end{figure}

Pictorially, we can depict the action of $\widetilde{b}^{\dagger}$ by 
figure \ref{figwidetilddeact}

\paragraph{\bf Union of all Constructions?}
The paper \cite{Almheiri:2013hfa} contained a further argument to try and account for the situation described in Fig. \ref{figwidetilddeact}. The argument was that, if we consider the ``union of all constructions'', we could get a contradiction with the expectation of  \eqref{tildealgebraexpected}. 

This argument was not spelled out in detail, but by this we understand the following: the operator $\widetilde{b}_{\omega}^{\dagger}$ provides a map between states of higher and lower energy, as shown in Fig. \ref{figwidetilddeact}. This map depends on the state. Perhaps, the authors of \cite{Almheiri:2013hfa} meant to suggest that by considering different states, and by consider the union of all these maps, we could obtain operators that satisfied \eqref{tildealgebraexpected} as an operator equation, rather than just on the states under consideration. 

\begin{figure}[!h]
\begin{center}
\resizebox{0.4\textwidth}{!}{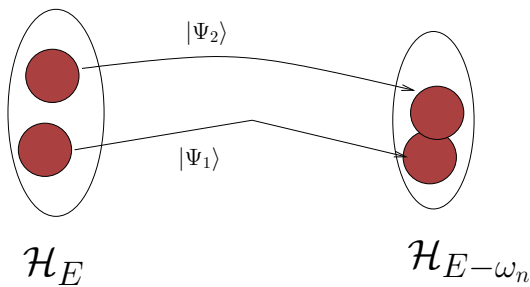}
\caption{\em Extending the domain of $\widetilde{b}^{\dagger}$ by taking the union of
the maps corresponding to different base states, leads to a many-to-one function.\label{figtildeovercover}}
\end{center}
\end{figure}

Here, we wish to point out that the ``union of all constructions'' does not help in this. If one tries to take the maps corresponding to different base states $|\Psi \rangle$, so as to completely cover the space with
energies in a band about $E$, then we invariably end up {\em over-covering} the space with energies in a band about $E - \omega_n$. This is shown in figure \ref{figtildeovercover}.

\subsection{The ``$N_a \neq 0$'' paradox \label{resolvenaneq0}}
We now turn to an argument made in \cite{Marolf:2013dba}, leading to
the apparent conclusion that AdS/CFT cannot describe the interior of the black hole.  First, we summarize the argument and then show why it fails
in our construction.

\paragraph{\bf Summary of the Marolf-Polchinski Argument:\\}
We start by defining two ``number operators''
\be
\label{numberops}
\begin{split}
&N_b = b^{\dagger} b, \\
&N_a = {1 \over 1 - e^{-\beta \omega_n}} \left[ \left(b^{\dagger} - e^{-\beta \omega_n \over 2 } \tb \right) \left(b - e^{-\beta \omega_n \over 2} \tb^{\dagger} \right) + \left( \tb^{\dagger} - e^{-\beta \omega_n \over 2} b \right) \left( \tb - e^{-\beta \omega_n \over 2} b^{\dagger} \right)  \right].
\end{split}
\ee
We see that $N_b$ measures the number of particles at frequency $\omega_n$
as seen by the Schwarzschild observer outside. The operator 
$N_a$ is the standard number operator as seen by the infalling observer
and the factors of $e^{- \beta \omega_n}$ come from the
standard Bogoliubov transformations between these two frames \cite{birrell1984quantum}.

Note that \eqref{numberops} is also relevant in Rindler space, where
$N_b$ could be the number operator measured by a Rindler observer
and $N_a$ the number operator measured by the Minkowski observer. However, now we come to a crucial difference between the Rindler and the AdS/CFT case. We see that
\be
[H_{\text{CFT}}, N_b] = 0 + \Or[\omega_{\text{min}}].
\ee
On the other hand commutator between the Minkowski Hamiltonian and the 
Rindler number operator clearly does not vanish. So, the CFT Hamiltonian
behaves like the Rindler Hamiltonian. 

As a consequence of the fact above, we can consider a set of eigenstates
of $N_b$, which we will denote by $|\bar{E}, n_b \rangle_i$, which have 
the property that
\be
H_{\text{CFT}} |\bar{E}, n_b \rangle_i = \bar{E} |\bar{E}, n_b \rangle_i + \Or[\omega_{\text{min}}], \quad N_b |\bar{E}, n_b \rangle_i = n_b |\bar{E}, n_b \rangle_i.
\ee
The two conditions above, which specify the energy up to an accuracy $\omega_{\text{min}}$ and the $N_b$ eigenvalue still leave an enormous degeneracy, and the index $i$ is meant to denote the different states that can satisfy this property. Now, consider the span of all such states that have mean
energy in some range
\be
{\cal S} = \text{span}\{|\bar{E}, n_b \rangle_i: \bar{E}_0 - \Delta \leq \bar{E} \leq \bar{E}_0 + \Delta \}.
\ee
It seems clear that no element of the {\em basis} of ${\cal S}$, that
we used above, has a smooth
horizon. If we reconstruct the bulk, for such a state, using the bulk-boundary map, and evaluate the stress-tensor as we approach what would have been the horizon, it will diverge. This is entirely consistent with the fact
that the states in this basis do not satisfy \eqref{noannihilgeneral}, and so  we cannot construct the mirror operators on them.  However, we can consider the following harder question: 
\begin{quote}
{\bf Question:} ``Consider a typical state
in  ${\cal S}$, picked with the Haar measure on this space. Does such a state have a smooth horizon, or not?''
\end{quote}

The authors of \cite{Marolf:2013dba} claim that for $\Delta \sim \Or[\beta^{-1}]$, the set ${\cal S}$ covers almost the entire microcanonical ensemble with width $\beta^{-1}$ centered on $\bar{E}_0.$  We do not entirely understand the basis for this estimate of the width, or the subtleties in determining whether ${\cal S}$ really contains almost all states in the microcanonical ensemble. Neither of these details are provided in the paper \cite{Marolf:2013dba}.  As a consequence, the reader should note that there 
may be a subtle difference between the question above, and the question
of whether a typical state in the microcanonical ensemble has a smooth horizon or not. 

The authors of \cite{Marolf:2013dba} argued that the answer to the
question above is negative. We will now review their argument, and then show that it fails for
state-dependent operators, and that typical states in the span of ${\cal S}$ do have a smooth horizon.

Let us say that some {\em state-independent} operator in the CFT
could tell us the particle number as measured by the infalling observer. We will call such an operator $N_a^{\text{univ}}$. Then, we could compute
\be
\langle N_a^{\text{univ}} \rangle = {1 \over \dim({\cal S})} \text{Tr}_{\cal S} \left(N_a^{\text{univ}} \right) = {1 \over \dim({\cal S})} \prescript{}{i}{\langle \bar{E}, N_b|} N_a^{\text{univ}} | N_b, \bar{E} \rangle_i = \Or[1].
\ee
The last equality follows because $N_a^{\text{univ}}$ is a positive operator. Moreover since
the state with $N_a^{\text{univ}} = 0$ has a thermal distribution of $N_b$, the expectation value of $N_a^{\text{univ}}$ in any $N_b$ eigenstate is $\Or[1]$.

This is consistent with the fact that typical states with a definite
Rindler energy are not regular as we cross the Rindler horizon.

\paragraph{\bf Failure of the argument for state dependent operators\\}
First, we point out the following simple fact. Consider a typical 
state $|\Psi \rangle \in {\cal S}$. With respect to the usual set of observables ${\cal A}$ defined in section \ref{tildescft}, we would expect such a state
to satisfy \eqref{noannihil}, and so we can define the mirror operators. Now, it is immediately clear from \eqref{tildedefcftactpsi} and \eqref{tildedefcftcomord} and the definitions \eqref{numberops} that
\be
N_a |\Psi \rangle = 0,
\ee
which follows from the simple observation that  both 
\be
\left(b - e^{\beta \omega_n \over 2} \tb^{\dagger} \right)|\Psi \rangle = 0, \quad \text{and} \quad\left( \tb - e^{\beta \omega_n \over 2} b^{\dagger} \right)|\Psi \rangle = 0.
\ee
The reason that the argument above fails is that our operator $N_a$ is state-dependent, and in fact, it is partly designed to ensure that $N_a = 0$
in a typical state $|\Psi \rangle.$ For such an operator, the change
of basis in the trace clearly fails.

Consider another simple example of this sort. Let us say that $\rho_{\psi} = |\Psi \rangle \langle \Psi |$
is the {\em density operator} corresponding to the state $|\Psi \rangle$.
Clearly, we have $\left(\rho_{\psi} - 1 \right)|\Psi \rangle = 0$,
and this is true for any state $|\Psi \rangle$.  On
the other hand $\tr_{\cal S}\left(\rho_{\psi} - 1\right) = 1 - \text{dim}({\cal S}) \neq 0$.  These two statements are not in any contradiction, 
because $\rho_{\psi}$ is a state-dependent operator precisely like 
our $N_a$.\footnote{One difference, of course, is that whereas $\rho_{\psi}$ is defined for all states $|\Psi \rangle$, including those that are 
eigenstates of $N_b$, our construction of mirror operators works for typical states that satisfy \eqref{noannihil}. As we mentioned above, for eigenstates of $N_b$, the
construction fails, and this is consistent with the physical understanding
that such states have no ``interior''.}

\subsection{Unitarizing Hawking Radiation with Small Corrections \label{resolvesmallcorr}}

We now address the claim that ``small corrections'' cannot unitarize 
Hawking radiation \cite{Mathur:2009hf}.  
Before, we address this claim, it is extremely important to specify what, precisely, is meant by ``small corrections.'' From our perspective, 
\begin{quote}
{\em The size of corrections is estimated by the size of corrections to low point correlation functions of light local operators compared to the results that we would get from ordinary effective field theory in the black hole background.}
\end{quote}
Thus, for example, if there is structure behind the horizon then we might expect large corrections to correlators involving insertions on either side of the horizon. Similarly, if the process of Hawking radiation is modified significantly, then we might expect large corrections even to correlators outside the black hole, because the state will not be
well approximated by the Unruh vacuum.

We stress that it is important to adopt the definition above, rather
than one that looks at, say, whether the full wave-function at
the end of Hawking evaporation is close to that predicted by the Hawking calculation. We can see the error in this kind of approach even if we 
consider the set of states that are dual to a large black hole in AdS. 
The wave-functions of these states differ widely, but from a geometric
perspective, or equivalently from the perspective of expectation values
of elements in the set ${\cal A}$, these states are almost impossible to 
distinguish.

With this prelude, we now consider two cases, and show how the Hawking
evaporation process is consistent with
\begin{enumerate}
\item
Small corrections outside the horizon.
\item
Small corrections across the horizon.
\end{enumerate}
We will phrase our arguments in this subsection in terms of the 
spin-chain toy-model of \eqref{dualspinchain}, since this is the
context in which the claim of \cite{Mathur:2009hf} was formulated.

\subsubsection{Small Corrections Outside the Horizon}

As we showed in our previous paper \cite{Papadodimas:2012aq}, it is perfectly consistent with unitarity for  correlators of local fields outside the black hole should be very close to their semi-classical values, as calculated in the Unruh vacuum. In the spin-chain model that we have described, this is the following simple statement. For a correlator made up of products of spin-operators, where the number of insertions does not scale with $K$, we have
\be
\langle \Psi | \paul_{i_1}^{a_1} \ldots \paul_{i_p}^{a_p} | \Psi \rangle = \tr\left(\rho_{i_1 \ldots i_p} \paul_{i_1}^{a_1} \ldots \paul_{i_p}^{a_p} \right)  = {1 \over 2^p} \tr \left( I_{2^p \times 2^p} \paul_{i_1}^{a_1} \ldots \paul_{i_p}^{a_p} \right) + \Or[2^{-\nc \over 2}],
\ee
where $\rho_{i_1 \ldots i_p}$ is the reduced density matrix for the spins $i_1 \ldots i_p$ in the state $|\Psi \rangle$.

The ordinary $\paul_{i}^a$ operators correspond to measurements made
outside the horizon.
So, the interpretation of this equation is as follows: for the purposes of computing correlation functions with a small number of insertions outside the black hole, it is always possible to use the thermal density matrix---which, in our toy-model, is just the identity matrix.

In our previous paper, we discussed this issue in a slightly different language by pointing out that a seemingly thermal density matrix could be unitarized by a correction matrix, that was exponentially suppressed. 

This means that it is possible to have a situation where the exact density matrix $\rho_{\text{exact}}$, which comes from unitary evolution 
differs from the
Hawking density matrix, which is just the identity here, by a correction matrix $\rho_{\text{corr}}$ whose elements, in some basis are very small.
\be
\label{unitarizingcorr}
\rho_{\text{exact}} = \rho_{\text{hawk}} + 2^{- N} \rho_{\text{corr}}.
\ee
This is consistent with the relation above. Correlators computed
in the two density matrices vary by a factor of $2^{-N \over 2}$, since
the typical contribution of the second term is $2^{-N \over 2}$.

In this context, we should mention that when $\Or[\nc]$ particles have
been emitted, it may look like the correction matrix is comparable
to the original Hawking matrix.  This is just an indication of the fact that the Hawking
computation cannot reliably predict the amplitude for any
given configuration of $\Or[\nc]$ emitted particles, since this
amplitude is exponentially suppressed. When we focus on correlators with a small number of insertions, then we invariably end up focusing on a reduced density matrix with a much smaller dimension, and the difference between the unitary and thermal density matrix
vanishes.

So, to conclude this subsection, small corrections to correlators of an O(1)
number of fields outside the black hole are completely
consistent with unitarity.  

\subsubsection{Small Corrections to Correlators Across the Horizon}
The proof of \cite{Mathur:2009hf} focused not just on the density matrix outside the horizon, but on the 
evolution of the wave-function of the theory during Hawking evaporation. 

The assumption is that the
full wave function evolves as

\be
\label{wavefuncevolve}
|\Psi \rangle_{t+1} = {1 \over \sqrt{2}} |U \Psi \rangle_t \otimes (|0 \rangle_B |1 \rangle_{\tilde{B}}  + |1  \rangle_{B} |0 \rangle_{\tilde{B}}. 
\ee
Here $|\Psi \rangle$ encodes the state of the black hole and the radiation at a given time. \eqref{wavefuncevolve} should be understood as the statement that the wave-function evolves by the addition of two entangled particles: of which one falls into the black hole and the other falls out, while the extant black hole and radiation evolve autonomously through the unitary matrix $U$. 

Now \eqref{wavefuncevolve} clearly cannot be taken literally in a theory of quantum gravity, especially if we are to take the lessons from AdS/CFT seriously. For example, as we explored above, a clear setting in which Hawking radiation
can be observed is given by the localized plasma-ball in AdS/CFT \cite{Aharony:2005bm}. As shown in Figure \ref{plasmaballconfig}, the plasma-ball is a localized lump of quark-gluon plasma that gradually evaporates via the emission of glueballs. What this teaches us is that Hawking radiation should be modeled as a process in which the black hole loses some of its energy to the emission of an external particle, while the remaining degrees of freedom reorganize themselves so that it {\em appears} that a particle has been created within the black hole. 

Hence, what we must demand is that, while the underlying dynamics may be quite complicated, the  wave-function {\em effectively} evolves as in \eqref{wavefuncevolve}. What does this mean in terms of correlators? 
In our spin-chain model of Hawking evaporation, after $k$ spins have
evaporated and we can measure measure operators $\paul_k^a$ on the emitted Hawking particle and there is an additional operator which effectively commutes with the spin-operators for the emitted Hawking quanta that is exactly correlated with measurements of $\paul_k^a$.   

The punch-line is that after $\nc$-steps for the purposes of low-point correlators involving the operators $\pault_n^a$ and $\paul_m^a$, the state effectively looks like a collection of $\nc$-Bell pairs, as shown below equation \eqref{tfdspinchain}. But, in reality, it is only an entangled state involving $\nc$, and not $2 \nc$, spins. 

We should emphasize two important factors in our construction, which were {\em not accounted} for in the construction of \cite{Mathur:2009hf}. One of them is that, loosely speaking, the interior particle may be constructed partly out of the previously emitted radiation. The precise version of this statement is, of course, given
by our operators $\pault_k^a$ above, which for $k > {\nc \over 2}$, must necessarily act on some of the first ${\nc \over 2}$ particles as well. 

The second point has to do with the state-dependence of our construction.  The papers \cite{Mathur:2009hf,Mathur:2011uj} described models of ``burning paper.'' 
In these models, the qubit $\widetilde{B}$ was identified with some specific qubit constructed in the remaining spin-chain. It is clear, that after ${\nc \over 2}$ bits have evaporated, 
there is no state-independent operator that can be perfectly correlated with the emitted spin. Even, for the first ${\nc \over 2}$ bits, the correlation between the qubit-$\widetilde{B}$ and the qubit $B$ can be maintained only by fine-tuning the state. 

Let us say this more precisely. Consider some {\em fixed} state-independent
operators $\underbar{s}_a$ and try and make this play the role of the 
mirror to the first spin. 
Then in some given state of the spin-chain,  we require
$\underbar{s}_a |\Psi \rangle = -\paul^1_a |\Psi \rangle$.  Clearly, for 
some fixed operator $\underbar{s}_a$, this condition will not be met
for a generic state $|\Psi \rangle$. So, for a generic state $|\Psi \rangle$, the correlator $\langle \Psi | \b{s}_a \paul^1_a |\Psi \rangle \approx 0$, whereas we would like it to have the value $-1$. 
This is what leads to the suggestion that in models of ``burning paper''
there are large corrections to correlators between operators ``inside'' 
and ``outside'' the black hole.

These conclusions {\em do not} hold for our state-dependent operators $\pault^i_a$. We clearly have $\pault^i_a |\Psi \rangle = -\paul^i_a |\Psi \rangle, \forall i$. Moreover, we see that {\em for a generic state}, we have
$\langle \Psi | \pault^i_a \paul^j_b | \Psi \rangle = -\delta^{i j} \delta_{a b}$, precisely as we need. So, with state dependent operators, we
can arrange to have small corrections across the horizon as well, while
remaining within a unitary framework.

Before, we conclude this section we would like to point out that the argument of \cite{Marolf:2013dba} can, in some sense, be understood to be a re-phrasing of the argument that small corrections cannot unitarize the black hole. In that discussion, the argument was that if we selected the operators corresponding to $\widetilde{b}_m$ and $\widetilde{b}_m^{\dagger}$ to be some {\em fixed} operators in the CFT Hilbert space, then it is clear that their action on the state will generically not be correlated with the action of $b_m$ and $b_m^{\dagger}$.   This argument fails for state-dependent operators as we showed above.

The conclusion is that provided local fields in the interior of the black hole are constructed in a state-dependent manner, we can consistently reconcile unitary evolution with small corrections to correlation functions in effective field theory. 

\paragraph{\bf Numerically Large Non-Locality \\}
At this point, we also briefly address another criticism made in \cite{Mathur:2013gua}. The ``non-localities'' that we mentioned do indeed spread out over the Page-sphere of the black hole. For a solar-sized black hole, the Page sphere is huge: $10^{77} \text{km}$. Nevertheless, we wish to emphasize that the non-locality is incredibly difficult to measure.

In the construction that we have described, we have to measure a correlation function involving or order $\exp{[10^{77}]}$ points, before we can detect this non-locality! So, to the extent that actual numbers are relevant to these conceptual issues, it is clear that we do not have any contradiction with either any observed or possible-to-observe physics!


%% file: statedependentfig.pdf_t
\begin{picture}(0,0)%
\includegraphics{statedependentfig.pdf}%
\end{picture}%
\setlength{\unitlength}{3947sp}%
\begingroup\makeatletter\ifx\SetFigFont\undefined%
\gdef\SetFigFont#1#2#3#4#5{%
  \reset@font\fontsize{#1}{#2pt}%
  \fontfamily{#3}\fontseries{#4}\fontshape{#5}%
  \selectfont}%
\fi\endgroup%
\begin{picture}(4740,2866)(1718,-2995)
\put(1876,-2836){\makebox(0,0)[lb]{\smash{{\SetFigFont{29}{34.8}{\rmdefault}{\mddefault}{\updefault}{\color[rgb]{0,0,0}${\cal H}_E$}%
}}}}
\put(5776,-2761){\makebox(0,0)[lb]{\smash{{\SetFigFont{29}{34.8}{\rmdefault}{\mddefault}{\updefault}{\color[rgb]{0,0,0}${\cal H}_{E - \omega_n} $ }%
}}}}
\end{picture}%

%% file: unionofall.pdf_t
\begin{picture}(0,0)%
\includegraphics{unionofall.pdf}%
\end{picture}%
\setlength{\unitlength}{3947sp}%
\begingroup\makeatletter\ifx\SetFigFont\undefined%
\gdef\SetFigFont#1#2#3#4#5{%
  \reset@font\fontsize{#1}{#2pt}%
  \fontfamily{#3}\fontseries{#4}\fontshape{#5}%
  \selectfont}%
\fi\endgroup%
\begin{picture}(4740,2866)(1718,-2995)
\put(3526,-436){\makebox(0,0)[lb]{\smash{{\SetFigFont{14}{16.8}{\rmdefault}{\mddefault}{\updefault}{\color[rgb]{0,0,0}$|\Psi_2 \rangle$}%
}}}}
\put(1876,-2836){\makebox(0,0)[lb]{\smash{{\SetFigFont{29}{34.8}{\rmdefault}{\mddefault}{\updefault}{\color[rgb]{0,0,0}${\cal H}_E$}%
}}}}
\put(5776,-2761){\makebox(0,0)[lb]{\smash{{\SetFigFont{29}{34.8}{\rmdefault}{\mddefault}{\updefault}{\color[rgb]{0,0,0}${\cal H}_{E - \omega_n} $ }%
}}}}
\put(3451,-1711){\makebox(0,0)[lb]{\smash{{\SetFigFont{14}{16.8}{\rmdefault}{\mddefault}{\updefault}{\color[rgb]{0,0,0}$|\Psi_1 \rangle$}%
}}}}
\end{picture}%

%% file: s_noneqscen.tex
\section{Non-Equilibrium Scenarios \label{noneqscen}}
So far, in this paper, we have discussed how to define the mirror operators
in an equilibrium state. In this section, we briefly discuss the 
non-equilibrium scenario, leaving a more detailed study to further work.

Let us phrase the question that we are interested in more precisely. Our construction of the previous section {\em already} gives us interesting time-dependent correlators of fields $\langle \Psi | \phi^{i_1}_{\text{CFT}}(t_1, \Omega_1) \ldots \phi^{i_m}_{\text{CFT}}(t_m, \Omega_m) |\Psi \rangle$. However, when any of these
operators is behind the horizon, we need to use the mirror operators to define it, and these mirror operators are defined in an equilibrium state. Now, we want to ask: let us say that someone gives us an out-of-equilibrium state $|\Psi'\rangle$, perhaps produced by exciting an equilibrium state with some sources.
 What, then is the correct way to define the mirror operators so as to get the
results expected from semi-classical field theory?

Stated briefly, our proposal is that to deal with a non-equilibrium state that is produced by turning on sources dual to a small number of local operators 
on an equilibrium state, we ``strip off'' the excitations that create this non-equilibrium
state from a thermal state. We now define the mirror operators on this 
thermal-state, and then use these {\em unchanged} operators in the non-equilibrium state. We describe this more precisely below.

Our construction automatically addresses a technical objection that has been made to state-dependent proposals,
which is sometimes called the ``frozen vacuum.'' This is simply the claim that defining the mirror operators using the rules for equilibrium
states always leads to a featureless horizon, even though one could manually excite the horizon by injecting some matter
before the infalling observer falls in. Our proposal below for non-equilibrium states does not lead to any such issue. 

It is true that one can, by hand, ensure that 
the infalling observer perishes at the horizon, by aiming a focused laser beam 
which intersects the observer just as he crosses it. In this section we show how to describe correlators
outside and inside the horizon in such a scenario. However, our 
construction makes another unambiguous connection. Just as one would
expect from semi-classical field theory, any such excitation soon shares the fate
of the observer and falls into the singularity in a short amount of time leaving behind a featureless horizon once again.
\begin{figure}[!h]
\begin{center}
\includegraphics[height=5cm]{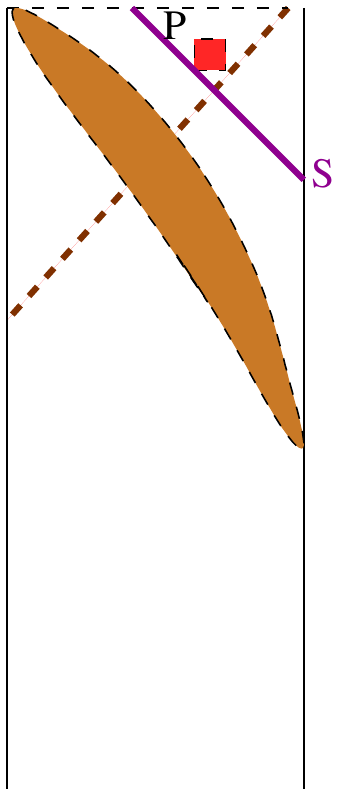}
\caption{\em A non-equilibrium state: a laser beam sent from the boundary point $S$ intersects the patch of interest at $P$ \label{shockwave}}
\end{center}
\end{figure}

\subsection{Detecting Non-Equilibrium States}
First, we discuss how to differentiate equilibrium states from
non-equilibrium states. Let us say we are given some state $|\Psi' \rangle$. Can we detect, by measuring expectation values in this state, whether
the state is in equilibrium or out of it?

The first point to note is that this, itself, is a manifestly time-dependent question. Consider a state that consists of superpositions of different energy
eigenstates. 
\be
|\Psi' \rangle = \sum_i c_{i} | E_i \rangle.
\ee
Now, consider an element $\al_p$ of ${\cal A}$. It is very natural that, in an interacting CFT like supersymmetric Yang-Mills, the elements
of $\al_p$ will obey the eigenstate thermalization hypothesis
\be
\label{eth}
\langle E_i | \al_p | E_j \rangle = A(E_i) \delta_{i j} + e^{-{1 \over 2} S\left({E_i + E_j \over 2} \right)}   R_{i j},
\ee
Here $S({E_i + E_j \over 2})$ is the density of states at the mean energy, and below we will write just $S$ for this quantity to lighten the notation. Note that $S \propto \nc$ for the systems that we have considered above.  $A$ is a ``smooth'' functions of their arguments, but $R_{i j}$ is a matrix comprising erratically varying phases but a smoothly varying magnitude. (In the papers \cite{srednicki1999approach,srednicki1994chaos}, sometimes
another function is introduced to capture this magnitude, but we have no need for it here.)

We will need a further technical assumption on the matrix $R_{i j}$. Note, that $\tr(R^{\dagger} R) = \Or[e^{2 S}]$ and $R$ has $e^{S}$ eigenvalues $r_1, \ldots r_{e^S}$,  we expect that the typical magnitude of each eigenvalue will be $|r_i| = \Or[e^{S \over 2}]$. We also need to assume that {\em no eigenvalue} of $R$ is much greater than this $|r_i| e^{-S \over 2} = \Or[1], \forall i$.  On the other hand, the phase of $r_i$ will generically be arbitrary.

This form is very natural, and follows from the following simple assumption. The eigenvectors of the operator ${\op}^i$ are not correlated with the exact energy eigenstates. This is quite common in interacting field theories. If
the two sets of eigenvectors are related by some ``random'' unitary transformation, then \eqref{eth} follows. 

For example, consider our ``regularized frequency modes''  ${\op}_{n,\vect{m}}$\footnote{Sometimes we omit the superscript ``$i$'' of the operators $\phi_{CFT}^i$ and ${\op}^i_{n,
\vect{m}}$ in order to lighten the notation.}
 which include a band of modes of width $\omega_{\text{min}}$, and 
are defined in \eqref{coarsening}. 
We see that between two energy
eigenstates, the following statements hold
\be
\begin{split}
&\langle E_i | {\op}_{\omega, \vect{m}} | E_j \rangle = \langle E_i| {\op}_{\vect{m}}(0) | E_j \rangle \delta(E_i - E_j - \omega), \\
&\langle E_i | {\op}_{n,\vect{m}} | E_j \rangle = 
{1 \over (\omega_{\text{min}})^{1 \over 2}} 
\langle E_i| {\op}_{\vect{m}}(0) | E_j \rangle \theta(E_i - E_j -(n-1)\omega_{\text{min}}) \theta(E_j + n \omega_{\text{min}} - E_i),
\end{split}
\ee
where ${\op}_{\vect{m}}(0) = \int {\op}(0,\Omega) Y^*_{\vect{m}}(\Omega) d^{d-1} \Omega$, which is a natural notation. 
The normalization factor of $\sqrt{\omega_{\text{min}}}$ ensures that the diagonal elements of the operator
\be
\langle E_i | {\op}_{n,\vect{m}} {\op}_{n,\vect{m}}^{\dagger} | E_i \rangle = \sum_{E_j = E_i + (n-1)\omega_{\text{min}}}^{E_i + n \omega_{\text{min}}}  |\langle E_i| {\op}_{n,\vect{m}} |E_j \rangle|^2 = \Or[1],
\ee
since the sum runs over $e^{S} \omega_{\text{min}}$ states, and each
term is of order $e^{-S} \omega_{\text{min}}^{-1}$, assuming
the ETH for the operator ${\op}_{\vect{m}}(0)$. So, we see that ${\op}_{n, \vect{m}}$ also obeys the ETH up to the additional normalization of $(\omega_{\text{min}})^{-{1 \over 2}}$, which is $\Or[1]$ in the accounting of \eqref{eth} and will not be important in the discussion below.

This analysis leads to
\be
\chi_p(t) = \langle \Psi' | e^{i H t} \al_p e^{-i H t} | \Psi' \rangle = \sum |c_i|^2 \left( A(E_i) + e^{-{S \over 2}} R_{i i}  \right) + \sum_{i \neq j} c_j c_i^* e^{-{S \over 2}} e^{i (E_i - E_j) t} R_{i j}.
\ee
The second term is manifestly time-dependent. Now note, that by the assumption on the maximum size of the eigenvalues of $R$ above, we see that, at most we can get an $\chi_p(t) - \chi_p(0) \leq \Or[1]$ time-dependence. However, in the generic situation where the coefficients $c_i$ and time $t$ are not carefully selected, the time-dependent term is $\chi_p(t) - \chi_p(0) = \Or[e^{-{S \over 2}}]$ --- exponentially suppressed in $\nc$. 

We can use this as a diagnostic of whether the state is in equilibrium
or not. As we mentioned above, this is a time-dependent question, and
in the CFT, we are interested in the issue of whether the state is in equilibrium from some starting time $t = 0$ to some other long time $t = \omega_{\text{min}}^{-1}$. Recall that we introduced $\omega_{\text{min}}$ 
to regulate the frequency modes of the CFT, and we can even take $\omega_{\text{min}} = e^{-\sqrt{N}}$, if we are interested in a big-black hole that
has a much longer lifetime.

So, to precisely evaluate whether a state is in equilibrium or not, 
we consider the following quantity
\be
{\nu}_p = \omega_{\text{min}}  \int_0^{\omega_{\text{min}}^{-1}}  |(\chi_p(t) - \chi_p(0))| d t. 
\ee
We will declare that a state is in equilibrium if 
\be
\label{equilibriumcrit}
\nu_p = \Or[e^{-{S \over 2}}], ~ \forall p,
\ee
i.e. this property holds for all observables in ${\cal A}$. Otherwise we will classify it as a non-equilibrium state. 

We emphasize that this is a much finer distinction than we require in practice. The lifetime of the black hole scales just polynomially with $\nc$ 
and, in practice, any state that does not vary appreciably over this
time is, for practical purposes, in equilibrium. In fact, when we 
consider small black holes in AdS, they have a lifetime that is only
polynomial in $\nc$. In that case, it may be useful to consider
a slightly modified definition of an equilibrium state, where these states
are effectively in equilibrium, while those that change over a much shorter
time are not. For simplicity here, however, we restrict ourselves to
large black holes in AdS.

\subsection{Near Equilibrium States}
We now discuss a class of states that we will call ``near equilibrium states''. These are states that are produced by acting on an equilibrium state $|\Psi \rangle$, with a unitary matrix produced by exponentiating a Hermitian element of $\al_p$
\be
\label{neareq}
|\Psi' \rangle = U |\Psi \rangle, \quad U = e^{i \al_p}.
\ee
where $|\Psi \rangle$ is in equilibrium and $\al_p$ is Hermitian. Recall
that although we have written the basis of the algebra ${\cal A}$ in
terms of Fourier modes, we are allowed to take arbitrary linear combinations. So, the set above includes states that are produced by coupling a 
source to the boundary field for a limited amount of time. For example, we could take $U =  e^{i\int J(t, \Omega) {\op}(t, \Omega)}$. We can also
consider sources that couple to stringy modes or brane probes.

Although our construction can be generalized to several other statistical-mechanics system, our presentation in this section will focus on the CFT.
In the CFT, it is true that if $|\Psi \rangle$ is in 
``equilibrium'' then any state $|\Psi' \rangle$ of the form \eqref{neareq} is out of equilibrium. The logic behind this claim is as follows. Consider 
turning on a source for some local operator in the CFT by adding $\int J(t, \Omega) {\op}(t, \Omega)$ to the Hamiltonian. (The
logic easily generalized to bi-local and k-local operators.)
It is possible to find another operator $\Pi(t,\Omega)$ so that
\be
\langle \Psi| [{\op}(t,\Omega), \Pi(t',\Omega')] |\Psi \rangle = k(t-t', \Omega, \Omega') \neq  0.
\ee
Now, we see that\
\be
\langle \Psi' | e^{i H t} \Pi(0, \Omega) e^{-i H t} | \Psi' \rangle = \langle \Psi | \Pi(t,\Omega) |\Psi \rangle + \int J(t', \Omega') k(t-t',\Omega, \Omega') d t' d \Omega' \, .
\ee
The second term above tells us that we have turned on the source, and can
be detected.

If we turn on sources for operators in {\em momentum space}, this is still possible, although it may be a little confusing. This follows from an 
examination of \eqref{coarsening}, which tells us that our momentum-space
operators are effectively defined over a time-range of length $\omega_{\text{min}}^{-1}$ in the CFT, and so turning on sources for these operators is just
like turning on a slow-acting source for a position space operator.

For example, consider a state 
\be
\label{excitingon}
|\Psi' \rangle = e^{i \lambda \left({\op}^i_{n, \vect{m}} + ({\op}^i_{n, \vect{m}})^{\dagger}\right)} |\Psi \rangle,
\ee
for some particular conformal primary, and some particular modes $n$ and $\vect{m}$, where $\lambda$ is a constant. In fact, such a state is out of thermal equilibrium since
we can see that
\be
\begin{split}
&\langle \Psi'|\big(  ({\op}^i_{n,\vect{m}})^{\dagger} - {\op}^i_{n, \vect{m}} \big)|\Psi' \rangle \\
&= \langle \Psi| e^{-i \lambda \left({\op}^i_{n, \vect{m}} + ({\op}^i_{n, \vect{m}})^{\dagger} \right)}  \left( ({\op}^i_{n,\vect{m}})^{\dagger} - {\op}^i_{n, \vect{m}}\right) e^{i \lambda \left({\op}^i_{n, \vect{m}} + ({\op}^i_{n, \vect{m}})^{\dagger} \right)} |\Psi \rangle \\
&= \langle \Psi | \left( ({\op}^i_{n,\vect{m}})^{\dagger} - {\op}^i_{n, \vect{m}}\right) + 2 \lambda | \Psi \rangle  + \Or[{1 \over \nc}]\ \\
&= 2 \lambda + \Or[e^{-{S \over 2}}].
\end{split}
\ee
Here we used the Baker-Campbell-Hausdorff lemma in going from
the second to the third line together with the fact that $[{\op}^i_{n,\vect{m}}, ({\op}^i_{n, \vect{m}})^{\dagger}] = 1 + \Or[{1 \over \nc}]$. However, now if
we evaluate $\chi(t)$, we can already see that by evolving for a time $t_0 \approx {i \pi \over n \omega_{\text{min}}}$ that $\chi(t_0) = -2 \lambda$. However, the long term value of $\chi(t) $ is $0$. Over time scales 
larger than $\omega_{\text{min}}^{-1}$, we see that the approximate
commutation relations between ${\op}^i_{n,\vect{m}}$ and $({\op}^i_{n,\vect{m}})^{\dagger}$
break down because the different oscillators in \eqref{coarsening} that
comprise these operators decohere.

It is very hard to detect that the state \eqref{excitingon} is out
of equilibrium partly because of the nature of the source that 
we turned on. In terms of local operators, this corresponds to a 
slow-acting source that acts over a time-scale of $\omega_{\text{min}}^{-1}$. 

We can consider a harder example: $|\Psi' \rangle = e^{i \lambda {\op}^i_{n,\vect{m}} ({\op}^i_{n,\vect{m}})^{\dagger}} |\Psi \rangle$ In fact, even here it is possible to detect the 
action of this source as {\em subleading order} in ${1 \over \nc}$. We need
to find an operator $\Pi$ so that $\langle \Psi | [{\op}^i_{n,\vect{m}} ({\op}^i_{n,\vect{m}})^{\dagger}, \Pi] |\Psi \rangle \neq 0$. In an interacting CFT such an operator should exist on general grounds, although we cannot write down its explicit form here without
knowing the OPE coefficients in detail. Given such an operator, we can
again use the logic above to detect this slow-acting source, and also
the fact that the state is slightly out-of-equilibrium.

To summarize, this discussion implies that if a non-equilibrium state can be
written in the form \eqref{neareq}, then $U$ is essentially uniquely fixed.
Any other $U'$ would only take the state out of equilibrium again. In some more detail: suppose that there are two different
equilibrium states $|\Psi_1\rangle,|\Psi_2\rangle$ such that we can write the near-equilibrium state $|\Psi'\rangle$ as $|\Psi'\rangle = e^{i A_1} |\Psi_1\rangle$ and also
$|\Psi'\rangle = e^{i A_2} |\Psi_2\rangle$. From these two we find $|\Psi_2\rangle = e^{-i A_2} e^{i A_1} |\Psi_1\rangle$. But we argued above that it is not possible for both $|\Psi_1\rangle$ and $|\Psi_2\rangle$ to be equilibrium states, unless $A_1=A_2$.

\subsection{Mirror Operators for Near-Equilibrium States}
We now describe how to construct mirror operators for non-equilibrium
states. Given a non-equilibrium state $|\Psi'\rangle$, we have described
above how we can detect that it is not in equilibrium and also 
find how it is related to the equilibrium state $|\Psi \rangle$ by
\[
|\Psi'\rangle = U|\Psi\rangle.
\]
We now define the action of the mirror operators by the following
modified recursive rules in the CFT
\begin{align}
\label{noneqtildedefcftactpsi}
\widetilde{\op}^i_{n,\vect{m}} |\Psi' \rangle &= U e^{-{\beta \omega_n \over 2}} ({\op}^i_{n, \vect{m}})^{\dagger} U^{\dagger} |\Psi' \rangle, \\
\label{noneqtildedefcftcomord}
\widetilde{\op}^i_{n, \vect{m}} \al_p  |\Psi' \rangle &= \al_p \widetilde{\op}^i_{n, \vect{m}} |\Psi' \rangle, \quad \forall \al_p \in {\cal A}.
\end{align}
As usual, the factor of $e^{-{\beta \omega_n \over 2}}$ in  \eqref{noneqtildedefcftcomord} must be corrected at subleading orders in ${1 \over N}$, but
the fact that the mirror operators commute through the ordinary operators
should hold at all order in perturbation theory. 

Keeping
this in mind, we can define the action of $\widetilde{\op}_{n,\vect{m}}$ on the
state $|\Psi' \rangle$ and its descendants in a single compact equation as
\be
\widetilde{\op}^i_{n,\vect{m}} \al_p |\Psi' \rangle = \al_p U e^{-{\beta \omega_n \over 2}} {\op}^i_{-n, \vect{m}} U^{\dagger} |\Psi' \rangle .
\ee

\subsection{The Frozen Vacuum}
We now address the ``frozen vacuum''  objection
 to state-dependent proposals  that was articulated by Bousso \cite{Bousso:2013ifa}. 
The argument of \cite{Bousso:2013ifa} was made in the context of the 
proposals of \cite{Verlinde:2013vja,Verlinde:2013uja,Verlinde:2012cy},
which also use state-dependent operators. We do not understand some of the details in  \cite{Bousso:2013ifa},
but we translate what we think is the relevant part of the argument, albeit in somewhat more prosaic language.

The point is simply that we cannot use the rules \eqref{tildedefcftactpsi} and  \eqref{tildedefcftcomord} in a non-equilibrium state like $|\Psi' \rangle$ in \eqref{neareq} and expect to get the right semi-classical
correlators. For example, as we saw above in section \ref{resolvenaneq0}, the rules \eqref{tildedefcftactpsi} tell us that the particle number observed
by the infalling observer is zero, if there are no additional excitations i.e.: $\langle \Psi | N_a |\Psi \rangle = 0$.

We do not expect this in non-equilibrium states. For a non-trivial $U$,
notice, for example, that generically we have
\be
\left( \tO^i_{n,\vect{m}} - e^{\beta \omega_n\over 2} {\op}^i_{-n, \vect{m}}  \right)  |\Psi' \rangle \neq 0,
\ee
and so with the operation of the mirror operators 
defined in \eqref{noneqtildedefcftactpsi} and \eqref{noneqtildedefcftcomord},
we generically obtain $\langle \Psi'| N_a | \Psi' \rangle  \neq 0$. The precise  expectation value depends on the kind of perturbation that we have
made to the state.

We wish to emphasize that even in the equilibrium construction of the previous
sections, or of our previous paper \cite{Papadodimas:2012aq}, it was 
perfectly possible to excite the horizon of the black hole. What we have done here is simply to explain how to construct the operators $\widetilde{O}_{n, \vect{m}}$ when the base-state that we are given is out of equilibrium. 
It is clear that 
our procedure of ``stripping off'' the $U$, and then restoring it,
gives us exactly the same answers as one would get from effective 
field theory in the bulk.

\paragraph{\bf Acting with a unitary behind the horizon? \\}
We should also mention a second issue raised by van Raamsdonk \cite{VanRaamsdonk:2013sza}. van Raamsdonk considers a case where an autonomous unitary transformation is made on the second sided CFT. In fact, in the eternal black hole, a small perturbation made early enough in the second CFT can lead to a highly boosted shock-wave just behind the horizon that separates regions I and II \cite{Shenker:2013pqa}.

van Raamsdonk's argument is not directly relevant to our construction since we do not really have a second side in the collapsing geometry. These
states do not really have an existence as states that are autonomously created by collapsing matter. 

This is important, since if we consider the state $e^{i J(n,\vect{m}) \widetilde{\op}_{n, \vect{m}}} |\Psi \rangle$, we do not have any way of 
detecting its departure from equilibrium by using the operators in $\al_p$. However, since in a collapsing geometry such a state can only be created
indirectly, by pumping in an excitation from the outside, we can detect
this excitation and use the more precise rules for the definition of
the mirror operators given above.

\subsection{An Example: a Beam from the Boundary}
Let us now consider an example in some detail, where we turn on a source
at the boundary dual to some operator. We wish to check the following
qualitative conclusions. First, the correlators across the horizon
should not be affected before the beam has time to reach. Then, the correlators should be affected for some time, in a way that is determined
by effective field theory. Finally, once we wait for a scrambling time,
the correlations should go back to their previous values. We wish to consider
\be
C_{12} = \langle \Psi' | \phi_{\text{CFT}}(t_1, \Omega_1, z_1) \phi_{\text{CFT}}(t_2, \Omega_2, z_2) | \Psi'\rangle, \quad 
|\Psi' \rangle = e^{i \int J(t,\Omega) {\op}(t, \Omega)} |\Psi \rangle,
\ee
where $J(t, \Omega)$ is a source that is sharply peaked around the 
origin of boundary coordinates ($t=0,\Omega=0$), the point $(t_1, \Omega_1,z_1)$ is in front of the horizon, and  $(t_2, \Omega_2,z_2)$ is {\em behind} 
the horizon, and $|\Psi \rangle$ is an equilibrium state.

Using the expansion \eqref{interiorexpansion}, we see that
\be
\begin{split}
C_{12} =  \sum_{\vect{m}} \int_{\omega>0} {d\omega \over  2 \pi} \,
 \Bigg[&\langle \Psi | U^{\dagger} \phi_{\text{CFT}}(t_1, \Omega_1, z_1) \left( {\op}_{\omega,\vect{m}} \,g_{\omega,\vect{m}}^{(1)}(t_2,\Omega_2,z_2)+ \text{h.c} \right) U |\Psi \rangle \\ + &\langle \Psi | U^{\dagger} \phi_{\text{CFT}}(t_1, \Omega_1, z_1) U \left( \widetilde{\op}_{\omega,\vect{m}}\, g_{\omega,\vect{m}}^{(2)}(t_2,\Omega_2 ,z_2) + \text{h.c} \right) |\Psi \rangle \Bigg],
\end{split}
\ee
where we used that $[U,\widetilde{\op}_{\omega,\vect{m}}] =0$, for the tildes defined with respect to the equilibrium state $\st$.

Now, we see that the properties of $C_{12}$ we inferred above, follow directly from the properties of {\em ordinary} local fields under conjugation by $U$. We know that $[{\op}(0,0), \phi_{\text{CFT}}(t_1, \Omega_1, z_1)] = 0$, when the bulk point $(t_1, \Omega_1, z_1)$ is spacelike separated from the origin of the boundary coordinates, and this commutator also becomes small when the point is in the {\em far future} of the origin. However this commutator is appreciably non-zero, when the bulk point is near
the light cone that extends from the origin of the boundary. The same
result holds for the commutator $\sum_{\vect{m}} \int_{\omega > 0} {d \omega \over 2 \pi} [{\op}(0,0), {\op}_{\omega,\vect{m}}\, g_{\omega,\vect{m}}^{(1)}(t_2,\Omega_2,z_2)+ \text{h.c}]$. These properties follow from an analysis of Green functions for perturbative fields in the bulk.

Just to clarify this point,  we remind the reader that in 4 dimensional flat space
the commutator for a scalar field $\psi$ of mass $m$ is \cite{bogoliubov1959itq}
\be
[\psi(\vect{x}), \psi(\vect{y})] =  {i \over 2 \pi} s(x^0 - y^0) \delta(\lambda) - {i m \over 4 \pi \sqrt{\lambda}} \theta(\lambda) s(x^0 - y^0) J_1(m \sqrt{\lambda}),
\ee
where $\lambda = (\vect{x} - \vect{y})^2$, and $s$ is the sign function
in this equation. This commutator always vanishes at spacelike separation. For a massless field, the commutator is non-zero only on the 
light-cone, but even for a massive field, this commutator vanishes
for large timelike separation as well. 
The explicit expressions in AdS-Schwarzschild are
much harder to write down, but the same qualitative properties hold. Note that this involves an interplay between the CFT commutators ${\op}_{\omega,\vect{m}}$ and the transfer function.

So, in 
the case where the bulk points are in the far future, or spacelike
separated from the source at the boundary, we can just commute the $U$
through the ordinary operators to annihilate the $U^{\dagger}$ and so $C_{12}$ reduces to the correlator in the state $|\Psi \rangle$. However,
when either of the bulk points are near the light-cone from the origin
of the boundary, we expect that this correlator will receive  appreciable
corrections. This is exactly what we had inferred.


%% file: s_tomitatakesaki.tex
\def\shil{{\cal H}_{\Psi}}
\section{Links with Tomita-Takesaki theory\label{tomitatakesaki}}

In this section we provide an additional (though mathematically equivalent) perspective to the construction of the mirror operators. We also discuss the relation of the current proposal to that of \cite{Papadodimas:2012aq}, which was
based on a coarse/fine decomposition of the Hilbert space. Finally we present some intriguing mathematical connections of
our construction with the Tomita-Takesaki theory of operator algebras. 

Many of the ideas described in this section have already been discussed in section \ref{sec:three}. We summarize them again for the 
convenience of the reader and slightly modify the presentation in order to connect with the Tomita-Takesaki theory.

\subsection{Another intuitive explanation of our construction\label{summaryint}}

In section \ref{sec:three} we emphasized that the  operators can be found by solving equations \eqref{tildedefgeneral}. We argued that since the number of equations is much smaller than the size of the Hilbert space, we can always find solutions, and we explicitly wrote down a solution in \eqref{talexplicit}. 
Here we expand on a slightly different perspective, which was already mentioned at the end of section \ref{tildegeneral}. This perspective leads
to a ``constructive'' definition of the mirror operators and is more suitable to make contact with the mathematical discussion of the next subsection.  

\begin{figure}
\begin{center}
\includegraphics[width=6cm]{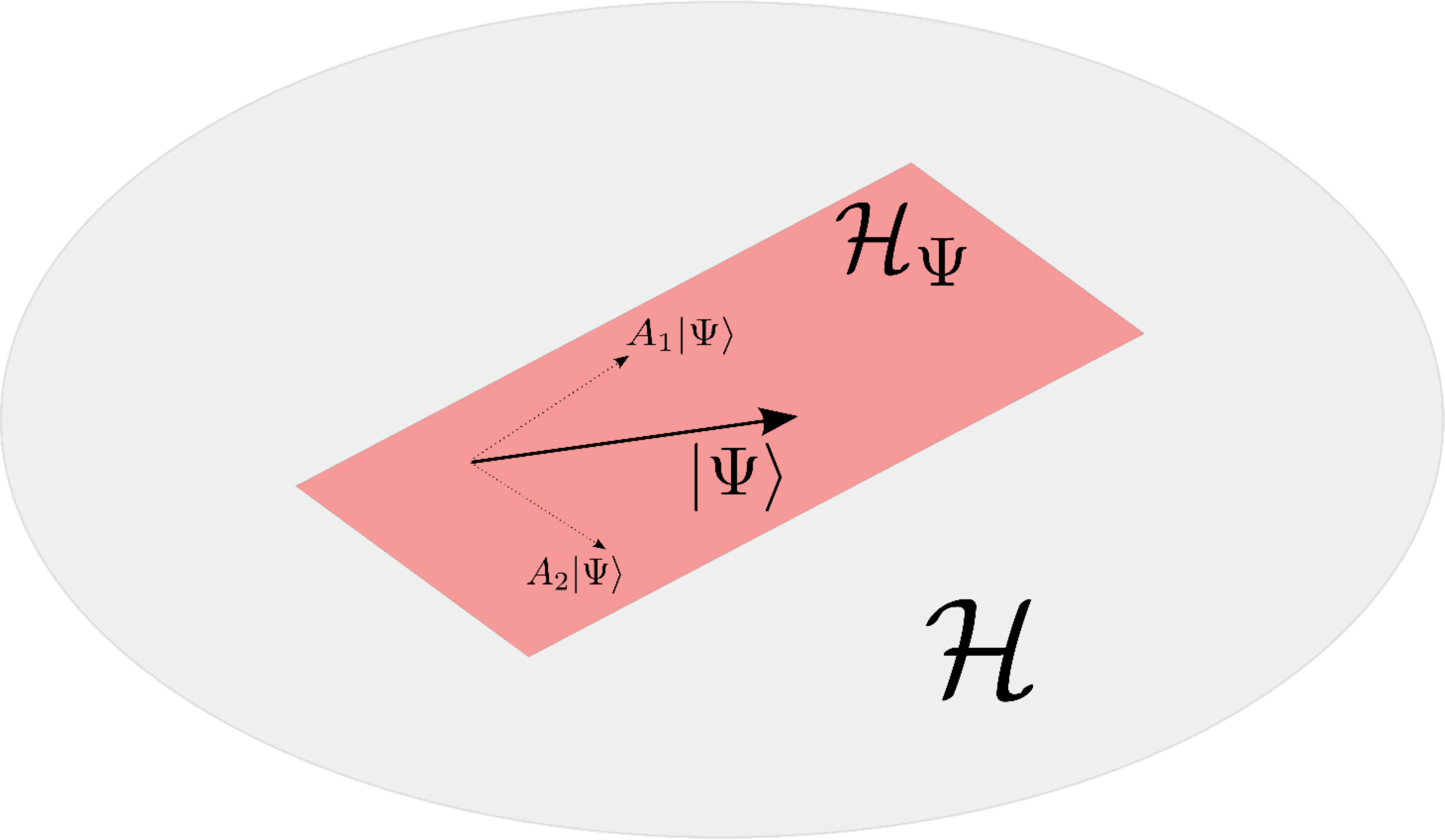}
\caption{A quantum system with Hilbert space ${\cal H}$ placed in the pure state $|\Psi\rangle$ is probed by a set of observables
$A_i \in {\cal A}$. We define the subspace ${\cal H}_{\Psi} = \{{\rm span\,\,of\,\, A\st\,,\,A\in {\cal A}} \}$ which is
relevant for computing correlation functions of observables in ${\cal A}$ on the state $|\Psi\rangle$ and the construction of the mirror operators.}
\label{subspacehpsi}
\end{center}
\end{figure}

The intuition is very simple. As mentioned several times in the previous sections of the paper, we imagine that we have a complicated quantum system with Hilbert space ${\cal H}$, which is
in a particular pure state $\st$. Also, we imagine that we can only probe the system by using a small set of observables
${\cal A}$. Since we will be computing correlation functions of these observables on the state $\st$, it is very natural to define
the span of states of the form
\be
A_i \st,
\ee
\be
A_i A_j \st,
\ee
\be
A_i A_j A_k \st,
\ee
\be
etc.
\ee
where $A_i \in {\cal A}$. We introduced the linear span of states of this form at the end of section \ref{tildegeneral} and we called it
\be
\label{smallHilbert}
{\cal H}_{\Psi} = \{{\rm span\,\,of\,\, }A\st,\qquad A\in {\cal A} \}.
\ee
The space ${\cal H}_{\Psi}$ is a subspace of the full Hilbert space ${\cal H}$, which obviously depends on the choice of the initial state $\st$. This is schematically depicted in figure \ref{subspacehpsi}.

The main shift in perspective from the discussion in section \ref{sec:three} is the following: in that section we constructed the operators $\tal_p$,
element by element, for each operators $\al_p$ in ${\cal A}$. Now, we 
will describe a more formal ``one shot'' construction of the mirror
operators.

We will see that we can define the mirror operators in a very natural way by concentrating on how they act on the subspace ${\cal H}_{\Psi}$. Their action on this subspace is extremely natural. Their action on the ``orthogonal subspace'' ${\cal H}_{\Psi}^\bot$ is not completely specified --- this is related to the fact that the equations \eqref{tildedefgeneral} have more than one solutions --- but this ambiguity has no effect on the computation of low-point correlation functions. We stress again, that this is equivalent to equations \eqref{tildedefgeneral}, the new perspective offered here provides some additional intuition and demonstrates a canonical solution of these equations.

As before, the starting point for the existence of the mirror operators is that: if the size of ${\cal A}$ is small relative to the size of the Hilbert space, then on general grounds we expect that a typical state {\it $\st$ cannot be annihilated by non-vanishing elements of the set ${\cal A}$} or, in equations, for $\al \in {\cal A}$ we have
\be
\label{separatingb}
\al \st = 0 \qquad \Leftrightarrow \qquad A = 0.
\ee
In the case of the big black hole in AdS/CFT, the idea is that ${\cal A}$ is the set of products of a small number of single trace operators, while
$\st$ is a typical Quark-Gluon-Plasma (QGP) microstate. It is clear that such a typical QGP microstate cannot be annihilated by a small number of single trace operators.

Equation \eqref{separatingb} expresses the point that the state $\st$ {\it looks entangled from the point of view of the algebra ${\cal A}$}. Usually we define entanglement in situations where the Hilbert space of the system has a bipartite structure, but here we  generalize the concept of entanglement, by discussing how the state {\it appears to be entangled} in terms of certain
{\it observables}. This is expressed by equation \eqref{separatingb}. We will argue that whenever we have such a situation, in which a quantum state looks (sufficiently) entangled when probed by a set of observables ${\cal A}$, then the set of observables ${\cal A}$ is ``doubled''. This doubling explains the origin of the dual modes behind the horizon.

We now start with these assumptions, i.e. that we have a big quantum system probed by a small number of observables ${\cal A}$, such that
\eqref{separatingb} is satisfied and we show how we define the mirror operators. 

The starting point is that there is a natural way to define the action of a {\it second copy} of the observables ${\cal A}$ acting on the subspace ${\cal H}_{\Psi}$. This can be achieved by defining, effectively, an action of observables in ${\cal A}$ ``from the right''. This can be compactly described by introducing an antilinear map
$
S : {\cal H}_{\Psi} \rightarrow  {\cal H}_{\Psi} 
$
defined by
\be
\label{smatrix}
S A |\Psi \rangle = A^\dagger |\Psi \rangle,
\ee
which obviously satisfies $S^2=1$ and also
\be
S \st = \st.
\ee
Notice that condition \eqref{separatingb} is crucial in order for \eqref{smatrix} to be well-defined.

Then it is easy to check that the operators defined by 
\be
\label{hatop}
\hat{A} = S A S,
\ee
satisfy the following two properties

i) Their algebra is isomorphic to that of operators in ${\cal A}$, since $S^2=1$.

ii) The hatted operators commute with operators in ${\cal A}$ when acting on elements of ${\cal H}_\Psi$. To see this, notice
that any vector in ${\cal H}_\Psi$ can be written as $C\st$ for some $C\in {\cal A}$ and we have
\be
\hat{A} B C \st= S A S B C\st = SA C^\dagger B^\dagger \st = B C A^\dagger \st = B \hat{A} C\st .
\ee
Hence
\be
[\hat{A},B] C \st = 0,
\ee
for all $A,B,C \in {\cal A}$.

Notice the following important point:  the subspace ${\cal H}_{\Psi}$ was defined as the span of states of the form $A|\Psi\rangle$. While
the operators $\hat{\cal A}$ commute with those in ${\cal A}$, they are still acting on the same space ${\cal H}_{\Psi}$!

The operators $\hat{A}$ can be extended in the full Hilbert space ${\cal H} = {\cal H}_{\Psi} \oplus {\cal H}_{\Psi}^{\bot}$. One 
naive possibility would be by defining them to be ``zero'' on the orthogonal subspace ${\cal H}_{\Psi}^\bot$, but there are many other possibilities. This issue was already discussed at the end of subsection \ref{tildegeneral}.

 In the previous steps we have identified a ``second copy'' $\hat{\cal A}$  of the observables acting on the space ${\cal H}$. This already captures the essence of the ``doubling''. However, to finalize the construction of the mirror operators and make contact with the conventional ``thermofield doubling'' it is convenient to perform a small redefinition of the operators $\hat{\cal A}$. The issue is that the mapping $S$ is --- in general --- not (anti)-unitary. Hence the ``normalization'' of the operators $\hat{\cal A}$ is not the same as those of the ${\cal A}$. In order to fix this we can rescale the magnitude of the antilinear operator $S$ by defining
\be
\label{polardecomp}
S = J \Delta^{1/2},
\ee
where $J$ is anti-unitary with $J^2=1$ and $\Delta$ positive and Hermitian. The precise definition of $\Delta$ will be discussed later. Then we can define the conventionally normalized mirror operators by
\be
\label{propertilde}
\widetilde{A} = J A J.
\ee
While it is obvious that the hatted operators \eqref{hatop} commute with elements of ${\cal A}$, it is less obvious that the $\widetilde{A}$'s commute with operators in ${\cal A}$, due
to the factors of $\Delta^{1/2}$. Nevertheless, it is a fact that they do commute and this  will be explained in more detail later.

Finally let us notice that in most situations ---and certainly in the case of the large $N$ gauge theory --- the ``set of observables'' ${\cal A}$ is not a closed algebra in the strict mathematical sense. For instance, if we attempt to define
${\cal A}$ as the set of ``small number of insertions of single-trace operators'', then this set is not strictly closed under operator multiplication. 

This point is at the heart of black-hole complementarity: while ${\cal A}$ is not an exact algebra, it behaves approximately like an algebra for certain low-point correlators. Hence the construction of the commuting mirror operators, as outlined above, approximately works for such low-point correlators. The existence of the mirror operators for low-point functions is sufficient in order to reconstruct the experience of the infalling observer.

On the other hand, if we act with too many of them, we will either ``get out of ${\cal A}$'', or we will have to allow the set ${\cal A}$ of ``accessible observables'' to be large enough so that it becomes an algebra and it contains all possible products. In this case the state $|\Psi\rangle$ is depleted of any entanglement with respect to ${\cal A}$  and condition \eqref{separatingb} is not satisfied any more. Then the dual-operator construction does not work and the black hole interior ceases to make sense.

This is all in agreement with the idea of complementarity and the validity of effective field theory in the bulk.

So far these ideas were motivated by physical considerations. Intriguingly, the mathematical language in which the dual-operator construction
was phrased above,  appears in surprisingly similar form in  the theory of operator algebras as we explain in the next subsection.

\subsection{Relation to the Tomita-Takesaki modular theory}

We now describe an extremely interesting link between our construction
of the mirror operators behind the horizon, and an area in the study of von Neumann algebras that goes by the name of Tomita-Takesaki theory. The existing reviews of this subject in the literature are somewhat formal, so we will
summarize the main ideas here. The reader interested in a more sophisticated mathematical discussion can refer to \cite{Haag:1992hx, bropalg, bropalgb}.

Exactly like in our physically motivated construction mentioned above, the Tomita-Takesaki construction, involves building the {\em commutant} of an algebra ${\cal A}$, and uses an appropriate state-vector to do so. For example, given
the set of operators on a finite-interval, one could use the construction to generate the operators outside the light-cone of this interval that, in a local quantum
field theory, should commute with the original algebra. Here, we will use it to construct operators ``behind'' the horizon, given the operators in front of it.

The Tomita-Takesaki construction starts with an algebra ${\cal A}$ acting on a Hilbert space ${\cal H}$ and a state-vector $|\Psi \rangle$ that is {\em cyclic} and {\em separating}. For a state to be {\em cyclic} means that the 
space ${\cal H}_{\psi} = {\cal A} |\Psi \rangle$ is dense in the Hilbert space ${\cal H}$. The statement that the state is separating is simply the condition \eqref{separatingb}: 
$\al | \Psi \rangle \neq 0,~ \forall \al \in {\cal A}.$

The reader can satisfy herself that these conditions are easily met, for example, in relativistic QFT, if one takes ${\cal A}$ to be the algebra of operators on an open ball of space time and $|\Psi \rangle$ to be the vacuum state. Part of this statement,  is the so-called Reeh-Schlieder theorem,
that we also discuss in Appendix \ref{appmeasurement}.  

Here, we are interested in a different situation. For us $|\Psi \rangle$ is a typical pure state that looks like it is close to thermality, whereas ${\cal A}$ is the {\em set} (not, necessarily, an algebra) of low-point correlation functions. Consequently, ${\cal H}_{\psi}$ is not dense in the larger Hilbert space ${\cal H}$, but this will not be an obstacle either, as we will now see. For the remaining of this section, and in order to state the Tomita-Takesaki theorem in simple form,  we will just assume that ${\cal A}$ is an algebra and we will think of ${\cal H}_{\Psi}$ as the entire Hilbert space, so that the assumptions that $\st$ is cyclic and separating are satisfied. In other words, in the following part we will imagine that ${\cal H}_\Psi$ plays the role of the entire Hilbert space ${\cal H}$ and will simply call it ${\cal H}$. We will discuss the important modifications necessary for the case of the large $N$ gauge theory later. We also assume that ${\cal A}$ is closed under the Hermitian 
conjugation operation.

This means that we have a von Neumann algebra ${\cal A}$ acting on a Hilbert space ${\cal H}$, which has a cyclic and separating vector $\st$. The Tomita-Takesaki theorem states that in this case the ``commutant'' ${\cal A}'$ of the algebra ${\cal A}$ can be constructed by an antilinear conjugation, which can be identified with the ``tilde'' mapping used in thermofield theory.

Like in the discussion of the previous section, the Tomita-Takesaki construction starts by constructing the {\em anti-linear} map $S$ 
that appeared above
\be
\label{antilineardef}
S \al | \Psi \rangle = \al^{\dagger} |\Psi \rangle.
\ee

We consider the polar decomposition of $S$ as
\be
S = J \Delta^{1/2},
\ee
where $J$ is anti-unitary, and $\Delta$ is Hermitian and positive. This can also be understood as follows. For an antilinear map $S$ we define the Hermitian conjugate as
\be
\label{sbardef}
(|A\rangle, S^\dagger |B\rangle ) \equiv ( |B\rangle, S |A\rangle),
\ee
where $(\,,\,)$ denotes the inner product. Then we have
\be
\label{deltadefss}
\Delta = S^\dagger S.
\ee
It is not too hard to prove the useful relations
\be
J \Delta^{1/2} = \Delta^{-1/2} J,\qquad J^2 = 1,
\ee
and
\be
S\st = J \st = \Delta \st = \st.
\ee
Finally, under the previous conditions, the Tomita-Takesaki theorem states that
\be
\label{ttfirst}
J {\cal A} J = {\cal A}',
\ee
and
\be
\label{ttsecond}
\Delta^{is} {\cal A} \Delta^{-is} ={\cal A},\qquad s\in {\mathbb R}.
\ee
Equation \eqref{ttfirst} implies that the commutant ${\cal A}'$ can be recovered by conjugating the operators in ${\cal A}$ with the antilinear map $J$.

To interpret equation \eqref{ttsecond}, let us first write $\Delta = e^{-K}$ where $K$ is a Hermitian operator. Then equation \eqref{ttsecond}  means that the the set ${\cal A}$ is ``closed under time evolution'' with respect to the ``modular Hamiltonian'' $K$. As we will see later in the case of the large $N$ gauge theory, and in the large $N$ limit, the analogue of the operator $K$ behaves like $\beta(H_{CFT} - E_0)$, where $E_0 = \langle \Psi|H_{CFT}|\Psi\rangle$. Hence by identifying $ s = t/\beta$ we see that this equation expresses the closure of the algebra ${\cal A}$ under time evolution.

In order to provide some additional intuition, let us consider the usual thermofield-double construction. We start with a quantum 
system with spectrum $H|E_i\rangle = E_i |E_i\rangle$. We consider the tensor product ${\cal H}_1 \otimes {\cal H}_2$ of two identical copies of this system and place it in a special entangled state
\be
\st_{\rm tfd} = {1\over \sqrt{Z}}\sum_{i} e^{-\beta E_i/2} |E_i,E_i\rangle,
\ee
where $Z=\sum_i e^{-\beta E_i}$. We call $H_1, H_2$ the two Hamiltonians. We also introduce the ``thermofield Hamiltonian'' defined as
\be
H_{\rm tfd} = H_1 - H_2,
\ee
which satisfies
\be
H_{\rm tfd} \st_{\rm tfd} = 0.
\ee
It should be an easy exercise for the reader to verify the following: if we take as our algebra of ``accessible observables'' ${\cal A}$ to be the operators acting on  system 1, then the conditions of the Tomita-Takesaki theorem are satisfied: i.e. the state $\st_{\rm tfd}$ is cyclic and separating for the algebra ${\cal A}$ in the Hilbert space ${\cal H}_1 \otimes {\cal H}_2$. We can thus define the operators $S,J,\Delta$ as described above. A few lines of algebra show that $J$ turns out to be the {\it antilinear} map that takes
\be
\label{ttthermo}
J: |E_i,E_j\rangle \rightarrow |E_j,E_i\rangle,
\ee
and
\be
\label{ttthermob}
\Delta = \exp(-\beta (H_1 -H_2)) = \exp(-\beta H_{\rm tfd}).
\ee
Hence, for any operator $A\in {\cal A}$, i.e. for any operator acting on the first copy of the system, the ``mirror operator'' $J A J$ given by the Tomita-Takesaki construction, is an operator acting on the second system and precisely coincides with what we would have defined as the dial via the usual thermofield doubling! The relation between the Tomita-Takesaki construction and the thermofield doubling has been noted before in the literature, for instance see \cite{Ojima:1981ma, Landsman:1986uw}.

Let us now consider the conformal field theory and consider the 
case where the elements of ${\cal A}$ are just modes of a generalized free-field. The last result that we wish to show here is that $\Delta$ really does reproduce the factors of $e^{-{\beta \w_n \over 2}}$ that we introduced above, 
at least for typical pure states. First consider the state $|\Psi'\rangle = {\op^i}_{\omega_n,\vect{m}}|\Psi\rangle$, where $\st$ is a typical equilibrium pure state. Using expression \eqref{deltadefss} we have
\be
\begin{split}
\langle \Psi'| \Delta |\Psi'\rangle &= \stl ({\op}^i_{\omega_n,\vect{m}})^\dagger \Delta {\op}_{\omega_n,\vect{m}} \st = 
\stl ({\op}^i_{\omega_n,\vect{m}})^\dagger S^\dagger S {\op}^i_{\omega_n,\vect{m}} \st  \\ &= 
\stl ({\op}^i_{\omega_n,\vect{m}})^\dagger S^\dagger ({\op}^i_{\omega_n,\vect{m}})^\dagger \st. 
\end{split}
\ee
Using the definition of the adjoint $S^\dagger$ of an antilinear operator given in \eqref{sbardef} we find
\be
\langle \Psi'| \Delta |\Psi'\rangle = \stl {\op}^i_{\omega_n,\vect{m}} ({\op}^i_{\omega_n,\vect{m}})^\dagger\st.
\ee
Now we remind the reader that typical equilibrium states in a large $N$ CFT satisfy the KMS condition, which for the modes of generalized free fields reads
\be
\stl {\op}^i_{\omega_n,\vect{m}} ({\op}^i_{\omega_n,\vect{m}})^\dagger\st   = e^{-{\beta \omega_n\over 2}} 
\stl ({\op}^i_{\omega_n,\vect{m}})^\dagger {\op}^i_{\omega_n,\vect{m}}\st.
\ee
This was extensively reviewed in \cite{Papadodimas:2012aq} where the reader can find more details. So all in all we find
\be
\langle \Psi'| \Delta |\Psi'\rangle = e^{-{\beta \omega_n \over 2}} \langle\Psi'|\Psi' \rangle.
\ee
Moreover if we have two different states of the form $|\Psi'_1\rangle = {\op}_{\omega_1,\vect{m}_1} |\Psi\rangle\,,\, |\Psi'_2\rangle
= {\op}_{\omega_2,\vect{m_2}} |\Psi\rangle$ with $1\neq 2$ (in hopefully obvious notation) we have $\langle \Psi_1'|
\Delta |\Psi_2'\rangle =0$.
The reader can easily verify that, using the KMS condition and the large $N$ factorization of the CFT, then for any two states
of the form
\be
|\Psi_1'\rangle = {\op}^{i_1}_{\omega_1,\vect{m}_1}...{\op}^{i_m}_{\omega_m,\vect{m}_m} \st,
\ee
and
\be
  |\Psi_2'\rangle = {\op}^{i_1'}_{\omega_1',\vect{m}_1'}...{\op}^{i_n'}_{\omega_n',\vect{m}_n'} \st.
\ee
We have\footnote{Notice that the equation \eqref{kmsmultiple} is consistent, even though it seems to break the symmetry between the number of insertions $m,n$ in the two states $|\Psi_{1,2}'\rangle$. The point is that in the large $N$ limit, both sides of the
equation are zero, unless $m=n$ and the frequencies/momenta of state $1$ are a permutation of those of $2$.}
\be
\label{kmsmultiple}
\langle \Psi_1' | \Delta | \Psi_2' \rangle = e^{-{\beta \over 2} \sum_{i=1}^m \omega_m}\langle \Psi_1' |  \Psi_2' \rangle  + \left({1\over N}\,\, {\rm corrections}\right).
\ee
If we are concerned with the action of $\Delta$ only in ${\cal H}_{\Psi}$, then this set of matrix elements completely specifies the operator. However, we see that the statement above is precisely the KMS condition
for the state $|\Psi \rangle$. So we see that in a state $|\Psi \rangle$, in which the correlators are close to being thermal, the operator $\Delta$ behaves precisely as $e^{-\beta (H_{CFT} - E_0)}$, where $E_0 = \stl H_{CFT} \st$, and this produces the 
$e^{-{\beta \w \over 2}}$ factors that we required above. 

We should caution the reader that in a real state in the CFT, which might correspond to a black-hole, the condition \eqref{kmsmultiple} might receive corrections at subleading order in ${1 \over N}$. These corrections
might have an effect on the eigenvalues of $\Delta$ in the case where the set ${\cal A}$ itself has a size that scales with $N$. We leave an investigation of these ${1 \over N}$ effects to further work.

\subsection{Finite-Dimensional Algebras \label{finitedalgebra}}

In this subsection we specialize to the case where ${\cal A}$ is a finite-dimensional closed subalgebra, which is acting
on a system with Hilbert space ${\cal H}$. This Hilbert space may be infinite-dimensional. We assume that the algebra ${\cal A}$ is closed under
Hermitian conjugation. We will find that the Tomita-Takesaki construction
reduces to the construction of the mirror operators defined in  \cite{Papadodimas:2012aq}. We {\it do not assume} that the system has necessarily a bipartite structure. 

The system is taken to be in a pure state $\st$. We consider the span ${\cal H}_{\Psi}$ of states of the form ${\cal A}\st$. If the dimensionality of the algebra ${\cal A}$ is $n$, then ${\cal H}_{\Psi}$ is an $n$-dimensional subspace of the full Hilbert space.

The interesting part of everything that follows will take place in this finite-dimensional space. While the algebra ${\cal A}$ is acting on ${\cal H}_{\Psi}$ it is clear
that there are many other operators which can act on the space ${\cal H}_{\Psi}$. In fact, the dimensionality of the algebra ${\cal A}$ is $n$ while the 
dimensionality of the algebra ${\cal B}(H_\Psi)$ of {\it all} operators acting on ${\cal H}_{\Psi}$ is $n^2$. We will argue that this is precisely related to the fact
that {\it on the same space ${\cal H}_{\Psi}$} we can naturally define the action of a second, commuting copy of the algebra ${\cal A}$, let us call it ${\cal A}'$
such that
\be
B(\shil) = {\cal A} \otimes {\cal A}'.
\ee
The construction proceeds exactly as before. In this case, all operators that we encounter are finite-dimensional, so it is very easy to 
check all steps in our argument explicitly, see appendix \ref{ttap} for technical details.

Again, we assume that the state $\st$ appears sufficiently entangled with respect to the algebra ${\cal A}$, which means that
\be
A\st =0 \quad\Leftrightarrow \quad A = 0.
\ee
This allows us to define the antilinear map 
$
S: \shil \, \rightarrow \,\shil
$
defined by
\be
S A \st = A^\dagger \st.
\ee
We also introduce its adjoint $S^\dagger$, which --- due to the fact that $S$ is antilinear --- is defined by
\be
(|A\rangle, S^\dagger |B\rangle) = (|B\rangle, S|A\rangle).
\ee
Using these two operators we consider the linear operator
$
\Delta :\shil \, \rightarrow \, \shil
$
defined by
\be
\Delta = S^\dagger S.
\ee
It is easy to show that $\Delta$ is a positive, Hermitian operator. It is related to what we would get from the 
polar decomposition of $S$ as 
\be
S = J \Delta^{1\over 2},
\ee
with $J$ anti-unitary. Equivalently we can just define the anti-linear operator
\be
J = S \Delta^{-{1\over 2}}.
\ee
As explained in appendix \ref{ttap} we can check $J$ satisfies
\be
J^2=1.
\ee
We also have the important relations
\be
S \st = J \st = \Delta \st = \st.
\ee
Now we define the mirror operators as operators acting on the Hilbert space ${\cal H}_{\Psi}$ by the relation
\be
\widetilde{A}_i = J A_i J.
\ee
Using the fact that $J^2=1$ we find that the antilinear ``tilde''-mapping
\be
\widetilde{~}: A \quad\rightarrow\quad \widetilde{A},
\ee
is an algebra *-isomorphism, i.e. the mirror operators satisfy the same commutation relations
as the original operators up to a conjugation of the structure constants, so if
\be
[A_i,A_j] = f_{i j}^k A_k,
\ee
then
\be
[\widetilde{A}_i,\widetilde{A}_j] = (f_{i j}^k)^* \widetilde{A}_k.
\ee
Moreover, as we demonstrate in appendix \ref{ttap}, the operators in ${\cal A}$ commute with the mirror operators
\be
[ A_i, \widetilde{A}_j] = 0.
\ee
Hence $ J {\cal A} J \in {\cal A}'$. We can also prove that any operator in $A'\in {\cal A'}$ can be written in the form
$A' \in J A J$ for some $A\in {\cal A}$. Hence we have that $J{\cal A}J = {\cal A}'$.

Let us now consider correlation functions. First, using $J^2=1$ and $J\st = \st$, we find that the dual-dual correlators are related to the original correlators by
\be
\stl \widetilde{A}_1\ldots \widetilde{A}_n\st = \stl A_1\ldots A_n \st^*,
\ee
and the mixed correlators obey
\be
\stl A \widetilde{B} \st = \stl A \Delta^{1 \over 2} B^{\dagger} \st.
\ee

The summary is that we can have on the subspace $\shil$ the action of the original algebra ${\cal A}$ together 
with the action of a second *-isometric copy $\widetilde{\cal A} = J {\cal A} J$, which commutes with ${\cal A}$. As discussed earlier, the mirror operators can be extended in the full Hilbert space ${\cal H}$ in more than one ways (for instance by taking
them to be ``zero'' on ${\cal H}_{\Psi}^{\bot}$.) The details of this extension do not affect correlation functions of ${\cal A}$'s and ${\widetilde{\cal A}}$'s evaluated on the state $\st$.

\subsubsection{Bipartite system}

We now demonstrate that, in the case where the system is bipartite, the $S,J,\Delta$ construction above is equivalent to the more direct construction of the mirror operators described \cite{Papadodimas:2012aq}.

Suppose we have a system which is bipartite ${\cal H}={\cal H}_A \otimes {\cal H}_{B}$, with $\dim{\cal H}_A \leq \dim {\cal H}_B$. We take the algebra ${\cal A}$ to be the algebra of operators
acting on ${\cal H}_A$. We take the system in a pure state $\st$, which is generally entangled. As before we assume that 
the entanglement is sufficiently large so that
\be
A\st =0 \qquad \Leftrightarrow \qquad A =0
\ee 
As will become clear below, this condition is equivalent to the condition that the reduced density matrix
\be
\rho_A = {\rm Tr}_B \left( \st \stl\right),
\ee
is of maximal rank.

\noindent {\bf Definition of mirror operators according to \cite{Papadodimas:2012aq}}: The pure state of the entire system can be expanded in a general orthonormal basis as
\be
|\Psi\rangle = \sum_{i j} c_{i j} |i\rangle_A \otimes |j\rangle_B.
\ee
We consider a (state-dependent) change of basis to bring the state in the Schmidt form
\be
\label{schmidtbasis}
|\Psi\rangle = \sum_{i=1}^{n_A} d_i |i\rangle_A^{\Psi} \otimes |i\rangle_B^{\Psi},
\ee
where we can take $d_i$ to be real and $\geq 0$. Here $n_a=\dim{\cal H}_A$. We have explicitly written the $\Psi$ superscript in  $|i\rangle_A^{\Psi},|i\rangle_B^{\Psi}$ to denote that these states depend on the choice of the pure state $\st$.

The reduced density matrix for system $A$ is
\be
\rho_A = \left(\begin{array}{ccc} d_1^2 & 0 &...\\ 0 & d_2^2 &... \\ ...\\ ... & ... & d_{n_A}^2 \end{array} \right).
\ee
We assumed that the entanglement of the original state $\st$ is sufficiently large, so that the matrix $\rho_A$ has maximal rank. Hence $d_i>0$.

In this case we defined the mirror operators as follows: for any operator acting on system $A$ of the form
\be
A = A_{ij} \,\,|i\rangle_A^{\Psi} \,^{\Psi}_A\langle j| \,\,\otimes \id_B,
\ee
we defined the mirror operator acting on $B$ as
\be
\label{tildeoldpaper}
\widetilde{A} = A_{ij}^* \,\id_A\,\otimes\,|i\rangle_B^{\Psi} \,_B^{\Psi}\langle j|.
\ee
Notice that this operator has non-vanishing matrix elements only along a $(\dim{\cal H}_A)$-dimensional subspace of the Hilbert space
${\cal H}_B$ --- i.e. it is a sparse operator. In the language of the previous subsections, this corresponds to the choice of taking the mirror operators to be zero on the subspace orthogonal to ${\cal H}_{\Psi}$.
\vskip10pt
\noindent{\bf The mirror operators according to the $S,J,\Delta$ Tomita-Takesaki construction}: Let us now see how the $S,J,\Delta$ construction leads to the same result. First we 
start with the state $\st$ and consider the linear space
\be
{\cal H}_{\Psi} = A \st.
\ee
This is an $n_A^2$-dimensional subspace of the full Hilbert space ${\cal H}_A \otimes {\cal H}_B$. Everything which will follow
will be defined on this space. First we define the antilinear map
\be
\label{modconj}
S A \st = A^\dagger \st,
\ee
and we also introduce $\Delta = S^\dagger S$. We then define the antilinear $J = S \Delta^{-1/2}$ and the mirror operators
as
\be
\widetilde{A} = J A J.
\ee
We will show that this definition coincides with the one above.

This construction is manifestly basis independent. We can thus apply it on a convenient basis. We select the Schmidt basis
\be
|\Psi\rangle = \sum_{i=1}^{n_A} d_i \,\,|i\rangle_A^{\Psi} \otimes  |i\rangle_B^{\Psi}.
\ee
Any operator $A$ acting on ${\cal H}_A$ acts as
\be
A|\Psi\rangle = \sum_{i , k = 1}^{n_A} d_i A_{ki} \,\,|k\rangle_A^{\Psi} \otimes  |i\rangle_B^{\Psi},
\ee
while the Hermitian conjugate acts as
\be
A^\dagger |\Psi\rangle = \sum_{i , k =1}^{n_A} d_i A_{ik}^* \,\,|k\rangle_A^{\Psi}\otimes  |i\rangle_B^{\Psi}.
\ee
We find that the antilinear operator $S$ which implements the modular conjugation \eqref{modconj} is
\be
S \,\, |i\rangle_A^{\Psi} \otimes |j\rangle_B^{\Psi} = {d_i \over d_j} \,\,|j\rangle_A^{\Psi}\otimes  |i\rangle_B^{\Psi},
\ee
and we see that
\be
S^\dagger \,\, |i\rangle_A^{\Psi} \otimes |j\rangle_B^{\Psi} = {d_j \over d_i} \,\,|j\rangle_A^{\Psi}\otimes  |i\rangle_B^{\Psi}.
\ee
Hence we find
\be
\Delta \,\, |i\rangle_A^{\Psi} \otimes |j\rangle_B^{\Psi} = {d^2_i \over d^2_j} \,\,  |i\rangle_A^{\Psi}\otimes  |j\rangle_B^{\Psi}.
\ee
Hence we notice that the states  $|i\rangle_A \otimes  |j\rangle_B$ in the Schmidt basis are eigenstates of $\Delta$. We can easily see
that we can express
\be
\Delta = \rho_A \otimes \rho_B^{-1},
\ee
and we have that $J$ is the antilinear map which is defined by
\be
J \left(|i\rangle^\Psi_A\otimes |j\rangle_B^\Psi\right) = |j\rangle_A^\Psi \otimes |i\rangle^\Psi_B.
\ee
Hence the mirror operators are defined by
\be
\widetilde{A} = J A J.
\ee
So again we find
\be
A = A_{ij} |i\rangle_A^\Psi \,^\Psi_A\langle j|\,\,\otimes \id_B.
\ee
We defined the mirror operator acting on $B$ as
\be
\widetilde{A} = A_{ij}^*  \,\,\id_A \otimes |i\rangle_B^{\Psi} \,_B^{\Psi}\langle j|.
\ee
This coincides with the definition \eqref{tildeoldpaper} according to \cite{Papadodimas:2012aq}.

\subsubsection{Construction in terms of projection operators\label{ttprojection}}

We also present one final (equivalent) way to look at this construction. Consider the algebra ${\cal A}$ with which we are probing the system. We would like to select a Cartan subalgebra ${\cal A}_a$, which we will use to ``label'' states by the collective eigenvalues $a$. Of course there are many possible ways to select the Cartan subalgebra. The point is that the way in which a state $|\Psi\rangle$ is aligned relative to the algebra ${\cal A}$, selects a particular preferred choice for the Cartan subalgebra, in which the ``entanglement is diagonalized''. 

Hence, for any given system and algebra ${\cal A}$, the preferred choice of the Cartan subalgebra depends on the state $\st$. This in turn depends on what we call a ``typical state'' i.e. what is the {\it ensemble} that we are considering. In particular, if we want to consider a microcanonical ensemble then it is the Hamiltonian which determines the ensemble and which finally selects the preferred orientation of the Cartan subalgebra. Hence the choice
of this Cartan subalgebra is a dynamical question. As we will see later, in the case of the large $N$ gauge theory, and if we think of ${\cal A}$ as the algebra of single trace operators, the {\it dynamics of the CFT} implies that the entanglement of a typical microstate selects as the preferred orientation of the Cartan subalgebra, the one generated by the ``occupation number operators'' $N_{\omega,\vect{k}}$ of the various modes.
\begin{figure}
\begin{center}
\includegraphics[width=7cm]{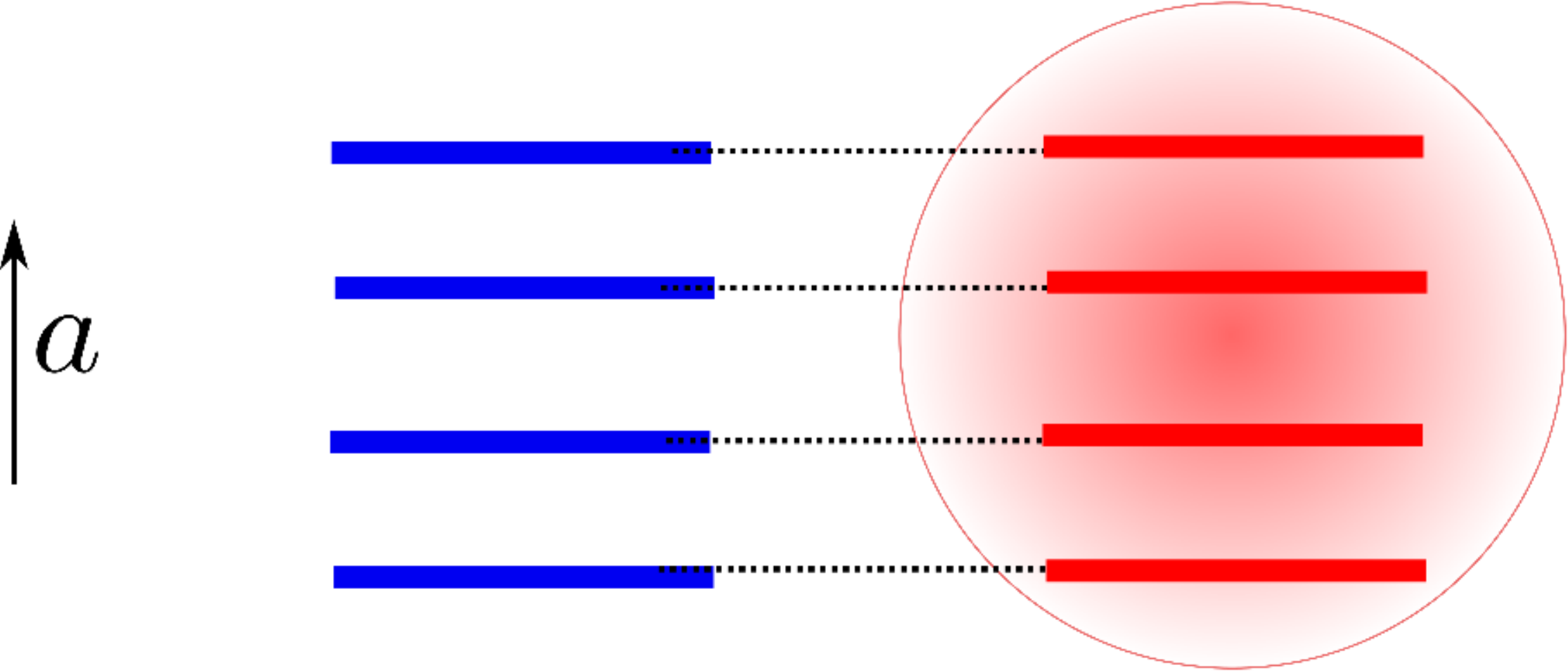}
\caption{A typical pure state $\st$ expanded in eigenstates of a Cartan subalgebra ${\cal A}_a$ of ${\cal A}$, selected so that the entanglement appears ``diagonal''. Here $a$ denotes the collective eigenvalues of ${\cal A}_a$.}
\label{tildeoperatorsproj}
\end{center}
\end{figure}
After these generalities, let us now see in more detail how the mirror operator construction works in this language. Suppose we select a particular Cartan subalgebra for ${\cal A}$. Consider projection operators $P_a$ on the eigenspaces of the Cartan subalgebra. The original state can be written as
\be
|\Psi\rangle = \sum_a P_a |\Psi\rangle = \sum_a d_a |a\rangle_{\Psi},
\ee
where\footnote{The fact that all $d_a>0$ follows from the assumption that $A\st \neq 0$ for $A \neq 0$.}
\be
|a\rangle_{\Psi} = {P_a |\Psi\rangle \over ||P_a |\Psi\rangle||}\qquad,\qquad d_a = ||P_a |\Psi\rangle||.
\ee
With this normalization, and since states of different eigenvalue $a$ are orthogonal, we have
\be
\,_{\Psi}\langle a| a' \rangle_{\Psi} =  \delta_{aa'}.
\ee

\begin{figure}
\begin{center}
\begin{subfigure}[t]{6cm}

\includegraphics[width=6cm]{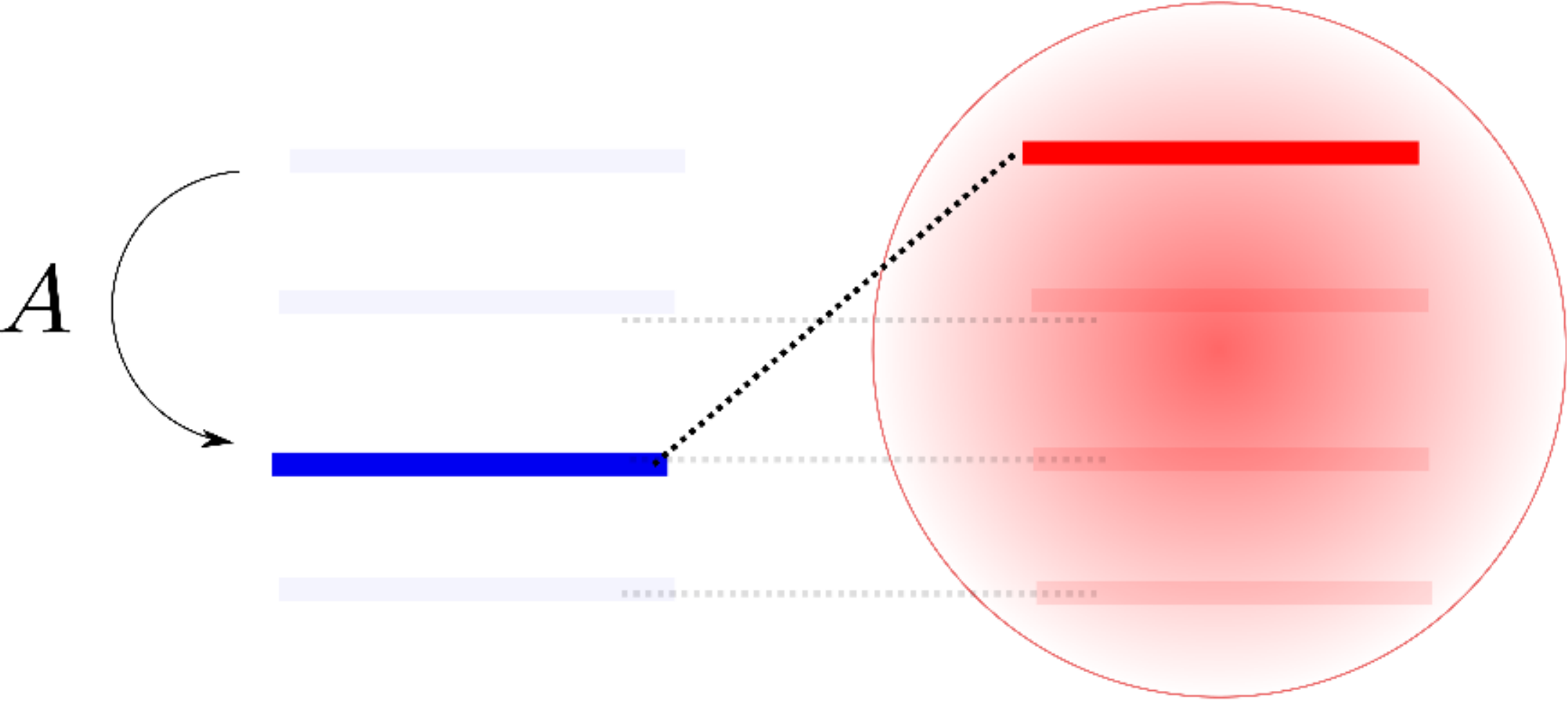}
 \label{fig:tildeproj}
\end{subfigure}
\qquad
\qquad
\begin{subfigure}[t]{6cm}
\includegraphics[width=6cm]{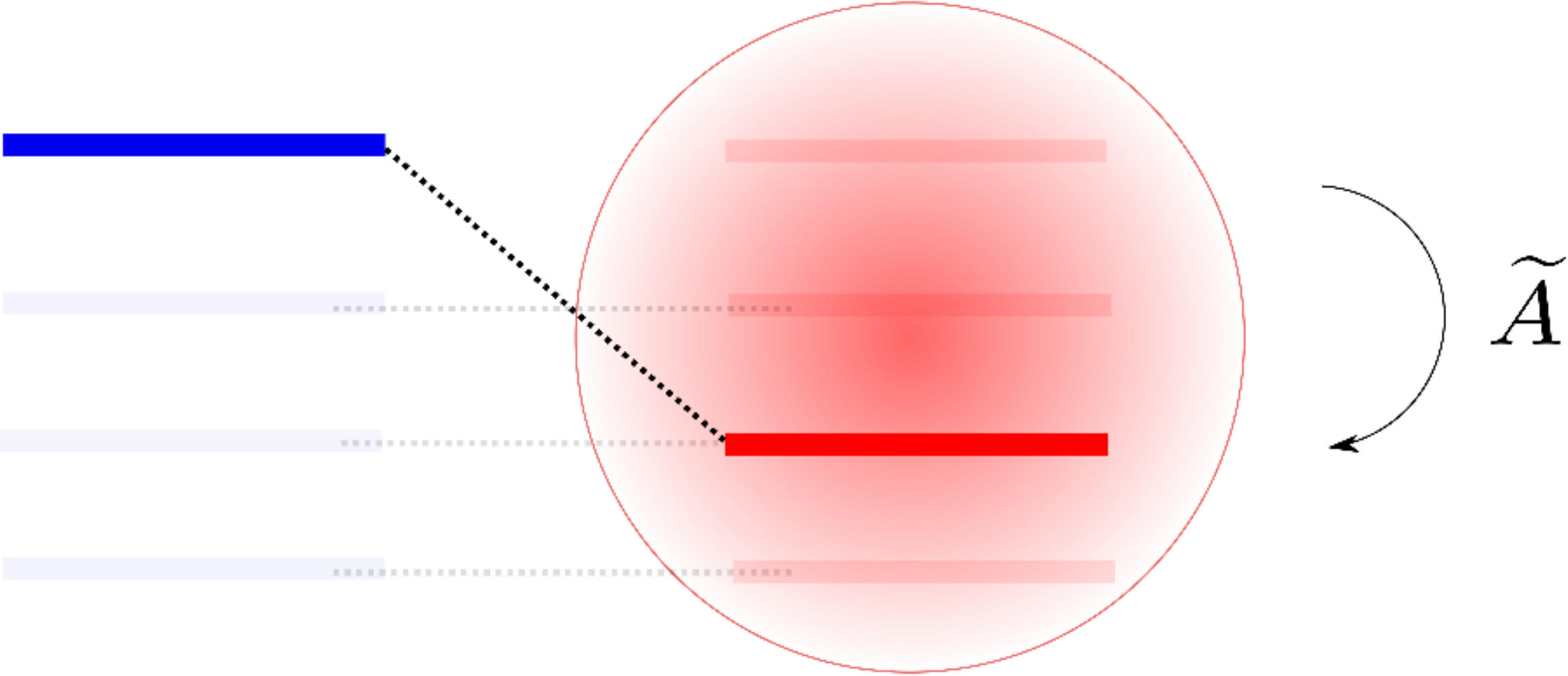}
\end{subfigure}
\caption{(Left) While we started with $n$ states, we can construct $n^2$ states by acting with operators in ${\cal A}$. (Right) The mirror operators can be defined as causing transitions between these $n^2$ states.  }
\label{tildeoperatorsprojb}
\end{center}
\end{figure}

It is clear that by acting with elements of the algebra ${\cal A}$, we can map a state of eigenvalues $a$ to a state with
eigenvalues $b$. This can be achieved by acting on the original state with an appropriate combination of operators from ${\cal A}$.
We call this combination of operators $T_{ba}$. Since we have assumed that $A\st =0$ implies $A=0$, we can see that for any possible transition $a\rightarrow b$, there is a unique choice of $T_{ba}$ (up to overall multiplicative constant). We then define the following set of states
\be
|b,a\rangle_\Psi = T_{ba} |a\rangle_\Psi.
\ee
We select the normalization of $T_{ba}$ so that all these states have unit norm. However, they are not necessarily orthogonal. In general we have
$
\,_{\Psi}\langle b,a| b', a' \rangle_{\Psi} = \delta_{bb'} f_{aa'}
$. 
The point now is that by a particular choice of the Cartan algebra, we can achieve that the the entanglement is ``diagonalized'' in the sense that $f_{aa'}= \delta_{aa'}$. Of course this problem is closely related to the Schmidt diagonalization. From now on we assume that we have aligned our Cartan algebra so that 
\be
\,_{\Psi}\langle b,a| b', a' \rangle_{\Psi} = \delta_{bb'} \delta_{aa'}.
\ee
In this case, the original pure state can be written as 
\be
 |\Psi\rangle = \sum_a d_a |a,a\rangle_{\Psi},
\ee
and schematically we see this in figure \ref{tildeoperatorsproj}. One can check that operators from the algebra ${\cal A}$ act on this state
as
\be
A = \sum_{a,a',b}A_{aa',b}  |a,\,b\rangle_{\Psi} \,_{\Psi}\langle a',\,b|.
\ee
We define the corresponding mirror operator as
\be
\widetilde{A} = \sum_{a,a',b}A_{aa',b}^*  |b,\,a\rangle_{\Psi} \,_{\Psi}\langle b,\,a'|.
\ee
We see a graphical representation of this in figure \ref{tildeoperatorsprojb}. We hope it is clear to the reader that this definition of the mirror operators is completely equivalent to the previous definitions.

\subsection{``Truncated Algebras'' and Complementarity}

In the previous subsections we described the doubling in the case where the set of ``accessible observables'' ${\cal A}$ forms
a closed algebra under multiplication. In that case the construction of the mirror operators was rather straightforward.

In this subsection we come to the more interesting case where the set of observables has a dual role: if it is truncated to a small subset, then we are naturally lead to the ``doubling'' and the introduction of  mirror operators for this subset. If it is not truncated, then the doubling is impossible and we can see that, what used to be the mirror operators before, can in reality be expressed as complicated combinations of the observables. Of course this is the operator-language version of the idea of black hole complementarity.

To be more precise, the case which is more relevant for us is when we have a set of ``basic observables'' $\{A_1,...A_n\}$ with
which we can probe the system. Since we want to compute correlation functions of these observables, it means that we also have to consider products of them. By considering unrestricted number of products, we do get a closed algebra generated by $A_i$.

But suppose that we do not want to consider this entire full algebra, but rather suppose that we want to consider the case where we probe the system only with a ``few insertions'' of the basic observables.  Hence we want to include in our observables products of the $A_i$'s as long as the number of factors does not get ``too large``. This requires some sort of regularization. A crude regulator would be to impose a hard cutoff $k$ in the number of insertions of the basic observables. For any 
choice of this regularization, the set ${\cal A}$ of allowed combinations of the basic observables is not a proper algebra, since it is not closed under multiplication in a strict mathematical sense.

In a large $N$ gauge theory, the large $N$ scaling provides us with a natural intuitive definition of these ideas: the ``basic observables'' are the single trace operators, and the allowed set of observables are products of $\Or[1]$ numbers of these operators. For any choice of the regulated set ${\cal A}$ we can define the mirror operators by following a slight modification of the procedure mentioned in the previous section. This leads to a definition of mirror operators which depends on the size of the regulated set ${\cal A}$, or relatedly, on the value of the cutoff $k$.

We need to be careful about the regime of validity of this construction. We want to take the cutoff $k$ to be large, but not too large --- otherwise we ''run out of space`` in the Hilbert space and we completely deplete the entanglement.
Whether this construction is sensible/useful  depends on the actual physical system under consideration. 
What we need to establish is that the mirror operators are ''robust`` under a change of the cutoff, when the cutoff is large but not too large.

To summarize, the realization of black hole complementarity in operator language is the statement that we have a set of  ``basic observables'' $\{A_1,...A_n\}$, which have the property that:

1. When we are considering low-point correlations of these observables, we can define the dual operators via the aforementioned construction.

2. When we are considering arbitrarily high order correlations, then we cannot --- this is due to the fact that the full algebra generated by arbitrary products of $A_i$  is so large, that with respect to this large algebra  the state $\st$ {\it does not  look entangled} any more, and there is no space/no need to define the dual operators. 

In that sense the dual operators can be understood as very complicated combinations of the basic observables $A_i$. This is in line with the paradigm of black hole complementarity.

\subsection{Large $N$ gauge theories}

In the case of the large $N$ gauge theory, we have  $\st$, a typical black hole microstate (i.e. a state of the CFT with energy of order $\Or[N^2]$). The set ${\cal A}$ is the vector space spanned by light operators. For example, as we mentioned above and in \cite{Papadodimas:2013b}, we could take the set to be spanned by the set of polynomials in the modes of ${\op}^i_{n,\vect{m}}$ with an upper bound on the energy, excluding the 
zero-modes of conserved currents, which we return to below.

 Alternately, we can also consider polynomials in just single-trace operators, and in an interacting theory, 
this should produce an equivalent set. As we have mentioned, in the $SU(N)$ theory, we can even consider a product of up to $N$-single trace operators.

The set ${\cal A}$ is not a proper algebra, because we have imposed the restriction that the number of insertions of single trace operators should not be too large. Let us call $k$ the effective cutoff in how many operators we allow. This defines the Hilbert space
$
{\cal H}^k_{\Psi} = \{\rm span \,\,: A |\Psi\rangle\}
$
we have included the superscript $k$ and the subscript $\Psi$ to indicate explicitly that this Hilbert space depends on the microstate $|\Psi\rangle$ and the cutoff $k$. It is a small subspace of the full Hilbert space ${\cal H}_{CFT}$.

We call $P^k$ the projection operator on ${\cal H}^k_{\Psi}$. By acting with this projection operators on the operators of ${\cal A}$ (i.e. replacing $A \rightarrow P^k A P^k$) we get a deformation of the set ${\cal A}$ into an algebra. Using this algebra we can define the $S,J,\Delta$ operators. It is clear that the matrix elements of $S,\Delta$ between states which do not carry
too many excitations relative to $\st$, are robust under scaling the cutoff $k$, and we will discuss this further in upcoming work.
In the large $N$ limit the modular operator $\Delta$ coincides with $e^{-\beta (H_{CFT}-E_0)}$ where $E_0 = \stl H_{CFT} \st$. Hence 
the correlators that we will get by following the Tomita-Takesaki
construction, are {\it to leading order in large $\nc$}, the same as the thermofield correlators.

\subsection{Conserved Charges}
Now, we describe how $S$ acts on conserved charges, including insertions
involving {\em polynomials} of charges, ${\cal Q}_{\alpha}$.  As usual,
by a conserved charge, we mean any operator that commutes exactly with the Hamiltonian, but we consider non-Abelian symmetries as well here.

As we noted in section \ref{cftnonabelian}, we can always move the charges so that they act directly on the state $|\Psi \rangle$. On such states, we define
\be
\label{sonqpoly}
S \al_{\alpha} {\cal Q}_{\beta} |\Psi \rangle  = \al_{\alpha}^{\dagger} S {\cal Q}_{\beta} |\Psi \rangle.
\ee
We emphasize that \eqref{sonqpoly} is valid only when the charges are immediately next to the state. We have not yet defined the action of $S$ on the space of states produced
by acting with the charge-polynomials on the base state $|\Psi \rangle$, which we discuss below.
However, since $S^2 = 1$, we see that even 
without specifying the action of $S$ on the charge-polynomial, we 
immediately obtain equivalence with \eqref{actiononn}.

First, let us check this fact: \eqref{sonqpoly} reproduces \eqref{actiononn}. We have
\be
\begin{split}
J O^i_{n,\vect{m}} J \al_{\alpha} {\cal Q}_{\beta} |\Psi \rangle &= S \Delta^{-{1 \over 2}} O^i_{n,\vect{m}} \Delta^{1 \over 2} S \al_{\alpha} {\cal Q}_{\beta} |\Psi \rangle \\  &= S \Delta^{-{1 \over 2}} O^i_{n,\vect{m}} \Delta^{1 \over 2} \al_{\alpha}^{\dagger} S {\cal Q}_{\beta} |\Psi \rangle \\ &= \al_{\alpha} \Delta^{1 \over 2}  (O^i_{n,\vect{m}})^{\dagger} \Delta^{-{1 \over 2}} {\cal Q}_{\beta} |\Psi \rangle.
\end{split}
\ee
Using the fact shown above that $\Delta \approx e^{-\beta (H_{\text{CFT}} - E_0)}$ at large $N$, we see that we precisely reproduce \eqref{actiononn}.

Now, we return to the definition of $S$ on the space produced by acting with charge-polynomials on $|\Psi \rangle$. We denote this space by ${\cal V}_{\cal Q} = \text{span}\{Q_{\beta_i} |\Psi \rangle \}$. We need to perform three checks on the action of the map $S: {\cal V}_{\cal Q} \rightarrow {\cal V}_{\cal Q}$.
 \begin{enumerate}
\item
On eigenstates, where ${\cal Q}_{\beta} |\Psi \rangle = Q_{\beta} |\Psi \rangle$, we have $S {\cal Q}_{\beta} |\Psi \rangle = Q_{\beta}^* |\Psi \rangle$.
\item
On null states, where ${\cal Q}_{n_i} |\Psi \rangle = 0$, we have 
$S {\cal Q}_{n_i} |\Psi \rangle = 0$.
\item
$S^2 {\cal Q}_{\beta} |\Psi \rangle =1$.
\end{enumerate}
In fact, these three conditions do not uniquely fix the action of $S$ on ${\cal V}_{\cal Q}$. 
For example, one possible definition of $S$ on ${\cal V}_{\cal Q}$ is as follows. Let the vectors 
\[
\{ |\Psi \rangle, {\cal Q}_{n_1} |\Psi \rangle \ldots {\cal Q}_{n_P} |\Psi \rangle, {\cal Q}_{b_1} |\Psi \rangle \ldots {\cal Q}_{b_M} |\Psi \rangle\},
\]
 form a basis for ${\cal V}_{\cal Q}$ so that
\be
{\cal Q}_{\beta} |\Psi \rangle = \kappa_{\beta}^1 |\Psi \rangle + \left(\sum_{i=1}^P \kappa_{\beta}^{n_i} {\cal Q}_{n_i} |\Psi \rangle \right) + \left( \sum_{i=1}^M \kappa_{\beta}^{b_i} {\cal Q}_{b_i} |\Psi \rangle \right).
\ee 
Then, we can define
\be
S {\cal Q}_{\beta} |\Psi \rangle = (\kappa_{\beta}^1)^* |\Psi \rangle + \left( \sum_{i=1}^M (\kappa_{\beta}^{b_i})^* {\cal Q}_{b_i} |\Psi \rangle \right).
\ee
This meets all three criterion above, but clearly we can redefine the $S$-map, by changing the basis. Such a redefinition does not affect the definition of the mirror operators.

\paragraph{Abelian Conserved Charges and Eigenstates\\}
We now specialize to Hermitian $U(1)$ charges, and to the situation
where the state $|\Psi \rangle$ is an eigenstate of such a charge. We include the Hamiltonian $H_{\text{CFT}}$ in this discussion, and so the
discussion here is also applicable to energy eigenstates. As we see below, the action of $S$ simplifies in this situation.

We start by considering a situation where the state $|\Psi \rangle$
satisfies
\be
H_{\text{CFT}} |\Psi \rangle = E_0 |\Psi \rangle.
\ee
Now, we find that
\be
\begin{split}
S H_{\text{CFT}} \al_{1} |\Psi \rangle = S \left([H_{\text{CFT}}, \al_{1}] + \al_1 H_{\text{CFT}} \right) |\Psi \rangle.
\end{split}
\ee
Without even assuming that $\al_1$ has a well defined energy,  we can 
write $[\al_1, H_{\text{CFT}}] = \al_1^H$, which, we assume, is in the set ${\cal A}$. So
\be
S  H_{\text{CFT}} \al_{1} |\Psi \rangle = \left(-(\al_1^H)^{\dagger} + \al_1^{\dagger} E_0 \right)|\Psi \rangle.
\ee
We see that this follows automatically if we use
\be
\label{actiononh}
S H_{\text{CFT}} \al_p |\Psi \rangle = \left(2 E_0 - H_{\text{CFT}}\right) S \al_p |\Psi \rangle; \quad J H_{\text{CFT}} \al_p | \Psi \rangle  = \left(2 E_0 - H_{\text{CFT}}\right) J \al_p |\Psi.
\ee
This has an interesting consequence. As we show
in Appendix \ref{ttap}, we have the relation
\be
J \Delta = \Delta^{-1} J.
\ee
Using the relation between $\Delta$ and the CFT Hamiltonian that
we described above $\Delta = e^{-\beta(H_{\text{CFT}} - E_0)}$, we see that this is precisely consistent with \eqref{actiononh}. 

We have the same relation between the map $S$ and any
other $U(1)$ conserved charge $\hat{Q}$, when the state is in
a charge eigenstate.
\be
S \hat{Q} \al_p |\Psi \rangle = \left(2 Q_0 - \hat{Q} \right) S \al_p |\Psi \rangle,
\ee
where $\hat{Q} |\Psi \rangle = Q_0 |\Psi \rangle$.
So, we see that our choice of gauge follows naturally in the Tomita-Takesaki construction.

Now, let us show directly that this is equivalent to \eqref{cftgaugeinvar}. Moving to the operators that transform simply under the charge, which were defined near \eqref{cftgaugeinvar}, we first note that
\be
\begin{split}
\tO^{i,q}_{n, \vect{m}} \al_{1} \hat{Q} \al_2 |\Psi \rangle &=  J {\cal O}^{i,q}_{n,\vect{m}} J \al_{1} \hat{Q} \al_2 |\Psi \rangle \\ &= J {\cal O}^{i,q}_{n,\vect{m}} J \left([\al_{1}, \hat{Q}]  + \hat{Q} \al_1\right)  \al_2 |\Psi \rangle \\
&=  J {\cal O}^{i,q}_{n,\vect{m}} J \left(\al_1^q  + \hat{Q} \al_1\right)  \al_2 |\Psi \rangle \\
&= J {\cal O}^{i,q}_{n,\vect{m}} \Delta^{1 \over 2}\left(\al_2^{\dagger} (\al_1^q)^{\dagger} + (2 Q_0 - \hat{Q}) \al_2^{\dagger} \al_1^{\dagger} \right) |\Psi \rangle,
\end{split}
\ee
where $\al_1^q = [\al_1, \hat{Q}]$.
Now, we use the fact that $[{\cal O}^{i,q}_{n, \vect{m}} \Delta^{1 \over 2}, \hat{Q}] = -q {\cal O}^{i,q}_{n, \vect{m}} \Delta^{1 \over 2}$, which is precisely how we
defined $q$ in \eqref{cftgaugeinvar}, and we have additionally used
that $\Delta$ commutes with $\hat{Q}$.

Substituting this relation above, we see that 
\be
\begin{split}
&\tO^{i,q}_{n, \vect{m}} \al_{1} \hat{Q} \al_2 |\Psi \rangle = J {\cal O}^{i,q}_{n,\vect{m}} \Delta^{1 \over 2}\left(\al_2^{\dagger} (\al_1^q)^{\dagger} + 2 Q_0 \al_2^{\dagger} \al_1^{\dagger} \right) |\Psi \rangle - J \hat{Q} {\cal O}^{i,q}_{n, \vect{m}} \Delta^{1 \over 2} \al_2^{\dagger} \al_1^{\dagger}  |\Psi \rangle \\
&= J {\cal O}^{i,q}_{n,\vect{m}} \Delta^{1 \over 2}\left(\al_2^{\dagger} (\al_1^q)^{\dagger} + (2 Q_0 - q) \al_2^{\dagger} \al_1^{\dagger} \right) |\Psi \rangle  - J \hat{Q} {\cal O}^{i,q}_{n, \vect{m}} \Delta^{1 \over 2} \al_2^{\dagger} \al_1^{\dagger} |\Psi \rangle \\
&= J {\cal O}^{i,q}_{n,\vect{m}} \Delta^{1 \over 2}\left(\al_2^{\dagger} (\al_1^q)^{\dagger}  - q \al_2^{\dagger} \al_1^{\dagger} \right) |\Psi \rangle  + \hat{Q} J  {\cal O}^{i,q}_{n, \vect{m}} \Delta^{1 \over 2} \al_2^{\dagger} \al_1^{\dagger} |\Psi \rangle \\
&= \left((\al_1^q) \al_2  - q \al_1^{\dagger} \al_2^{\dagger}  \right) \Delta^{1 \over 2} ({\cal O}^{i,q}_{n,\vect{m}})^{\dagger}  |\Psi \rangle  + \hat{Q} \al_1 \al_2  \Delta^{1 \over 2} ({\cal O}^{i,q}_{n, \vect{m}})^{\dagger} |\Psi \rangle \\
&= A_1 \big(\hat{Q} - q \big) A_2 \Delta^{1 \over 2} ({\cal O}^{i,q}_{n, \vect{m}})^{\dagger} |\Psi \rangle,
\end{split}
\ee
which agrees precisely with \eqref{cftgaugeinvar}. Note, in particular,
that the terms involving  $Q_0$ have canceled, and are not of
significance for the definition of the mirror operators, which always
come with {\em two} $J$'s.

\paragraph{Link with the Hartle-Hawking description of the State}
Before we close this section, let us take this opportunity to make a 
link to the usual Hartle-Hawking state for the bulk modes. The 
Hartle-Hawking state is sometimes written as an entangled state of
free-field ``modes'' outside and inside the horizon, and we would like to make this precise here.

First, note that given the single-trace operators ${\op}^i_{n,\vect{m}}$ in 
the set ${\cal A}$, we can form the number operators: $N_{n,\vect{m}} =  G^{-1}_{\beta}(n,\vect{m}) {\op}^\dagger_{n,\vect{m}} {\op}_{n,\vect{m}}$,
precisely as we did in section \ref{resolveleftinv} and \ref{resolvenaneq0}, where $G$ has the same meaning as there. 

These operators effectively commute with the Hamiltonian: $[N_{n, \vect{m}}, H] \approx \Or[{1 \over \nc}]$, and also with each other $[N_{n,\vect{m}}, N_{n',\vect{m}'}] \approx \Or[{1 \over \nc}].$ In a sense, these operators are describing the excitation of ``particles'' above the black-hole state.

The eigenvalues of these operators $N_{n,\vect{m},}$ are integral, and for each
such eigenvalue $p$, we can construct the projector $P_{n,\vect{m}}^p$,
which projects onto a state with a definite eigenvalue\footnote{We cannot {\em simultaneously} project all the $N_{n,\vect{m}}$ onto their eigenstates. There are $\Or[\omega_{\text{min}}^{-1}]$ different regularized frequencies, 
and a simultaneous projection requires us to multiply this many projectors, where ${1 \over \nc}$ effects become very important.\label{toomanymodes}} of $N_{n,\vect{m}}$
\be
|p,n,\vect{m} \rangle_{\Psi} = {{P_{n,\vect{m}}^p} \st \over || P_{n,\vect{m}}^p \st||}.
\ee
These states $|p,n,\vect{m} \rangle$ do {\em not} satisfy \eqref{noannihilgeneral} or \eqref{separatingb}. We cannot construct the mirror operators on these states, and these are precisely the {\em firewall} states.

States of different eigenvalues $p$ are approximately orthogonal. The original pure state can be written as
\be
\st = \sum_{p} P_{n,\vect{m}}^p \st = \sum_{p} d_{p} \, |p,n,\vect{m}\rangle_{\Psi}.
\ee
It is important that this superposition of states has an interpretation
as a smooth geometry although the individual states in the sum above
do not.

It is also useful to estimate the spread of $p$. 
At large $\nc$, we expect that with $Z = \sum_p e^{-\beta \sum p \omega_n}$
the coefficients $d_{p}$ satisfy
\be
|d_{p}|^2 = {1\over Z} e^{-\beta \omega_n p}.
\ee

Now, if we take $p$ to be very large,  say $p = N$, then
the formula for $d_{p}$ is not really valid, but this formula suggests that while we expect $d_p$ to be exponentially suppressed it should still be non-zero. 

Put another way, we expect that the original state $|\Psi \rangle$
contains a spread of ``number eigenvalues'' that is rather large. This has an immediate implication for \eqref{noannihilgeneral}. For example, if we try and annihilate the state $|\Psi \rangle$
by acting with a polynomial in the number operator: $\prod_{j=1}^{j_{\text{max}}} (N_{n,\vect{m}} - j)$, 
then we find that we must take $j_{\text{max}}$ to scale with $\nc \propto N^2$ before the polynomial annihilates the state.

Let us also briefly mention the link to the usual Hartle-Hawking state,
which is often written as an entangled state of free-field modes. As we mentioned in footnote \ref{toomanymodes}, we can still approximately diagonalize some $\Or[1]$ set of modes centered around frequencies $\omega_1, \ldots \omega_p$. 

As above, we construct mirror operators for $b_{n_1,\vect{m}_1} \ldots b_{n_p, \vect{m}_p}$ and  for $b^{\dagger}_{n_1,\vect{m}_1} \ldots b^\dagger_{n_p, \vect{m}_p}$. Then, for these modes, (ignoring their interaction outside this set) the state
of the CFT {\em appears} to be in the Hartle-Hawking state 
\be
|\Psi \rangle_{HH} = {1 \over \sqrt{Z}}\sum_{p_i}  e^{-\beta \omega_i p_i} |\{p_i\} \rangle | \{ \widetilde{p}_i \}\rangle.
\ee

We have been careful to restrict to an $\Or[1]$ set of modes, to avoid
complications that occur with the interaction of these modes within themselves. However, as we mentioned in section \ref{decoupledsho}, for the Hawking gas produced by an evaporating black-hole there is a description in terms of a Fock space of an $\Or[N]$ set of modes. It is clear, in that case, that the Hilbert space is not large enough to literally allow for the existence of a mirror-operator for each mode. But, as we have discussed many times above, these mirror operators exist in a state-dependent sense and have
precisely the correct properties unless we look at correlators with
too many insertions.


%% file: s_conclusions.tex
\section{Conclusions and Discussion \label{conclusions}}
In this paper, we have shown that if we allow the mapping between boundary
operators, and local bulk operators to depend on the state of the theory,
then {\em all} the recently articulated arguments in favour of structure
at the horizon are effectively resolved.

We described in section \ref{secneedtildes}, that the issue of whether
the black hole interior is smooth or not, could be reduced to an issue
of whether the light degrees of freedom of a single CFT could be effectively doubled in a thermal state. We showed explicitly how this could be done.

Our construction relies on the simple philosophy that only low-point correlators of light operators (where the number of insertions does not
scale with $\nc$) could be interpreted in terms of correlators of local
perturbative fields. So, the ``doubled'' operators that we need also
need to have the correct behaviour only {\em within} such correlators. 

We showed in section \ref{sec:three} that this imposes a set of 
linear constraints on the operators, that is much smaller than the
dimension of the Hilbert space that we are working within. These 
constraints lead to a set of consistent linear equations in 
a state that is close to a thermal state, 
since such a state cannot be annihilated by the action of
a small number of single-trace operators. We wrote down 
an explicit solution to these equations in \eqref{talexplicit}. Hence, it is 
possible to effectively double the number of degrees of freedom. 

As we showed in section \ref{resolveallpar}, this completely resolves
all issues that might suggest the presence of structure at the horizon.
We showed how to resolve the strong subadditivity paradox, while
making the commutators of operators inside and outside the horizon 
vanish exactly within low-point correlators.  We also
explained why the ``creation''
operators did not need to have a left-inverse inside the horizon,
by pointing out that their commutation relations with the 
corresponding ``annihilation'' operators had to be obeyed within correlation functions, and not necessarily as operator relations. We
also showed that our construction allowed an explicit computation 
of the expectation value
of $N_a$ --- the particle number, as observed by the infalling observer ---
 with the result that $N_a = 0$. The argument of \cite{Marolf:2013dba} 
breaks down for state-dependent operators. 

We can already study time-dependent correlators about equilibrium
states with our construction, including those where the horizon of a black hole is excited. 
 However, we also showed how to extend
our construction to cases where the mirror operators
are built directly on top of non-equilibrium states and showed
that this gave us results that were completely consistent with
semi-classical intuition.

We also pointed out that our construction, modulo some technical features
having to do with the presence of conserved charges in the CFT, was
the same as the well-known Tomita-Takesaki construction that has played
an important role in the mathematical quantum field theory literature.

We are left with the issue of whether state-dependence must be 
allowed, even in principle. Although various other authors
have explored subtleties in these arguments, which may eventually
invalidate them, the arguments of \cite{Marolf:2013dba} 
strongly suggest that it is not possible to find state-independent
operators behind the horizon, 

In this paper, we have investigated how these arguments
break down, if we allow a state-dependent mapping between
the bulk and boundary operators. As we mentioned 
the state-dependence of our
operators is somewhat similar to the state-dependence of the density
matrix in a given state: $\rho = |\Psi \rangle \langle \Psi |$. 
The density matrix can be treated as an ordinary operator, and given
the density matrix for some state $|\Psi \rangle$, nothing prevents us
from considering its action on another state $|\Psi'\rangle$, or 
evaluating $\langle \Psi'| \rho | \Psi' \rangle$. In this sense
the density matrix is a usual operator in the Hilbert space. However, it has a useful physical interpretation in a given state $|\Psi \rangle$.  

The situation in our case is certainly a little unusual, in that the local
quantum field in the bulk $\phi_{\text{CFT}}(x)$ itself seems to depend on the
state $|\Psi \rangle$. We should point out that 
as we pointed out in section \ref{secneedtildes}, that if we 
consider the bulk-boundary mapping outside the horizon then this is 
also naively state-dependent, since the ``transfer function''
is different in the vacuum, and the thermal state.  At the least, it is clear that the ${1 \over \nc}$ expansion of the mapping depends on the state. The authors of \cite{Marolf:2013dba} have suggested (at the discussion in the ``Fuzz, Fire or Complementarity'' 
workshop in August 2013 at KITP) that it may be possible
to write down a state-independent operator that has the correct behaviour
in a given state, presumably along the lines of the gauge-invariant
relational observables of \cite{Giddings:2005id}. However, it would
be nice to see a precise formulation of this statement, including an
analysis that we can make such a construction stable with respect
to quantum corrections. 

We leave a deeper analysis of state-dependence to further work. 
However, in this paper, we have tried to analyze this state-dependence
as carefully as possible, and we have found that it does not
contradict any expectations from quantum mechanics. 

We should mention that state
dependent operators have also appeared, in parallel with our work, in \cite{Verlinde:2013vja, Verlinde:2013uja,Verlinde:2012cy}. The reader should
consult those papers for an alternate perspective.

Another important direction for future work has to do with the ``uniqueness'' of our construction. This issue also exists outside the horizon,
where it is possible to write down a mapping between boundary
and bulk operators. It has been suggested \cite{Kabat:2012av} that
bulk-locality uniquely fixes this mapping, but it would be nice to
put this on a firmer footing. We explore this question, to some extent
in Appendix \ref{tildealternate} but this issue of uniqueness is even more acute
for operators behind the horizon and deserves further investigation.

We would also like to address some philosophical issues regarding
the relation of our approach to previous perspectives on this problem. 
Our perspective in this paper has been that, low-point correlators in the conformal field theory can be reinterpreted in terms of correlators on a semi-classical spacetime. If we make the number of insertions too large, scaling with the central charge of the theory,  then the picture of semi-classical spacetime breaks down. This is the fundamental limitation that must be respected according to us.

This delineates our perspective from some previous approaches to this problem. Earlier perspectives on black hole complementarity posited a picture of ``observer complementarity'', where the infalling observer and the observer outside the horizon saw different ``realities'', except that they could
never communicate to obtain a contradiction. Some recent modifications of 
this approach have attempted to suggest that each different light-cone
might admit its own ``reality.'' In our opinion, these perspective are not
 entirely tenable, and we have not used them at all in this paper.

Our perspective is that there is a {\em global picture} of reality, which 
can be accessed by a super-observer in the CFT. As we have shown, this picture is consistent with correlation functions, where the number of insertions does not scale with $\nc$. If we exceed this bound
it is approximate locality that breaks down. This perspective
on black hole complementarity --- where we do not attempt to 
find semi-classical bulk interpretations for $\nc$-point correlators ---
resolves many of the confusions surrounding the information paradox in AdS/CFT.


%% file: app_alternatepure.tex
\section{ ${1 \over \nc}$ Corrections, Alternate Purifications and Uniqueness  \label{tildealternate}}

There are two parts to
our construction of the mirror operators behind the horizon. One of them
is \eqref{tildedefcftcomord}, which tells us that the mirror operators
commute with the ordinary operators. The second is \eqref{tildedefcftactpsi}, which tells us their correlations with ordinary operators. As we 
have mentioned many times above, we expect \eqref{tildedefcftcomord} to
hold unchanged when ${1 \over \nc}$ corrections are included, but
\eqref{tildedefcftactpsi} should receive corrections at first non-trivial
order in ${1 \over \nc}$. 

To compute these corrections is a formidable task, even for simple
Witten diagrams outside the black hole particularly in a state with
energy that scales with $\Or[\nc]$. Unlike vacuum Witten diagrams,
where ${1 \over \nc}$ corrections correspond to bulk-loops, here, we
also have to be careful about the ensemble (canonical vs microcanonical) that we are working in. Nevertheless, in the first part of this section, we discuss how we would modify our prescription if someone were to {\em give us} the right answer. This brings up the issue of the uniqueness of our
construction that we discuss next.

\subsection{Accounting for ${1 \over \nc}$ corrections}
In principle  one could compute ${1 \over \nc}$ corrections to the Bogoliubov coefficients that translate between quantization on the Schwarzschild slices, and the nice slices. To our knowledge, no such computation has actually been performed
in anti-de Sitter space. However, let us say that such a computation of
Bogoliubov coefficients in the {\em bulk} tells us that we 
should use not the thermofield doubled state, but the state
\be
\label{exactpur}
|\Psi \rangle_{\text{doub}} = \sum_{i, j} C_{ i j}  |E_i \rangle_{\text{out}} | \widetilde{E}_j \rangle_{\text{in}},
\ee
to do bulk computations. 
This indicates that the state $| E_i \rangle_{\rm out}$ in the Hilbert space
of the field theory of the outside Schwarzschild observer is entangled
with the state $|E_j \rangle_{\rm in}$ in the field theory of the inside observer.

Here $C_{ij}$ is a matrix that tells us the entanglement between the two sides. Obviously, if we take $C_{ij} ={1\over \sqrt{Z}}
e^{-{\beta E_i \over 2}} \delta_{i j}$,
we get the thermofield doubled state. We are interested in matrices $C_{ij}$
that are close to this form, and differ from it by ${1 \over \nc}$ corrections. However, below, we will not assume anything 
about $C_{ij}$ except that it is invertible. 

Note, however, that given a pure state in the theory $|\Psi \rangle$, and
a set of observables ${\cal A}$, we cannot mimic the state $|\Psi \rangle_{\text{doub}}$ for any arbitrary matrix $C_{ij}$. We have the very important
consistency condition, that for correlators of ordinary operators in ${\cal A}$
\be
\label{doubconsistency}
\prescript{}{\text{doub}}{\langle \Psi|} \al_p |\Psi\rangle_{\text{doub}} = \langle \Psi | \al_p | \Psi \rangle, ~ \forall \al_p \in {\cal A},
\ee
where these elements of ${\cal A}$ can, of course, be products of smaller elements. We see why the thermofield doubled state is generically a good
choice to leading order in ${1 \over \nc}$. In this state
\be
\prescript{}{\text{tfd}}{\langle \Psi|} \al_p |\Psi\rangle_{\text{tfd}} = \tr\left(e^{-\beta H} \al_p\right) = \langle \Psi | \al_p | \Psi \rangle + \Or[{1 \over \nc}],
\ee
for almost any equilibrium state $|\Psi \rangle$. However the $\Or[{1 \over \nc}]$ corrections above also tell us that in general, at subleading
orders in ${1 \over N}$, this consistency condition requires us to use a more general state of
the form \eqref{exactpur}. We show how to now correct \eqref{tildedefcftactpsi} for a state of the form \eqref{exactpur}.

Just as in \eqref{tfdactpsi} and \eqref{almatelem}, as usual, for each operator,  $\al_p |E_i \rangle = \left(\al_p\right)_{j i} |E_j \rangle$, we have the 
mirror operator, which acts on the other side: $\al_p^{\text{doub}} |\widetilde{E}_i \rangle =  \left(\al_p\right)_{j i}^* |\widetilde{E}_j \rangle$. This is the operator, 
that {\em in a physical sense} acts in the same way on the other side,
because the $|E_i \rangle$ form a privileged energy eigenbasis. For example,
we could go to the Schmidt basis, in which the entanglement
is diagonal and then ask for the operators that act on the other side
of the Schmidt basis in the same way, as we did in equation \ref{tildeoldpaper}. The reader should note that we are asking a slightly different question here.

Now, note that, in the state $|\Psi \rangle_{\text{doub}}$,  we can convert the action of $\al_p^{\text{doub}}$, which
acts only on the tilde-states, to an action of operators that act only
on the ordinary states
\be
\begin{split}
&\al_p^{\text{doub}} |\Psi \rangle_{\text{doub}} =  C_{i j} (\al_p)^*_{k j} |E_i \rangle |\widetilde{E}_k \rangle = C_{i j} (\al_p)^*_{k j} (C^{-1})_{k l} C_{l m} |E_i \rangle |\widetilde{E}_m \rangle \\
&= \left(C^{-1} \al_p^{\dagger} C \right)_{i l} C_{l m} |E_i \rangle |\widetilde{E}_m \rangle = \breve{\al_p} |\Psi \rangle_{\text{doub}},
\end{split}
\ee
where all repeated indices are summed 
\be
\breve{\al}_p = C^{-1} \al_p^{\dagger} C,
\ee
and $C^{-1}_{i l} C_{l m}  = \delta_{i m}$.

Now, to mimic the action of the mirror operators in a state $|\Psi \rangle_{\text{doub}}$, we expand the set of observables ${\cal A}$ to include the 
observables $\breve{\al_p}$ and then 
we simply define our tilde operators to satisfy
\be
\label{tildeactdef1}
\tal_p |\Psi \rangle = \breve{\al_p} |\Psi \rangle.
\ee
Once we can define the tildes to have an action on the state of some product of ordinary operators, simply by commuting them to the right
\be
\label{tildeactdef2}
\tal_{p_1} \al_{p_2} \ldots \al_{p_m} |\Psi \rangle = \al_{p_2} \ldots \al_{p_m} \widetilde{\al}_{p_1} |\Psi \rangle,
\ee
and use this, by induction, to define the action of a product of tildes as well. 

We can again check that this definition works correctly to reproduce
correlators of products of simple operators. We see, from an application
of the rules above, that
\be
\tal_{p_1} \tal_{p_2} |\Psi \rangle  =  \breve{\al}_{p_2} \breve{\al}_{p_1} |\Psi \rangle.
\ee
On the other hand, we can also check that
\be
\begin{split}
&\al_{p_1}^{\text{doub}} \al_{p_2}^{\text{doub}} |\Psi \rangle_{\text{doub}} =  (\al_{p_1})^*_{l k} (\al_{p_2})^*_{k j} C_{j i} |E_i \rangle |\widetilde{E}_l \rangle \\
&= C_{ i j}(\al_{p_1})^*_{l k} (\al_{p_2})^*_{k j} C^{-1}_{l t} C_{t m} |E_i \rangle |\widetilde{E}_m \rangle = C \al_{p_2}^{\dagger} \al_{p_1}^{\dagger} C^{-1} |\Psi \rangle_{\text{doub}}.
\end{split}
\ee
So
\be
\langle \Psi | \tal_{p_1} \tal_{p_2} | \Psi \rangle = \langle \Psi |\breve{\al}_{p_2} \breve{\al}_{p_1} | \Psi \rangle = \prescript{}{\text{doub}}{\langle \Psi |} \al_{p_1}^{\text{doub}} \al_{p_2}^{\text{doub}} |\Psi\rangle_{\text{doub}}.
\ee
By an extension of this to higher products we can check that, just 
as desired,
\be
\label{rightcorrdoub}
\langle \Psi |\tal_{p_1} \ldots \tal_{p_m}  \al_{p_{m+1}} \ldots \al_{p_n} | \Psi \rangle = \prescript{}{\text{doub}}{\langle \Psi|} \al_{p_1}^{\text{doub}} \ldots \al_{p_m}^{\text{doub}} \al_{p_{m+1}} \ldots \al_{p_n} | \Psi \rangle_{\text{doub}}. 
\ee

\subsection{Uniqueness}
This discussion brings up another point. If we are given a state, and
bulk correlators, how do we fix the matrix $C$. For an equilibrium
state, it is reasonable to choose $C$ to be diagonal in the energy
eigenbasis. 

Geometrically, this is the following statement. Consider a black hole
that has reached thermal equilibrium. If the black hole was formed
from the collapse of a state with a narrow band of energies,
it {\em may not} be well represented by the thermofield doubled state. 
However, it should still be well represented  by the state
\be
\label{diagonalc}
\psi_{\text{doub}} = \sum_i C_{i i} |E_i \rangle  |\widetilde{E}_i \rangle.
\ee
For example, if the original black hole is well represented
by the microcanonical ensemble, then we could take $C_{ii}$
above to be constant for a given range of energies, and zero outside.
Geometrically this also corresponds to an ``eternal black hole'', but
where the entanglement corresponds to the microcanonical ensemble. This
geometry differs at $\Or[{1 \over \nc}]$ from the canonical eternal
black hole geometry.

Both these geometries share the property that they are invariant
if we evolve forward in time on the right, and backward in time on the
left. In the bulk, this is an isometry which rotates a spacelike slice
passing through the bifurcation point. 

If we do make the assumption that $C$ is diagonal in the energy eigenbasis, then our tilde operators are essentially fixed. This is because
the eigenvalues $C_{ii}$ can be set by measuring expectation values
of ordinary operator $\al_p$ in the state $|\Psi \rangle$ and demanding \eqref{doubconsistency}.

However, a note of caution is in order here. Even if $|\Psi \rangle$
is in equilibrium, as defined in \eqref{noneqscen}, and \eqref{doubconsistency} hold, it is {\em not} necessary for $C$ to be diagonal. 
This is because we see
\be
\prescript{}{\text{doub}}{\langle \Psi|} e^{i H t} \al_p e^{-i H t} |\Psi\rangle_{\text{doub}} = C_{i j} C^*_{j k} \al_{i k} e^{i (E_i - E_k) t}.
\ee
Now, it is easy to see that even for a generic matrix $C$, that 
satisfies $\tr(C^{\dagger} C) = 1$,  the time-dependence above is extremely small. 

Note that this question could also be raised about the correspondence
between the eternal black hole and the thermofield doubled state. What 
sets the precise form of the entanglement there, to be diagonal in the
energy eigenbasis. One answer would be that the bulk theory
has the isometry above, where we can rotate a spacelike slice about
the bifurcation point. However, this isometry exists to excellent
precision even if we change the structure of the entanglement.
This issue is related to the issue of the uniqueness of our construction. We leave a more detailed study to further work.

%% file: app_choosegauge.tex
\section{Choice of Gauge \label{gaugechoice}}
We now briefly discuss our choice of gauge in \eqref{cftgaugeinvar}. The 
construction of local operators corresponding to charged fields was also discussed in recent papers \cite{Heemskerk:2012np, Heemskerk:2012mn, Kabat:2012av}.

First, we briefly remind the reader of the non-local commutators that result
from working in a fixed gauge. We take the example of scalar QED in curved space, although
non-Abelian gauge theories lead to similar results, and we believe that our
qualitative conclusions should also hold for gravity.

Let us put the metric in the standard ADM d+1 form:
\be
ds^2 = -N^2 dt^2 + h_{i j} (d x_i + N_i d t) (d x_j + N_j dt).
\ee
With this split, we have:
\be
\sqrt{-g} = \sqrt{h} N,
\ee
and the components of the inverse metric are
\be
\label{inverse31}
g^{00}  = -1/N^2; \quad g^{i j} = h^{i j} - N^i N^j/N^2; \quad
g^{0 i} = {N^i \over N^2}.
\ee
The Lagrangian density for scalar QED is given by:
\be
{\cal L} = -{1 \over 4} F_{\mu \nu} F^{\mu \nu} - J_{\mu} A^{\mu}  + {\cal L}_{\text{matter}},
\ee
where $J^{\mu}$ is composed of the matter fields, but we are not interested in the matter Lagrangian here, except for the Poisson brackets it will induce with the
matter field.

 We see that we can write
\be
\label{expandfmunu}
{1 \over 4} F_{\mu \nu} F^{\mu \nu} = {1 \over 2} F_{0 i} F_{0 j} \left(g^{00} g^{i j} - g^{0 j} g^{0 i} \right) + {1 \over 2} F_{0 i} F_{k l} \left(g^{0 k} g^{i l} - g^{0 l} g^{i k} \right) + {1 \over 4} F_{m n} F_{k l} g^{m k} g^{n l},
\ee
where all Latin indices run only over the spatial direction. This can be simplified by using the form of the inverse metric given above
\be
{1 \over 4} F_{\mu \nu} F^{\mu \nu} = -{h^{i j} \over 2 N^2} F_{0 j} F_{0 i} + {N^k \over N^2}  h^{l i} F_{k l} F_{0 i} + {1 \over 4} F_{k l} F_{m n} g^{k l} g^{l n}.
\ee

Now, we go over to the Hamiltonian formalism to make contact with quantum mechanics. We use $i$ for the spatial directions only. We find that:
\be
\Pi^i (x) = {\partial L \over \partial \left( \partial_0 A_i (x) \right)} = 
 - F^{0 i}(x).
\ee
Note that we have
\be
\begin{split}
F^{0 i} &= g^{0 \mu} g^{\rho i} F_{\mu \rho} = \left(g^{0 0} g^{j i} - g^{0 j} g^{0 i} \right) F_{0 j}  + g^{0 k} g^{l i} F_{k l} \\
&= -{h^{i j} \over N^2} F_{0 j} + {N^k \over N^2}  h^{l i} F_{k l},
\end{split}
\ee
which is entirely consistent with the expansion of the Lagrangian above in \eqref{expandfmunu}.

Just from the structure of the Lagrangian, we have the ``primary'' constraint
\be
\label{constraint1}
\phi_1 = \Pi^0 = 0.
\ee
Following Dirac \cite{dirac2001quantum},  we will use the notation $\phi_n$ to denote the various constraints that will arise. 

We proceed to work out the Hamiltonian. As usual, the sign of the term quadratic in $F_{0 i}$ is reversed and the the term linear in $F_{0 i}$ drops out. We see that
\be
\begin{split}
&\Pi^i \partial_0 A_i +  {1 \over 4} F_{\mu \nu} F^{\mu \nu}  = \Pi^i \left(F_{0 i} + \partial_i A_0 \right) +  {1 \over 4} F_{\mu \nu} F^{\mu \nu} = {1 \over 2} {h^{i j} \over N^2} F_{0 j} F_{0 i} + {1 \over 4} F_{k l} F_{m n} g^{k l} g^{l n}  + \Pi^i \partial_i A_0 \\
&=  {1 \over 2} h_{i j} \left(\Pi^i + {N^k \over N^2}  h^{l i} F_{k l}\right) \left(\Pi^j + {N^m \over N^2}  h^{n j} F_{m n}\right) + {1 \over 4} F_{k l} F_{m n} g^{k m} g^{l n} + \Pi^i \partial_i A_0.
\end{split}
\ee
Using the form of the inverse metric given in \eqref{inverse31}, we find
\be
\begin{split}
g^{k m} g^{l n} F_{m n} F_{k l} &= \left(h^{k m} h^{l n}  - {1 \over N^2} h^{k m} N^{l} N^{n} - {1 \over N^2} h^{l n} N^k N^m + N^l N^k N^m N^n \right)F_{m n} F_{k l} \\
&= F_{m n} F^{m n} - {2 \over N^2} F_{m n} h^{k m} N^l N^n F_{k l},
\end{split}
\ee
where as usual, the spatial indices have been raised using $h$.
We see that the second term above precisely cancels with the term that appears when the whole square involving the momentum in the Hamiltonian is expanded out. So, we find that finally
\be
\Pi^i \partial_0 A_i +  {1 \over 4} F_{\mu \nu} F^{\mu \nu} = {1 \over 2} \Pi_i \Pi^i + \Pi^i N^k F_{k l} + {1 \over 4} F_{m n} F^{m n} + \Pi^i \partial_i A_0.
\ee
This leads to the Hamiltonian density
\be
\begin{split}
{\cal H}_0 &= \left[\Pi^i \partial_0 A_i - {\cal L} + U_1 \Pi^0  \right] d^3 x \\
&= {1 \over 2} \Pi_i \Pi^i + \Pi^i N^k F_{k l} + {1 \over 4} F_{m n} F^{m n} + \Pi^i \partial_i A_0 + J^{\mu} A_{\mu}.
\end{split}
\ee
Of course, the Hamiltonian is given by $H_0 = \int \sqrt{-g} {\cal H}_0$.

We have called this Hamiltonian density ${\cal H}_0$, since we will have to modify it systematically to get consistency with the constraints as laid down in Dirac's procedure. To start with,  we also need to include a term $U_1 \Pi^0$ for the constraint, as specified by Dirac. Here $U_1$ can be an arbitrary function of the $A_i$ and the conjugate momenta $\Pi^i$. After adding this term, we have
the modified Hamiltonian ${\cal H}_1 = {\cal H}_0 + U_1 \phi_1$.

 Now, to preserve the constraint we require
\be
\{\Pi^0, H_1 \} = 0.
\ee
Recall that when we compute the Poisson brackets, by definition, we have
\be
\begin{split}
&\pb[A_0(x), \Pi^0(x')] = {1 \over \sqrt{-g}} \delta^d(x-x'); \\
&\pb[A_i(x), \Pi^j(x')] = {1 \over \sqrt{-g}} \delta^j_i \delta^d(x - x'),
\end{split}
\ee
where we have suppressed the time-coordinate, which is always equal in
the quantities in Poisson bracket.
The additional factor of $\sqrt{-g}$ appears because of the way we defined
our Lagrangian, without the $\sqrt{-g}$.

So, we see that the Poisson bracket above immediately leads to the Gauss law
\be
\label{constraint2}
\phi_2 = {1 \over \sqrt{-g}} \partial_i \left[\sqrt{-g} \Pi^i\right] + J^0 = 0.
\ee
So, we have obtained $\phi_2$ as a secondary constraint. 
We see that \eqref{constraint2} does not lead to any further constraints because
\be
\label{piwithflm}
\{\partial_i \sqrt{-g}(x) \Pi^i(x),  F_{l m} (y) \} = \partial_{x^i}  \sqrt{-g}(x) \left[\delta^i_m \partial_{y^l} {1 \over \sqrt{-g}(y)} \delta^d(x-y) -  \delta^i_l \partial_{y^m} {1 \over \sqrt{-g}}\delta^d(x-y) \right] = 0,
\ee
since we can convert the $(-g)^{-{1 \over 2}}(y)$ to a $(-g)^{-{1 \over 2}}(x)$, using the delta function, pull it out of the derivative and then cancel
it with the $(-g)^{1 \over 2}(x)$ that accompanies the momentum. 

As a result, since we see that we have
\be
\{\phi_2, H_1 \} = 
\left(\partial_i \sqrt{-g} J^i + \{\sqrt{-g} J_0, H_1\}  \right)= 0,
\ee
where we have not displayed terms that vanish because of \eqref{piwithflm}.
We are implicitly assuming that when we write down the matter Lagrangian, it gives rise to 
\be
\{\sqrt{-g} J^0, H_1\} + \partial_i \sqrt{-g} J^i = 0,
\ee
 as an identity.

We now write a second Hamiltonian as
\be
\begin{split}
H_2 &= \int \sqrt{-g}  \left({\cal H}_1 +  U_2 \phi_2\right),
\end{split}
\ee
where $U_2$, for now, is another arbitrary parameter.

However, we see that we cannot fix the Hamiltonian uniquely, and that
$U$ and $U_2$ are left undetermined. This is because \eqref{constraint1} and \eqref{constraint2} have zero Poisson bracket with each other, so they are {\em first class} constraints.

At this point, in principle, we could restrict ourselves to
only gauge invariant operators. In this language, the analogue of local fields 
would be fields with Wilson lines attached to them. Here, we will take a slightly cruder approach of simply fixing gauge, since that is more convenient
from the point of view of constructing local bulk observables. 

\subsection{Gauge Fixing}
To convert these first class constraints into second class constraints, we will consider a set of ``algebraic gauges'' which are fixed by imposing
\be
\label{gaugefixing}
\phi_3 = A_{a}  = 0; \quad \phi_4 = \Pi_a + {N^k \over N^2}  F_{k a} - \partial_a A_0 = 0.
\ee
The second constraint is meant to impose $F_{a 0} + \partial_a A_0 = 0$.
For example, in flat space, we could take $a = 3$ to get axial gauge. 
The reader should keep in
mind that $a$ is not a dummy index in this section, but is fixed to be 
the index of some particular {\em spatial} coordinate.

Note that we now have the following matrix of Poisson brackets between the
constraints
\be
\begin{split}
&C_{m n}(x,y) = \{ \phi_m(x), \phi_n (y) \} \\
&= \begin{pmatrix} 
0&0&0&{1 \over \sqrt{-g}(x)}{\partial \over \partial y^a} \delta^d(x - y) \\
0&0&-{1 \over \sqrt{-g}(y)} {\partial \over \partial x^a} \delta^d(x-y)&0\\
0&{1 \over \sqrt{-g}(x)} {-\partial \over \partial y^a} \delta^d(x-y)&0&{h_{aa} \over \sqrt{-g}} \delta^d(x-y)\\
{-1 \over \sqrt{-g} (y)} {\partial \over \partial x^a} \delta^d(x-y)&0&{-h_{aa} \over \sqrt{-g}} \delta^d(x-y)&0\\
\end{pmatrix}.
\end{split}
\ee
We write down a third Hamiltonian:
\be
\label{h3}
H_3 =H_2 + \int \sqrt{-g} \left(U_3 \phi_3  + U_4 \phi_4 \right) d^d x.
\ee

For consistency, we need to ensure that \eqref{constraint1}, \eqref{constraint2},\eqref{gaugefixing} are all consistent with the Hamiltonian \eqref{h3}. 
\be
\label{allweakconstraints}
\{\phi_m, H_3 \} = 0,~\text{for}~m=1,2,3,4.
\ee
where all the equations have to hold in a weak sense.

The main thing to calculate is
\[
\{\phi_4, H_3 \} 
\]
We note that
\be
\begin{split}
\{ \Pi_a(x), F_{k l}(y) \} &= h_{a b}(x) \{ \Pi^b(x), F_{k l}(y) \} \\&= h_{a b} (x) \left[\delta^b_l {1 \over \sqrt{-g}(x)} {\partial \over \partial y^k} \delta^d(x-y) + \delta^b_k {1 \over \sqrt{-g}(x)} {\partial \over \partial y^l} \delta^d(x-y) \right] \\
&= {h_{a l}(x)  \over \sqrt{-g}(x)} \partial_{y^k} \delta^d(x - y) + {h_{a k}(x) \over \sqrt{-g}(x)} \partial_{y^l} \delta^d(x-y).
\end{split}
\ee
Consequently,
\be
\begin{split}
&\{ \Pi_a(x), \int \sqrt{-g}(y) F_{k l}(y) F^{k l}(y) \} = -4 {h_{a l}(x) \over \sqrt{-g}(x)} \partial_k \sqrt{-g} F^{k l}(x), \\
&\{ \Pi_a(x) \int \sqrt{-g} N^k \Pi^l F_{k l} d y \} = -{h_{a l}(x) \over \sqrt{-g}(x)} \partial_k \sqrt{-g} \left(N^k \Pi^l - \Pi^l N^k \right).
\end{split}
\ee
Putting all this together, we see that
\be
\begin{split}
\{\phi_4, H_3\} &=  -{h_{a l}(x) \over \sqrt{-g}(x)} \partial_k \sqrt{-g} F^{k l}(x) - {h_{a l}(x) \over \sqrt{-g}(x)} \partial_k {\sqrt{-g} \over N^2}  \left(N^k \Pi^l - \Pi^l N^k \right) - J_a \\ &+  {N^k  \over N^2} \left(\partial_k \left(\Pi_a  + {N^p  \over N^2} F_{p a} + \partial_a A_0 \right) - \partial_a \left(\Pi_k + {N^p \over N^2} F_{p k} + \partial_k A_0 \right) \right) \\ &- \partial_a U_1 - U_3.
\end{split}
\ee
We also see see that 
\be
\{\phi_3, H_3 \} = \phi_4  - \partial_a U_2 - h_{a a} U_4, \quad
\{\phi_1, H_3 \} = \partial_3 U_4, \quad
\{\phi_2, H_3 \} = \partial_3  U_3.
\ee
We can solve the equations above as follows
\be
U_2 = 0, \quad U_3 = 0, \quad U_4 = 0. \quad U_1 = \int_{z_0}^z \{\phi_4(\zeta), H_3 \} d \zeta.
\ee
In the last line, we have explicitly displayed the dependence of the 
quantities on the spacetime coordinates.
These solutions are {\em not unique}. For example,
in the first line above, we set $U_2 = 0$, although, technically we could still set $U_2$ to be a function of only the $x,y$ coordinates; this is a symptom of the residual gauge invariance after our gauge fixing.

So, finally we end up with a non-local Hamiltonian, which is just $H_3$ with the solutions for $U_1 \ldots U_4$ substituted.
\be
\begin{split}
H_4 = \int N \sqrt{h} \Big[&{1 \over 2} \Pi_i \Pi^i + \Pi^l {N^k \over N^2} F_{k l} + {1 \over 4} F_{m n} F^{m n} + \Pi^i \partial_i A_0 + J^{\mu} A_{\mu} \\ &+ \Pi^0 \int^z_{z_0} \{ \phi_4(\zeta), H_0 \} d \zeta
\Big] d^d x.
\end{split}
\ee

\subsubsection{Quantization}
Finally, we turn to the non-local commutators we get by quantizing this
theory. 
To do this
we have to use Dirac's prescription. First, we need to find the
inverse of the matrix $C_{m n}$. Note that this is defined by 
\be
\int D_{m n}(x - y) C_{n p}(y - z)d y  = \delta_{m p} \delta^d(x - z).
\ee
We need solutions to the following differential equations
\be
\begin{split}
&{\partial \over \partial x^a} \left({1 \over \sqrt{-g}} {\cal A}(x,y) \right) =  \delta^d(x - y), \\
&{\partial \over \partial x^a} {1 \over \sqrt{-g}} {\cal B}(x,y) = {h_{a a} \over \sqrt{-g}} {\cal A}(x,y).
\end{split}
\ee

We see that 
\be
\begin{split}
D_{m n}(x-y) &= \delta^d(x_1 - y_1) \delta^d(x_2 - y_2) 
\begin{pmatrix}
0&{\cal B} (x,y) &0& {\cal A} (x,y)  \\
-{\cal B}(x,y)&0&{\cal A}(x,y)& 0 \\
0 & -{\cal A}(x,y) & 0 & 0 \\
{\cal A}(x,y) &0 &0 &0 
\end{pmatrix}.
\end{split}
\ee
A solution to these differential equations is given by
\be
\begin{split}
{\cal A}(x,y) &= \sqrt{-g} \left( \theta(x^a - y^a) + A(\hat{x}, \hat{y}) \right) \\
{\cal B}(x,y) &= \sqrt{-g} \left(\int d x^a h_{a a} {1 \over \sqrt{-g}} {\cal A}(x,y) + B(\hat{x}, \hat{y}) \right).
\end{split}
\ee

where the dependence on $\hat{x}$, and $\hat{y}$ means that  $A$ and $B$ do {\em not} depend on $x^a$. 
So, we see that we have two arbitrary functions $A$ and $B$. This is because our gauge-fixing condition does not completely fix the gauge. We will ignore these functions for now.

Now, the Dirac prescription is to consider Dirac brackets given by
\be
[{\cal F},{\cal G}]_{\text{D.B.}} = \{{\cal F},{\cal G}\} - \{{\cal F}, \phi_m\} D^{m n} \{\phi_n, {\cal G}\}.
\ee 

We are finally in a position to compute commutators between the electric field
and the scalar field. When we write down the matter Lagrangian, we
get a current that should satisfy
\be
\{J^0(x), \Phi(y) \} = {1 \over \sqrt{-g}} \delta^d(x-y) \Phi(x).
\ee
So, the interesting commutator that we want to investigate is
\be
\begin{split}
[\Pi^a(x), \phi(y)]_{\text{D.B.}} &= -\int d z_1 d z_2 \{\Pi^a(x), \phi_3(z_1) \} D^{3 2} (z_1, z_2) \{\phi_2(z_2), \Phi(y) \} \\ 
&= \theta(x^a - y^a) + A(\hat{x}, \hat{y}).
\end{split}
\ee
This is a simple and universal result. However, notice
that the choice of the function $A(\hat{x}, \hat{y})$ gives us some freedom
in choosing the exact commutator, as we point out below.

Now, we apply all this to the case of the AdS black brane. In the region
outside the brane, we choose the gauge $A_z = 0$. Recall that
this is the gauge that must be chosen close to the boundary in any case
to get the usual relationship between bulk and boundary correlators.
We choose the function $A = -1$, and this leads to the commutators
\[
\deb[\Pi^z(t,x_1), \phi(t,x_2)] = -\theta(z_1 - z_2) \delta^{d-1}(x_1 - x_2).
\]
The physical interpretation of this commutator in terms of Wilson lines
is simple. We think of the field $\phi$ as being attached to a Wilson
line that goes all the way to the boundary at $z = 0$, along a 
path of constant spatial coordinates. 
So, if the electric field operator is placed at a {\em smaller} value
of $z$ (closer to the boundary), it intersects the Wilson line, leading
to the non-zero commutator above. 

Now consider the region behind the horizon. In this region, $z$ becomes
a time-like coordinate and $t$ becomes a spatial coordinate. We now
choose the gauge $A_t = 0$. Now, we find the commutator 
\[
\deb[\Pi^t(z,x_1), \phi(z,x_2)] = -\theta(t_1 - t_2)  \delta^{d-1}(x_1 - x_2),
\]
where the the last $\delta^{d-1}$ excludes the $\delta$-function in $z$, of course, which is now playing the role of a time coordinate.

This formula suggests an amusing interpretation in terms of Wilson lines. 
The operators behind the horizon, have Wilson lines that extend deeper into the black hole, and eventually re-emerge near the boundary through
a wormhole! So, their charge can be measured at infinity, but not their position.

\subsection{Mirror Operators Below the Hawking-Page Temperature?}
We now discuss another issue that has sometimes been raised. We expect
the tilde operators to exist in any thermal state, including one where
the temperature is low enough that the dual state is represented by a gas of gravitons, rather than a black hole. 
What is the significance of mirror operators below the Hawking-Page temperature? 

In fact, the issue of gauge invariance helps us here. The mirror operators
cannot be used in any region that is connected to the boundary. First, note
that the form of the equal-time commutator between a conserved current and a charged
local operator is fixed by locality. We must have
\be
[j^{0}(t,\Omega), {\cal O}(t,\Omega')] = q {\cal O}(t,\Omega) \delta^{d-1}(\Omega - \Omega'),
\ee
where the delta function is understood to be correctly normalized on the sphere.
Clearly, the commutation relations that we have imposed in \eqref{cftgaugeinvar} are not of this form. So, the $\tO$-operators {\em cannot} be 
understood to be local operators on the boundary.

Now, this also implies that they cannot be appear in 
fields, in a region that is not causally separated from the boundary. We could, otherwise, take a limit as these operators tend to the boundary,
and the commutator would have the wrong form for a local field. This fact, by itself, implies that the mirror-operators $\tO$ do not appear in 
expressions for local bulk fields below the Hawking-Page temperature.

To conclude, {\em the existence of the $\tO$-operators is necessary
to construct fields behind the boundary. However, just because they
can be defined in a state in the CFT does not mean that they
appear in the expressions for bulk fields.}

\subsection{Constructing the Other Side?}
For the same reason, it is clear that the mirror operators do not 
really represent a region III of the Black Hole. Note that,
in the eternal black hole geometry if $\hat{Q}$ is the charge
that we are measuring near the boundary of the first CFT, and $\tO^i(t,\Omega)$ are operators in the second CFT then we would have $[\hat{Q}, \tO^i(t,\Omega)] = 0.$
The commutation relations \eqref{cftgaugeinvar}, which allow
us to measure the charge of the $\tO$-operators from the boundary, tell us that boundary
of the CFT always covers the region that the $\tO$ operators live in. This seems to suggest that we cannot really reconstruct region III from our
construction.


%% file: app_measurement.tex
\section{The ``Measurement'' Argument \label{appmeasurement}}

The first AMPS paper contained the statement ``We can therefore construct operators acting on the early radiation, whose action ...  is equal to that of a projection operator onto any given subspace of the late radiation.''

This statement by itself does not lead to any paradox. The apparent paradox
appears, when we also consider the radiation behind the horizon and this 
leads to a seeming violation of the strong subadditivity of entropy. We have already discussed this issue in section \ref{resolvestrong}. There, we
also discussed in detail how it was not possible, while remaining
within a framework that is described by semi-classical spacetime, for the
observer to distill the part that is entangled with the late radiation.

However, even if the authors of \cite{Almheiri:2012rt} did not intend this,
the statement above has led to a  misunderstanding that the existence of these operators acting on the early radiation, that can project the late radiation onto a given state, {\em ipso facto}, implies that the horizon may
have a firewall.

In this appendix, we clarify some of these basic issues.

\paragraph{\bf ``Acting'' with Operators \\}
The basic fact to realize is that even operators 
that are localized to a region ``acting'' on a state can produce funny effects far away from the region. In fact, the Reeh-Schlieder theorem tells
us the following. Consider a local quantum field theory, and the set
of local operators acting on some open set $M$, which we denote by $\Phi(M)$, and {\em any state} of finite energy $|\Omega \rangle$, which may
even be the vacuum.  Then, the set $\Phi(M) |\Omega \rangle$ is dense in the full Hilbert space ${\cal H}$. 

\begin{figure}
\begin{center}
\includegraphics[width=4cm]{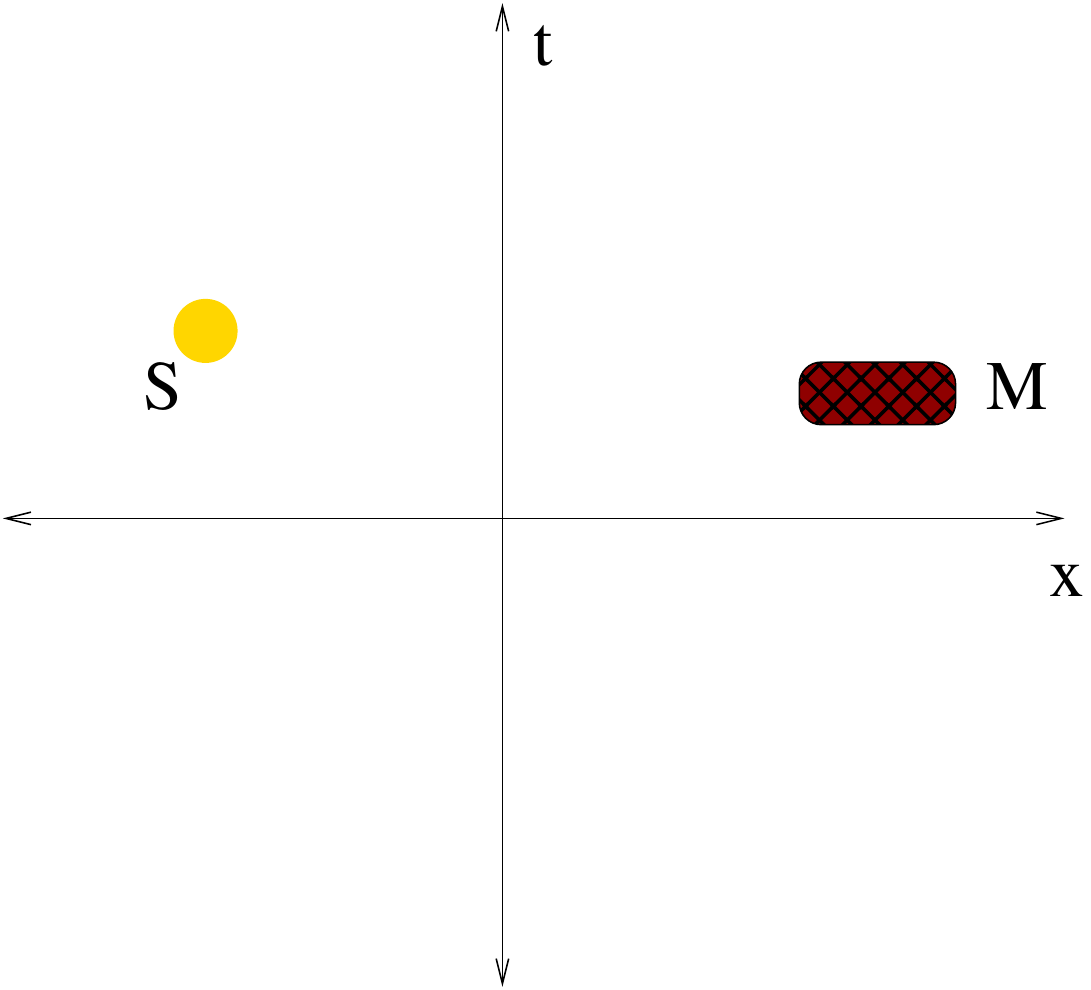}
\caption{\em {\bf Implication of the Reeh-Schlieder Theorem:} A spacetime
diagram shows how operators in the bounded region $M$ can create
the sun $S$ at large spacelike separation!\label{reehschleider}}
\end{center}
\end{figure}

For details and references about this theorem see \cite{Haag:1992hx}. This theorem has
to do with the fact that even the vacuum state has long-range entanglement. So, by delicately manipulating even a localized region of this state,
we can create whatever we want even in the causal complement of the region. The physical implication of this theorem in shown in Fig \ref{reehschleider}.

This theorem {\em does not} imply any violation of locality. In
fact, from a physical perspective, we can modify the Hamiltonian
and cause the state to undergo unitary evolution, but we cannot ``act''
with some arbitrary operators on a state. 

\paragraph{\bf Complicated Measurements \\}
Now, it is also well known that very complicated ``measurements'' do fall
into the class of actions that can be obtained through unitary evolution. 
We could turn on a Hamiltonian that would entangle a local quantum field theory with another much larger system and permit us to measure some quantity with great accuracy. The next simple point we want to make is
that if we actually perform such a measurement, it can disturb the
system and again lead to funny effects.

As we have already explained in section \ref{resolvestrong}, if we try and measure the early radiation to distill the part that is entangled with
the late radiation, there is no reason to expect that this operation
will have a simple semi-classical interpretation because it will
involve insertions of operators with energies that scale with $\nc$.

However, even some measurements that may have spacetime interpretations
can disturb the system enough to create a firewall. This is particularly
true of measurements that are extremely sharp i.e. measurements
where the associated projection operators project onto an extremely
low-dimensional subspace of the Hilbert space.

Here, we point out how
this phenomenon can be seen even in the flat space {\em Minkowski vacuum.}
\begin{figure}
\begin{center}
\includegraphics[width=0.5\textwidth]{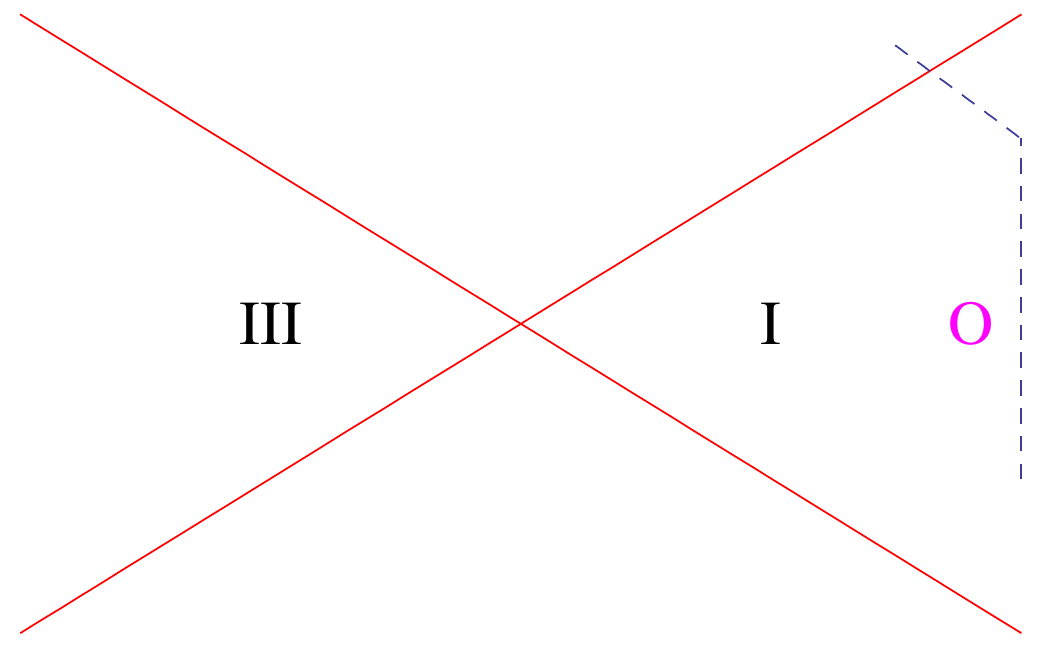}
\caption{An observer who measures the Rindler energy, and then tries to cross
the Rindler horizon, encounters a ``firewall.'' \label{rindlerfig}}
\end{center}
\end{figure}
We will show, how by a very special measurement in the Minkowski vacuum, we can create a firewall at the Rindler horizon. Consider quantizing a massless scalar field $\Box \phi = 0$, in $d+1$-dimensional spacetime, with metric 
\[
ds^2 = -dt^2 + dz^2 + d \vect{x}^2,
\]
where $\vect{x}$ is a $(d-1)$-dimensional vector.
In region I (as shown in figure \ref{rindlerfig}), we transform to the coordinates
\[
t = \sigma \sinh \tau, \quad z = \sigma \cosh \tau,
\]
so that the metric becomes
\[
ds^2 = -\sigma^2 d \tau^2 + d \sigma^2 + d \vect{x}^2.
\]
We can quantize the field in region I (as shown in the figure) using the expansion
\[
\phi(\tau,\sigma,\vect{x}) = \int_{\omega>0}{d\omega d^{d-1}\vect{k} \over (2 \pi)^d} \left[{1\over \sqrt{2\omega}} a_{\omega,\vect{k}} e^{-i\omega \tau+i \vect{k} \vect{x}}{2 K_{i\omega}(|\vect{k}| \sigma)
\over |\Gamma(i\omega)|} + {\rm h.c.}
\right].
\]
For region III, we use the coordinate transformation $t= -\sigma \sinh \tau,\,\,z=-\sigma \cosh \tau$ and expand the field as
\[
\phi(\tau,\sigma,\vect{x}) = \int_{\omega>0}{d\omega d^{d-1}\vect{k} \over (2 \pi)^d} \left[{1\over \sqrt{2\omega}} \ta_{\omega,\vect{k}} e^{i\omega \tau-i \vect{k} \vect{x}}{2 K_{i\omega}(|\vect{k}| \sigma)
\over |\Gamma(i\omega)|} + {\rm h.c.} \right].
\]

Now, it is well known, that that in this expansion, even the Minkowski vacuum appears as an entangled state:
\be
\label{omegaentang}
|\Omega \rangle_{\text{Mink}} = \sum_{E} e^{-\pi E} |E \rangle_{I} \otimes | E \rangle_{III},
\ee
where the sum over $E$ runs over the entire Fock space and $E$ is the energy of the state in this Fock space. 

Now, consider an observer who lives in region I for a long time, and makes an
accurate measurement of the Rindler energy.  At the end of this process, s/he
is entangled with a superposition of states in the Fock space, that have 
the specific energy corresponding to the result of his measurement. However, the stress-tensor in a state with a specific energy diverges
at the Rindler horizon \cite{Candelas:1978gg}. Hence, for the observer $O$ shown in Figure \ref{rindlerfig}, this creates a firewall as s/he crosses the Rindler horizon. Obviously this does not mean that the Minkowski vacuum has such a firewall, but merely that a very sharp measurement,\footnote{In this case, unlike the AMPS scenario, the measurement merely has to be sharp and not even fine-tuned.} which effectively involves entangling the system with an extremely large measurement ``apparatus'' 
can disturb the system enough to create funny objects.

\paragraph{\bf State-Dependence of the AMPS projection: \\}
Finally, we would like to mention  that the projection that AMPS
consider \cite{Almheiri:2012rt} is state-dependent. We are merely
pointing out this fact, and do not attach
any special significance to this issue, since we feel that state-dependent operators like the density matrix or the mirror operators that we have been considering are useful. 

Note, that AMPS would like to consider a measurement (in their notation) where the state of the black hole 
after the Page time is
\be
|\Psi\rangle = \sum_i |\psi_i\rangle_E\otimes |i\rangle_L,
\ee
where $E$ indexes the early radiation and $L$ indexes the late radiation.
Since ${\rm dim}{\cal H}_{E} \gg {\rm dim}{\cal H}_L$. AMPS point out that for any state $|i\rangle_L$ of the late radiation and the corresponding projection operator $P^i = |i\rangle_L \langle i|_L$ we can define another projection operator $\hat{P}^i$ written in terms of the early radiation such that
 \be
 \label{measuramps}
 \hat{P}^i |\Psi\rangle \approx P^i |\Psi\rangle = |\psi_i\rangle_E \otimes |i\rangle_L.
 \ee
We can write
\be
\hat{P}^i = |\Psi_i\rangle_{E} {\langle \Psi_i|_E} \otimes I_L.
\ee
However, since the precise state $|\Psi_i\rangle$ that is correlated
with $|i \rangle_L$ depends on the state $|\Psi \rangle$, this projector $\hat{P}^i$ must be correlated with the state $|\Psi \rangle$ to perform the action \eqref{measuramps}.


%% file: app_tomitatakesaki.tex
\section{Technical Details of the Tomita-Takesaki Construction \label{ttap}}

In this appendix we present some proofs of statements made in section \ref{tomitatakesaki}. For illustrative purposes we concentrate
on the case of a finite-dimensional algebra acting on a finite-dimensional space $\shil$, because then the proofs are easy and we do not need to worry about issues of convergence. Of course, in the finite-dimensional case the quickest way to prove these statements
is by working in an appropriate Schmidt basis, as was done in subsection \ref{finitedalgebra}. Here we provide an alternative presentation which may help the reader in
following the more elaborate proofs for the infinite-dimensional case, which can
be found in the mathematical literature \cite{bropalg}.

Remember that we have the finite-dimensional algebra ${\cal A}$ acting on the Hilbert space $\shil$. We assume that if $A \in {\cal A}$ then 
$A^\dagger \in {\cal A}$. Clearly ${\cal A}$ is a von Neumann algebra. We also assume that states of the form $A \st\,,\,A\in {\cal A}$ span the entire Hilbert space $\shil$, which means that the vector $\st$ is {\it cyclic} for the algebra ${\cal A}$ and also that $A\st =0$ implies $A=0$, which means $\st$ is {\it separating}. Below we present the proofs of various technical statements that enter in the construction of the mirror operators in the language of the Tomita-Takesaki framework.
 
First we define the commutant ${\cal A}'$ (the set of operators acting on $\shil$ which commute with all elements of ${\cal A}$), which is also a von Neumann algebra. The von Neumann bi-commutant theorem guarantees that
\be
({\cal A}')' = {\cal A}.
\ee
For a proof of this classic theorem we refer the reader to \cite{bropalg}.
\vskip20pt
Then we show that if $\st$ is cyclic and separating for ${\cal A}$ then it is also cyclic and separating for ${\cal A}'$. 

\noindent {\bf Proof:} i) First we will prove that the vector $\st$ is separating for the algebra ${\cal A}'$. Suppose we have an operator $A'\in {\cal A}'$ such that
\be
A' \st = 0.
\ee
We will show that implies that $A'=0$ as an operator. Consider any other vector $|A\rangle$ in $\shil$. Since (by assumption) $\st$ is cyclic for the algebra ${\cal A}$, it means that we can find an element $A \in {\cal A}$ such that $|A\rangle = A |\Psi\rangle$.
We have
\be
A' |A\rangle = A' A \st = A A'\ \st =0,
\ee
where we used that $[A,A']=0$ and the assumption that $A'\st=0$. From this equation we find that $A'$ actually annihilates every vector in $\shil$, hence we find the operator equation
\be
A' = 0.
\ee

ii) Then, we will prove that the vector $\st$ is cyclic for the algebra ${\cal A'}$, which means that for every vector $|\Psi'\rangle \in \shil$, there is $A' \in {\cal A}'$ such that $|\Psi'\rangle = A' \st$. Define the space
\be
{\cal H}'_{\Psi} = {\cal A}' \st.
\ee
The space ${\cal H}'_{\Psi}$ is a subspace of ${\cal H}_{\Psi}$. What we need to prove is that actually ${\cal H}'_{\Psi} = {\cal H}_{\Psi}$. Define as $P$ the projection operator on ${\cal H}_{\Psi}'$. It is clear that $P$ commutes with all elements of ${\cal A}'$, hence $P \in ({\cal A}')' = {\cal A}$. This means that we have
\be
(\id - P)\st = 0,
\ee
and since $(\id-P) \in {\cal A}$ and since, by assumption, $\st$ is separating for ${\cal A}$ we find that $P = \id $, or ${\cal H}'_{\Psi} = \shil$, which shows that $\st$ is cyclic for the algebra ${\cal A}$.

\vskip20pt

\noindent Under the previous assumptions (i.e. that ${\cal A}$ is a von Neumann algebra acting on ${\cal H}_{\Psi}$ and that
$\st$ is cyclic and separating) we define the antilinear operator
$
S: {\cal H}_{\Psi}\,\rightarrow \, {\cal H}_{\Psi}
$
by
\be
S A \st = A^\dagger \st.
\ee
It is clear that 
\be
S^2 =1.
\ee 
and also 
\be 
S|\Psi\rangle = |\Psi\rangle.
\ee 
Since $S$ is antilinear, the Hermitian conjugate operator is defined by
\be
(|A\rangle, S^\dagger |B\rangle) = (|B\rangle, S |A\rangle).
\ee
\vskip20pt
\noindent We will now prove that for all $A' \in {\cal A}'$ we have
\be
\label{defsadjoint}
S^\dagger A' \st = (A')^\dagger \st.
\ee

\noindent {\bf Proof:}
Consider any state $|B\rangle\in \shil$ and multiply both sides  of \eqref{defsadjoint} with $\langle B|$. Since $\st$ is cyclic we can write $|B\rangle = B \st$ for $B \in {\cal A}$ and we have
\be
\langle B| S^\dagger A' \st = ( B\st, S^\dagger A' \st) = (A' \st, S B\st) = \stl (A')^\dagger B^\dagger \st
\ee
\be
= \stl B^\dagger (A')^\dagger \st = \langle B| (A')^\dagger \st.
\ee
which is indeed true.

\vskip20pt

\noindent  From the previous item it follows that
\be
S^\dagger |\Psi\rangle = |\Psi\rangle.
\ee
We define the linear operator $\Delta:{\cal H}_{\Psi}\,\rightarrow \, {\cal H}_{\Psi}$ by
\be
\Delta = S^\dagger S.
\ee
From the previous results it is obvious that
\be
\Delta \st = \st.
\ee
We show that $\Delta$ is Hermitian and positive (all eigenvalues strictly $>0$).

\noindent {\bf Proof:} Consider any two states $|A\rangle,|B\rangle$ of $\shil$ which can be written as $|A\rangle = A \st\,,\,
|B\rangle = B \st$, with $A,B\in {\cal A}$. We have
\be
\langle A| \Delta |B\rangle = (A \st, S^\dagger S B\st) = (A \st, S^\dagger B^\dagger \st) = (B^\dagger \st , S A \st) =
(B^\dagger \st A^\dagger \st),
\ee
or to summarize
\be
\langle A| \Delta |B\rangle = \stl B A^\dagger \st.
\ee
Similarly
\be
\langle B | \Delta |A\rangle = \stl A B^\dagger \st .
\ee
From which we see that $\Delta$ is Hermitian. To prove that $\Delta$ is {\it strictly} positive we notice that any state in $\shil$ can be written as $|A\rangle = A \st$ for some $A\neq 0$. We have
\be
\langle A | \Delta |A\rangle  = \stl A A^\dagger \st = ||A^\dagger \st||^2 >0,
\ee
since by assumption that $\st$ is separating, $A^\dagger \st \neq 0$.

\vskip20pt
\noindent Since $\Delta$ is Hermitian and strictly positive, it means we can define the inverse $\Delta^{-1}$ and all other powers $\Delta^z\,,\,z\in{\bf C}$.
\vskip20pt

\noindent Now we show that for any $A\in {\cal A}$  and any $A' \in {\cal A}'$ we have
\be
\label{sas}
S A S \in {\cal A}',
\ee
\be
\label{sdasd}
S^\dagger A' S^\dagger \in {\cal A}.
\ee
{\bf Proof:} Since $\st$ is cyclic for ${\cal A}$, any vector $|B\rangle$ in $\shil$ can be written as $|B\rangle = B \st$ with $B\in {\cal A}$. Consider any operator $C\in {\cal A}$. We have
\be
(S A S) C |B\rangle = SA B^\dagger C^\dagger \st = C B A^\dagger \st ,
\ee
and also
\be
C (S A S) |B\rangle = C S A B^\dagger \st = CB A^\dagger \st.
\ee
Hence
\be
[S A S,C]= 0,
\ee
as an operator, for all $C\in {\cal A}$, or $S A S \in {\cal A}'$. Similarly we prove $S^\dagger A' S^\dagger \in {\cal A}$. 
\vskip20pt

\noindent We consider the polar decomposition of $S$ as 
\be
\label{defsj}
S = J \Delta^{1/2}.
\ee
Here we define $\Delta^{1/2}$ to have positive eigenvalues.
We will now prove that
\be
\label{defsjb}
J \Delta^{1/2} = \Delta^{-1/2} J.
\ee

\noindent {\bf Proof:}
Since we defined $J = S \Delta^{-1/2}$ this can also be written as
\be
\label{alternativepolar}
\Delta^{1/2} S = S \Delta^{-1/2}.
\ee
Multiplying both sides from the left by $S^\dagger$ and using $S^\dagger S=\Delta$ we find that we have  to prove equivalently
\be
S^\dagger \Delta^{1/2}S = \Delta^{1/2}.
\ee
We argued before that the RHS of this equation is a positive operator. We briefly show that the LHS is also positive. For any
non-vanishing state $|A\rangle = A \st$ we notice that $S |A \rangle = A^\dagger \st$ is also non-vanishing. Hence
\be
(|A\rangle, S^\dagger \Delta^{1/2}S |A\rangle) = (A\st, S^\dagger \Delta^{1/2} A^\dagger \st) = 
( \Delta^{1/2} A^\dagger \st, S A \st) = \stl A \Delta^{1/2} A^\dagger \st >0,
\ee
since $\Delta^{1/2}$ is a strictly positive operator. This demonstrates that both sides of the equation \eqref{alternativepolar} are 
strictly positive. Hence to prove that equation, we can just check that the square of the equation is true. The square of the LHS is
\be
S^\dagger \Delta^{1/2}SS^\dagger \Delta^{1/2}S =S^\dagger \Delta^{1/2}\Delta^{-1} \Delta^{1/2}S  = S^\dagger S = \Delta,
\ee
which is the square of the RHS, as we wanted to prove.

\vskip20pt

\noindent Now we prove that for any $A \in {\cal A}$ we have
\be
\Delta A \Delta^{-1} \in {\cal A}.
\ee
{\bf Proof:} We have
\be
\Delta A \Delta^{-1}  = S^\dagger S A S S^\dagger  =  S^\dagger (S A S) S^\dagger .
\ee
From the relation \eqref{sas} we find that $S A S = A'$ for some $A' \in {\cal A}'$. But then from \eqref{sdasd} $S^\dagger A' S^\dagger \in {\cal A}$, as we wanted to prove.

\vskip20pt

\noindent By induction we can prove that
\be
\Delta^m A \Delta^{-m} \in {\cal A} \qquad m = 0,1,2,...
\ee

\vskip20pt

\noindent Actually we will now prove that
\be
\Delta^z A \Delta^{-z} \in {\cal A} \qquad, \qquad z\in {\bf C}.
\ee

\noindent{\bf Proof:} To do this, we will show that for any $z\in {\bf C}$ 
the operator $\Delta^z A \Delta^{-z}$ commutes with all elements of ${\cal A}'$ and hence it belongs to $({\cal A}')' = {\cal A}$.
Consider any elements $A' \in {\cal A}'$. We will prove that the commutator $[\Delta^z A \Delta^{-z},A']$ vanishes. Notice, we
have already proved that it vanishes when $z = {\rm positive\,\, integer}$. Consider the matrix elements of this commutator on any two states $|\Psi_1\rangle \,,|\Psi_2\rangle$. We define the function
\be
f(z) = {1\over ||\Delta||^{2z}}\langle \Psi_1| [\Delta^z A \Delta^{-z},A'] |\Psi_2\rangle.
\ee
Here we defined the norm $||\Delta||$ of the operator. Since $\Delta$ is a finite-dimensional (positive) matrix, the function $f(z)$ is a holomorphic function of $z$. It is zero at $z=0,1,2,...$ and does not grow too fast at infinity. Then by Carlson's theorem it is identically equal to zero. Hence for any $z$ and any $A' \in {\cal A}'$ we have $[\Delta^z A \Delta^{-z},A'] = 0$ and hence $\Delta^z A \Delta^{-z}
\in ({\cal A}')' = {\cal A}$, as we wanted to prove.
\vskip20pt 

\noindent This shows in particular that
\be
\label{dad}
\Delta^{1/2} A \Delta^{-1/2} \in {\cal A}.
\ee

\vskip20pt

\noindent If we remember  equations \eqref{defsj}, \eqref{defsjb}, we can write $J = \Delta^{1/2}S = S \Delta^{-1/2} $. Hence
\be
J A J  = \Delta^{1/2} S A S \Delta^{-1/2} \in {\cal A}.
\ee
by combining \eqref{sas} and \eqref{dad}
\be
J A J  \in {\cal A}'.
\ee
and similarly
\be
J A' J \in {\cal A}.
\ee
So if we define the mirror operators as
\be
\tilde{A} = J A J,
\ee
then we see that they commute with the original operators.
 
\noindent Moreover we can easily show that any element in ${\cal A}'$ can be written as $J A J$ for some $A\in {\cal A}$ hence the previous inclusions are actually equalities
\be
J {\cal A} J = {\cal A}',
\ee
\be
J {\cal A}' J = {\cal A}.
\ee
Let us also summarize the other important result we derived above
\be
\Delta^z {\cal A} \Delta^{-z} = {\cal A},
\ee
\be
\Delta^{z} {\cal A}' \Delta^{-z} = {\cal A}'.
\ee

\noindent The latter equations can be interpreted as follows. If we write $\Delta = e^{-K}$ these equations show that
\be
e^{i K t} {\cal A} e^{-i K t} = {\cal A}\qquad,\qquad e^{i K t} {\cal A}' e^{-i K t} = {\cal A}',
\ee
so the two algebras ${\cal A}, {\cal A}'$ are closed under ``time-evolution'' using the modular Hamiltonian $K$.


%% file: app_program.tex
\section{Numerically Computing the Mirror-Operators in the Spin Chain \label{appnumerical}}
 In the main text, we have proved that the mirror operators exist
under the appropriate conditions. Nevertheless, it is still fun to see this in an explicit numerical computation. The spin-chain provides us with a nice toy-model, in which we numerically compute the mirror-operators and examine their matrix elements.

We include a Mathematica program ``spinchaintildes.nb''
with the arXiv source of this paper, that performs this computation. Here, we provide a few comments to help the reader understand this program.

\paragraph{The Numbers Involved: \\}
In the text, we have proved that these mirror operators 
exist in the spin-chain provided that we take the number of insertions $K$,
so that we have
\be
\label{condexist}
{\cal D}_{\cal A} = \sum_{j=0}^K \begin{pmatrix} \nc \\ j \end{pmatrix} 3^j
 \leq 2^{\cal N}.
\ee

Note, that, in the text, we have always taken the dimension of the set ${\cal A}$ to be much smaller than that of the Hilbert space: ${\cal D}_{\cal A} \ll {\cal D}_{\cal H}$ to avoid issues with edge effects. However, as we see here, we actually need a much weaker condition, and in the case of the
spin-chain the precise condition is specified in \eqref{condexist}.

It is interesting to examine the numbers that are involved here. Even for $K = 2$, the first value of ${\cal N}$ for which ${\cal D}_{\cal A} \leq 2^{\cal N}$ is ${\cal N} = 9$. For ${\cal N} = 9$ and $K = 2$,
we have ${\cal D}_{\cal A} = 277$ compared to $2^9 = 512$. 

If we want to take $K = 3$, we see that we must take ${\cal N} \geq 14$. With ${\cal N} = 14$, we have ${\cal D}_{\cal A} = 10690$ compared to $2^{14} = 16384$. Since the algorithm below involves the inversion of a ${\cal D}_{\cal A} \times {\cal D}_{\cal A}$ matrix, we see that it rapidly
becomes expensive to find the tildes for higher values of $K$. 

\paragraph{Algorithm \\}
Now, we briefly describe the algorithm used to do the numerical computation.

First, we compute the set of all possible products of $\pault^i_a$
operators, up to $K$ operators. These products are put together in
an array, that we can call $\ell_\alpha$ here. It is clear that the index
$\alpha$ ranges from $1 \ldots {\cal D}_{\cal A}$.

Now, we take a state 
\be
|\Psi \rangle = \sum_B \alpha_B | B \rangle,
\ee
where the $\alpha_B$ are chosen to be arbitrary complex coefficients, 
satisfying $|\alpha_B|^2 = 1$.

Then, we generate the set of vectors
\be
|v_\alpha \rangle = \ell_\alpha |\Psi \rangle
\ee

Now, we have two choices. We can either compute the anti-linear map $S$, and 
then generate the mirror-operators, or else just solve the equations \eqref{rule1} and \eqref{rule2}. Computing $S$ might seem a little more efficient, because  we need to compute this anti-linear map only once,
and then evaluate \eqref{hatop}. (For the spin chain, $\Delta = 1$, and so $S = J$.)  However, since we need to consider $S \paul_a^i S \ell_i$, we
see that we need to compute the action of $S$ on a product of $K+1$-spin-operators. As we pointed out above, increasing $K$ is expensive, 
so in this program we simply compute the mirror-operators for
each $a,i$ separately.

Consider some particular $i_0, a_0$. So, we need to solve the equations
\be
\pault^{i_0}_{a_0} v_i = -\ell_i \paul^{i_0}_{a_0} |\Psi \rangle \equiv | u_i \rangle.
\ee
Now, precisely as in \eqref{tildedefgeneral} we consider the ``metric'' defined by
\be
g_{i j} = \langle v_i | v_j \rangle ,
\ee
and ``invert'' this metric to get $g^{j k}$ satisfying
\be
g^{j k} g_{k i} = \delta^j_i.
\ee
This is the numerically expensive step because this matrix is ${\cal D}_{A}$ dimensional.
In terms of this metric, precisely as in \eqref{talexplicit}, now we have simply
\be
\pault^{i_0}_{a_0} =  g^{j k} | u_j \rangle \langle v_k |.
\ee

The reader can experiment with this program. These explicit
numerical computations show, for example, that the commutator
of the mirror-operators with the ordinary-operators can be a 
rather complicated matrix, 
but precisely annihilates the state and its descendants produced
by acting with some number of ordinary operators.
